# Anomalous point-gap interactions unveil the mirage bath

Yue Sun,[1,2] Tao Shi,[3,4,*] and Ying Hu[1,2,†]

[1] *State Key Laboratory of Quantum Optics Technologies and Devices, Institute of Laser Spectroscopy, Shanxi University, Taiyuan 030006, China*

[2] *Collaborative Innovation Center of Extreme Optics, Shanxi University, Taiyuan 030006, China*

[3] *CAS Key Laboratory of Theoretical Physics, Institute of Theoretical Physics, Chinese Academy of Sciences, Beijing 100190, China*

[4] *School of Physical Sciences, University of Chinese Academy of Sciences, Beijing 100049, China*

Non-Hermitian topology has revolutionized our understanding of energy gaps and band topology, unveiling phases that do not exist within the Hermitian framework. Nonetheless, its fundamental implications for quantum interactions in open quantum systems remain largely unexplored. Here, we uncover an interaction mechanism by examining a quantum-optical system where quantum emitters interact through the photonic band gap of a dissipative photon bath with periodic boundaries, described by a nonreciprocal Su-Schrieffer-Heeger model. Although localized photons within the gap should inhibit interactions between emitters in certain regimes, we find that interactions emerge even over long distances, defying conventional expectations. These anomalous interactions are mediated by a "mirage bath"—a virtual bath that unfolds onto a distinct layer of the Riemann surface. This mirage bath generates emitter dynamics identical to those produced by the physical bath but possesses distinct band topology. Crucially, the interactions inherit the topology of the mirage bath, not the physical one. This bath duality is universal to any dissipative bath with point gaps, leading to a mechanism for interactions and correlations across the multilayered Riemann surface, unseen in traditional settings. We demonstrate that the topological mismatch between the dual baths arises necessarily from their equivalent observable dynamics. We further show how such dynamical equivalence enforces the unification of non-Hermitian topologies with different boundaries. Our findings open avenues in quantum optics, many-body quantum simulations, and offer fresh insights into non-Hermitian topology.

## I. INTRODUCTION

Interactions mediated by quantum baths are fundamental to a myriad of intriguing quantum many-body phenomena. Prominent examples include the electron-mediated Ruderman-Kittel-Kasuya-Yosida (RKKY) interaction [1–3]—a cornerstone of quantum magnetism—and phonon-mediated electron pairing that underpins BCS superconductivity [4]. The advent of nanophotonics [5,6] has extended this paradigm to horizons, where meticulously tailored photonic baths provide unprecedented opportunities for engineering interactions between quantum emitters (QEs), such as atoms or artificial atoms. This progress has unlocked regimes of physics involving long-range couplings between spins or photons, spanning applications from many-body quantum simulations [7–11] and quantum optics [12–16] to quantum information processing [5,17].

A leading development in this field involves QEs coupled to topological nanophotonic lattices [18–28], where atomic interactions are governed by the photonic bath's band topology. The key mechanism lies in the photonic band gap: An atom can localize [21,23,29–33] a topological photon mode within the gap, which mediates range-tunable interactions [6,7] with other atoms that inherit the photonic band topology (cf. Fig. 1). Such range-tunable topological interactions enable classes of many-body phenomena, like symmetry-protected frustrated magnetism [21,25], which are unattainable in conventional setups.

Recently, extensive studies in non-Hermitian topology of open systems [34–41] have unveiled topological phenomena arising from the complex-energy spectrum, beyond the Hermitian paradigm. Outstanding examples include the non-Hermitian skin effect [42], anomalous bulk-edge relations [43–49], and directional amplification [50]. The topological origin of these phenomena is intrinsically linked to the so-called point gaps in the complex spectrum under periodic boundary conditions (PBCs) and band winding on the complex-energy plane [41,51–54], which redefine traditional concepts like band gaps and expand the classifications of topological bands [39,55]. Given the pivotal role of the excitation spectrum in determining the physical properties of quantum matter, understanding how this spectral topology influences interacting open quantum systems is a critical next step toward uncovering quantum phenomena. Despite significant theoretical and experimental progress [28,47,56–63], how non-Hermitian topological properties can fundamentally alter quantum interaction mechanisms between particles remains largely unexplored.

*Contact author: tshi@itp.ac.cn

†Contact author: huying@sxu.edu.cn

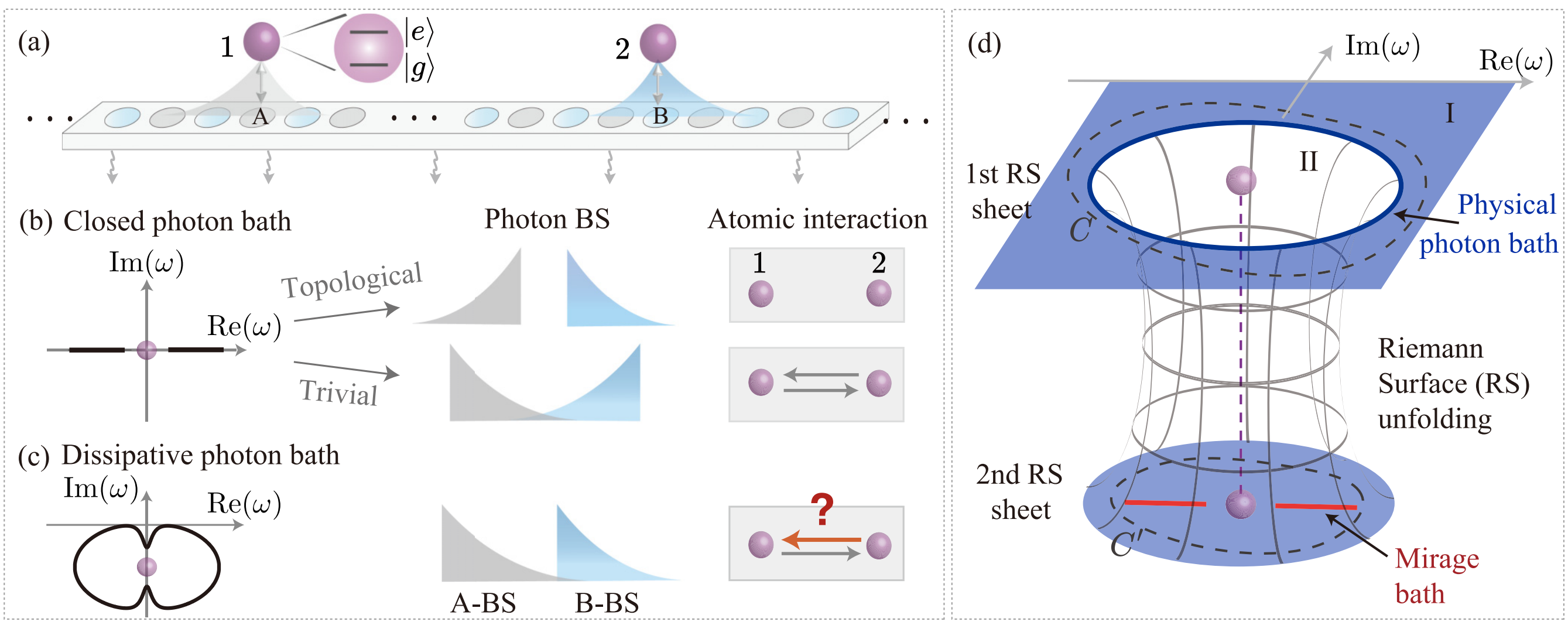

FIG. 1. Anomalous photon-mediated atomic interactions and the mirage bath. (a) Setup schematic: Two-level QEs, such as atoms or artificial atoms, are coupled to an engineered, dissipative photonic bath with PBCs effectively described by a nonreciprocal SSH model with sublattices $A$ and $B$. (b) Conventional topological interaction mediated by a closed bath (i.e., without dissipation). A closed photonic SSH bath exhibits real-energy bands. A QE embedded in the photonic gap seeds a chiral BS, where the photonic mode is localized asymmetrically around the atom. Depending on the QE coupled to sublattice $A$ or $B$, the resulting $A$-BS (gray) and $B$-BS (blue) exhibit opposite chirality, reflecting the bath's topology. These photon BSs mediate interactions with other QEs, which inherit photonic band topology. (c) Anomalous interaction in the nonreciprocal photon bath whose spectrum features point gaps. BSs in the point gap are homodirectional, i.e., irrespective of the sublattices—which should prevent the second QE from interacting with the first one. However, interactions emerge (red arrow). (d) Mirage bath on the multilayered RS of the complex frequency plane. In the first RS sheet, the self-energy due to the physical bath has distinct expressions inside (white region II) and outside (blue region I) of the branch loop (blue circle). The emitter dynamics are determined by the Fourier transformation of the Green function in region I. Through analytic continuation, the multilayered RS in the complex-energy plane is unfolded, and the physical self-energy in region I transitions to region II of the second RS layer (blue region). The branch cut on the second layer (red lines) defines the mirage bath, which produces the same emitter dynamics [cf. Eq. (11)], but with distinct topology.

Here, we demonstrate that non-Hermitian topology fundamentally reshapes bath-mediated interactions in open quantum systems with an engineered bath. We focus on a quantum-optical system of two-level QEs embedded within the complex band gap of a one-dimensional (1D) dissipative photonic lattice [Fig. 1(a)] described by a nonreciprocal Su-Schrieffer-Heeger (SSH) model with PBC. Remarkably, even though localized photonic modes within the gap should inhibit atomic interactions in certain regimes, we find the emergence of interactions that defy conventional expectations [Figs. 1(b) and 1(c)].

We find that this anomalous interaction is mediated by virtual photons from what we term a "mirage bath"—an effective bath that produces exactly identical emitter dynamics to the physical bath but possesses a drastically distinct topology. Crucially, the interactions inherit the topology of this mirage bath rather than that of the physical one. Through analytic continuation, we show how the mirage bath unfolds onto a different layer of the Riemann surface (RS) in the complex frequency plane [Fig. 1(d)]. The mirage bath is translationally invariant, yet it is topologically equivalent to the physical one with open-boundary conditions (OBCs) that has skin modes. Moreover, we demonstrate that the topological mismatch between the mirage and physical baths emerges necessarily from their identical observable dynamics.

We show that this duality between the physical and mirage baths is *universal* for any dissipative bath with point gaps, independent of the presence of QEs. It results in an interaction mechanism across multilayered RS not present in conventional (closed) settings; namely, nonlocal interactions and correlations are governed by virtual excitations of the mirage bath on the RS. This intrinsic multilayered nature of the interaction, rooted in the point-gap topology of the bath's non-Hermitian spectrum, fundamentally distinguishes it from traditional interaction mechanisms through a closed bath, opening avenues for exotic quantum matter from interacting spins, photons, or phonons.

Beyond interest in the interaction mechanisms *per se*, our work has remarkable implications for non-Hermitian topology, quantum many-body physics, and experiments:

(1) We show that preserving identical observable dynamics enforces the unification of non-Hermitian PBC and OBC topologies—through a pole selection rule that retains only those poles governing physical dynamics. Existing studies typically rely on non-Bloch theories [39,41,42,52,54,64] to address the open-boundary systems, as conventional bulk-edge correspondence breaks down due to skin effects. Here, we demonstrate that their distinct topologies are necessitated by their dynamical equivalence, thus enabling prediction of OBC topological invariants solely from PBC properties.

(2) Our work introduces a highly efficient approach to study few- and many-body physics in open quantum systems with non-Hermitian topological properties. Harnessing non-Hermitian topology in genuinely quantum regimes demands a fully quantum treatment of long-time dynamics of interacting open quantum systems, which is challenging, particularly under OBC and with external driving. The mirage bath offers a systematic solution to this challenge by circumventing complexities of non-Bloch methods, allowing direct applications of established concepts (e.g., density of states) and tools from many-body physics and field theory to the study of quantum non-Hermitian phenomena. In our work, we demonstrate this advantage explicitly by studying the generation of sub-Poissonian light in a driven nonlinear emitter coupled to a dissipative photonic SSH lattice. Here, the quantum correlated statistics are characterized by the second-order correlations of photons.

(3) Our finding indicates the intriguing possibility to experimentally probe distinct non-Hermitian topologies under different boundaries—all within a *single* setup. Namely, a single QE coupled to PBC bath probes the PBC topology, whereas interactions between two QEs encode the OBC topology despite the PBC bath. This dual ability is highly appealing for ongoing experiments aimed at exploring topological effects in synthetic quantum materials with tunable dissipations [47,59,62,63,65–68].

## II. BACKGROUND: TOPOLOGICAL INTERACTION MEDIATED BY CLOSED BATHS

To motivate our discussion, we begin by briefly summarizing the key physics for interactions between two-level QEs embedded in the band gap of a closed, photonic SSH bath [6,21]. The total Hamiltonian is given by

$$H_0 = H_{\mathrm{a}} + H_{\mathrm{b}} + H_{\mathrm{ab}}, \tag{1}$$

where the bath Hamiltonian $H_{\mathrm{b}} = \sum_j (J_1 b^\dagger_{A,j} b_{B,j} + J_2 b^\dagger_{B,j} b_{A,j+1} + \mathrm{H.c.})$ represents a photonic SSH model under PBC. Here, $b_{A/B,j}$ annihilates the photonic mode at sublattice $A/B$ in unit cell $j$, and $J_1$ and $J_2$ denote alternating couplings. For simplicity, we assume $J_{1,2} \geqslant 0$ henceforth. This bath exhibits gapped energy bands, $\omega_b(k) = \pm\sqrt{(J_1 + J_2\cos k)^2 + J_2^2 \sin^2 k}$, with the quasimomentum $k \in [-\pi, \pi)$. The bath is topologically nontrivial for $J_1/J_2 < 1$ and trivial for $J_1/J_2 > 1$, with a phase transition occurring at the gap-closing point $J_1/J_2 = 1$. The QEs are described by $H_a = \Delta \sum_{m=1}^{N_a} \sigma^m_{ee}$, where $\sigma^m_{\mu\nu} = |\mu\rangle_m\langle\nu|$ operates on the internal states of the $m$th QE and $\Delta$ is the detuning of the QE's transition frequency relative to the central frequency of the bath. The emitter-bath coupling is $H_{\mathrm{ab}} = \Omega \sum_m (\alpha^\dagger_{j_m} \sigma^m_{ge} + \mathrm{H.c.})$, where $\Omega$ is the Rabi coupling strength and $\alpha_{j_m} \in \{b_{A,j_m}, b_{B,j_m}\}$ annihilates the photon in the unit cell $j_m$ to which the $m$th QE is coupled.

For a single QE in the single-excitation sector, when $\Delta$ lies within the photonic band gap, a chiral BS is formed [21,23], Appendix A, whose photonic component is localized predominantly to the left or right of the QE depending on the bath's topology [Fig. 1(b) and Appendix A]. The photon BS for a QE coupled to sublattice $A/B$ (denoted as $A/B$-BS, respectively) possesses two key properties [Fig. 1(b)]: (1) The chirality of the $A$-BS and $B$-BS are always opposite, and (2) both reverse chirality at the phase-transition point $J_1/J_2 = 1$.

These chiral BSs facilitate interactions with other QEs that inherit a topological nature. For instance, consider two QEs with $\Delta = 0$, where they interact only if coupled to different sublattices. For one QE at $j_1 = 0$ on sublattice $A$ and another at $j_2 = d > 0$ downstream on sublattice $B$, their effective interaction is given by [6,21]

$$H_{\mathrm{int}} = \Sigma^{AB}_d \sigma^1_{eg} \sigma^2_{ge} + \mathrm{H.c.}, \tag{2}$$

with the interaction strength [Appendix A]

$$\Sigma^{AB}_d = \begin{cases} -\frac{\Omega^2}{J_1}\left(-\frac{J_2}{J_1}\right)^d, & J_1 > J_2, \\ 0, & J_1 < J_2. \end{cases} \tag{3}$$

This represents a topological interaction whose presence or absence depends on the bath's band topology. For a chain of two-level QEs, such topological interactions give rise to exotic many-body phases [21].

In summary, QEs in a closed SSH bath seed topological photon BSs within the photonic band gap, which, in turn, mediate topological interactions between QEs.

## III. ANOMALOUS BAND-GAP INTERACTION IN A DISSIPATIVE BATH

Here, we uncover an anomalous interaction between QEs embedded in a dissipative SSH bath, which challenges the conventional bath-mediated interaction mechanism described previously.

Our system comprises QEs coupled to an SSH photon bath, which itself interacts with an external environment and dissipates [see Fig. 1(a)]. The density matrix $\rho$ of the QEs and the photon bath as a whole obeys the master equation [69]:

$$\dot{\rho} = -i[H_0, \rho] + \sum_j \frac{\gamma_b}{2}\mathcal{D}_{\mathrm{b}}[l_j]\rho + \sum_m \frac{\gamma_a}{2}\mathcal{D}_a\big[\sigma^m_{ge}\big]\rho. \tag{4}$$

Here, the first dissipator, $\mathcal{D}_b[l_j]\rho = 2 l_j \rho l^\dagger_j - \{l^\dagger_j l_j, \rho\}$, models intrinsic dissipation in the bath, where $l_j = -i b_{A,j} + b_{B,j}$ dissipatively couples intracell photonic modes with rate $\gamma_b/2$. Such dissipative coupling has been recently realized in both photonic and atomic setups [59,62,70–74]. Additionally, we include the free-space emission of QEs at rate $\gamma_a/2$, represented by the second dissipator $\mathcal{D}_a$, to account for typical free-space emission in realistic systems.

The QEs' dynamics, governed by Eq. (4), can be exactly solved using the Green function or the resolvent approach [6,61], which are determined solely by the effective Hamiltonian $H_{\mathrm{eff}} = H'_a + H'_{\mathrm{b}} + H_{\mathrm{ab}}$. Here, $H'_a = \Delta' \sum_m \sigma^m_{ee}$ with the effective detuning $\Delta' = \Delta - i\gamma_a/2$. The effective bath Hamiltonian $H'_{\mathrm{b}} = H_b + (\gamma_b/2)\sum_j [-i b^\dagger_{A,j} b_{A,j} - i b^\dagger_{B,j} b_{B,j} + (b^\dagger_{A,j} b_{B,j} - \mathrm{H.c.})]$ represents the well-studied nonreciprocal SSH model with PBC [42,44,46]. Note that $H_{\mathrm{eff}}$ commutes with the number operator of excitations.

Crucially, the nonreciprocal photonic SSH bath exhibits non-Hermitian Bloch bands:

$$\omega'_{\mathrm{b}}(k) = -i\frac{\gamma_b}{2} \pm \sqrt{(J_1 + J_2\cos k)^2 + \left(J_2 \sin k + i\frac{\gamma_b}{2}\right)^2}, \tag{5}$$

which form loops in the complex frequency plane, giving rise to two distinct types of band gaps [40,41]: the point gap, encompassing the spectral area enclosed by a loop, and the line gap, acting as a line separating different bands in the frequency plane. While the line gap bears similarities to energy gaps in Hermitian bands, the point gap is uniquely a non-Hermitian property.

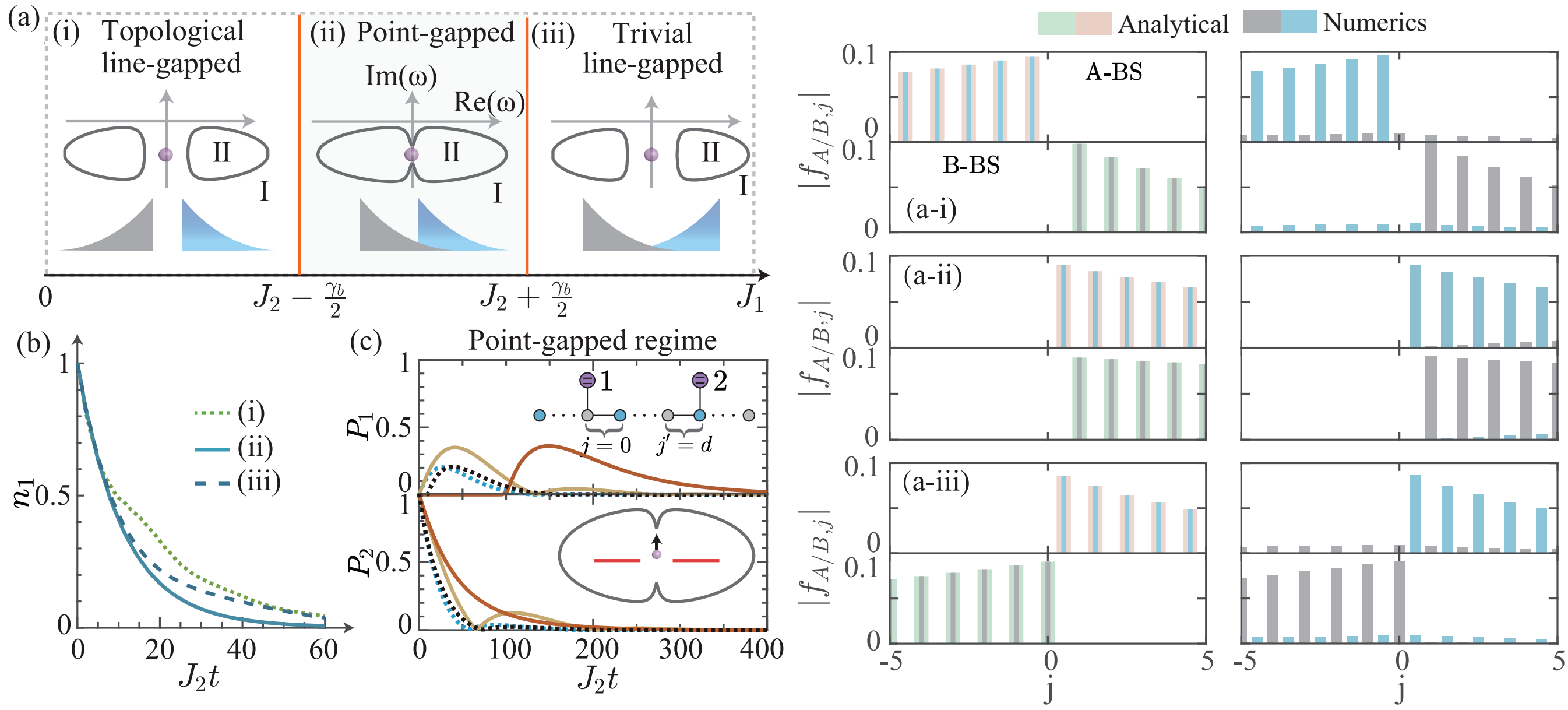

FIG. 2. Photon BSs and point-gap interaction in a dissipative photon bath. (a) Depending on whether the QE is embedded in the line gap or point gap, the bath is divided into three topologically distinct phases, including the point-gap phase that has no counterpart in closed baths. Photon BSs in the complex band gap remain to inherit the bath topology. Panels (a-i)–(a-iii) show the photonic distribution $|f_{A/B,j}|$ of the photon BS, for $\Delta = 0$ in the left column and $\Delta \neq 0$ in the right column, respectively, when $\gamma_a/J_2 = \gamma_b/J_2 = 0.1$ and $\Omega/J_2 = 0.1$. Results are shown for (a-i) $J_1/J_2 = 0.9$ in the topological line-gapped regime (i), (a-ii) $J_1/J_2 = 1.03$ in the point-gapped regime (ii), and (a-iii) $J_1/J_2 = 1.1$ in the trivial line-gapped regime (iii). Both $A$-BS and $B$-BS are illustrated. Analytical results for $\Delta = 0$ are derived via the Green function approach (see Table II). Numerical results are obtained through diagonalizing the effective Hamiltonian (B4) with a single QE and a nonreciprocal SSH bath of size $N_b = 500$ under PBC. (b) Single-QE dynamics. Numerical results of the probability $n_1 = |G_0(t)|^2$ to find a single QE in the excited state are shown for $J_1/J_2 = 0.7$ in regime (i), $J_1/J_2 = 1.02$ in regime (ii), and $J_1/J_2 = 1.1$ in regime (iii). Other parameters: $\gamma_a/J_2 = \gamma_b/J_2 = 0.05$, $\Omega/J_2 = 0.2$, and $\Delta/J_2 = 0$. (c) Dynamics of two QEs in the point-gap regime. The inset depicts the configuration of QEs, coupled to sublattices $A$ and $B$, respectively, and separated by $d$ unit cells. Initially, the second QE is excited while the first is in the vacuum. The top (bottom) panel shows the probability amplitude $P_{1(2)}$ to find the first (second) QE in the excited state. Dynamics are exactly computed using the Green function approach. Results are shown for $\gamma_a/J_2 = 0.005$, $J_1/J_2 = 1.01$ when $d = 0$ (yellow curves) and $d = 100$ (red curves), and for $\gamma_a/J_2 = 0.05$, $J_1/J_2 = 1.02$ when $d = 0$ (blue curves) and $d = 10$ (black curves). For other parameters in panel (c), $\gamma_b/J_2 = 0.05$, $\Omega/J_2 = 0.2$, and $\Delta/J_2 = 0$.

Thus, different from the aforementioned closed bath, the dissipative photon bath is classified into three topologically distinct phases [Fig. 2(a)]: (1) Topological line-gapped phase ($J_1 < J_2 - \gamma_b/2$), (2) point-gapped phase ($J_2 - \gamma_b/2 < J_1 < J_2 + \gamma_b/2$), and (3) trivial line-gapped phase ($J_1 > J_2 + \gamma_b/2$). In the line-gapped phases (1) and (3), the two bands are separated by a line gap, and the band topology is characterized by the Zak phase, similar to the closed bath. The point-gapped phase (2), however, lacks a Hermitian counterpart, where the two bands merge to form a single point gap. The gap-closing points, $J_1 = J_2 \pm \gamma_b/2$, correspond to the transitions between the line- and point-gapped phases.

We aim to study the behavior of QEs within the band gap of this dissipative photonic bath. To this end, we assume $\Delta' = -i\gamma_a/2$ without loss of generality, which lies either in the line gap or the point gap depending on the value of $J_1/J_2$ [Fig. 2(a)]. This represents the simplest situation capturing all essential physics, and we refer readers to Appendix B for other cases.

We first consider a single excited QE. In the single-excitation sector, the spontaneous emission dynamics can be exactly solved through the Fourier transform of the Green function [Appendix B]

$$G_0(\omega) = \frac{1}{\omega - \Delta' - \Sigma_0(\omega)}, \tag{6}$$

where $\Sigma_0(\omega) = \Omega^2 \int_{-\pi}^{\pi} \frac{dk}{2\pi} \frac{\omega + i\gamma_b/2}{(\omega - \omega'_{b+})(\omega - \omega'_{b-})}$ is the self-energy due to the dissipative photon bath. Given the photonic spectrum in Eq. (5), Eq. (6) exhibits branch loops [58,61], $\omega = \omega'_b(k)$. Importantly, the self-energy exhibits distinct expressions [Appendix B] outside and inside the branch loop [denoted as I and II, respectively; see Fig. 2(a)]. In region I, we find

$$\Sigma_0(\omega) = \frac{\Omega^2\left(\omega + \frac{i\gamma_b}{2}\right)\mathrm{sign}(|z_-| - |z_+|)}{\sqrt{\left[\left(\omega + i\frac{\gamma_b}{2}\right)^2 - \sigma_1\right]^2 - J_2^2\left(4J_1^2 - \gamma_b^2\right)}}, \quad \omega \in \mathrm{I}, \tag{7}$$

where $\sigma_1 = J_1^2 + J_2^2 - \gamma_b^2/4$ and $z_\pm$ denote the poles of $\Sigma_0(\omega)$. However, in region II, we have $\Sigma_0(\omega) = 0$. The real-time dynamics of a QE is finally given by

$$G_0(t) = \int_C \frac{d\omega}{2\pi} G_0(\omega) e^{-i\omega t}, \tag{8}$$

where $C \in \mathrm{I}$ is a contour outside the branch loops [cf. Fig. 1(d)]. Figure 2(b) shows typical results of the probability $n_1(t) = |G_0(t)|^2$ for an initially excited QE across phases (1)–(3) of the dissipative SSH bath, which appears insensitive to photonic band topology.

However, the BSs within the band gap still faithfully manifest the band topology—similar to the closed-bath case—despite it being fundamentally shaped by dissipation [Fig. 2(a)]. The BS energy, $\omega_{\mathrm{BS}}$, is determined by the pole of Eq. (6), $G_0^{-1}(\omega_{\mathrm{BS}}) = 0$. The corresponding eigenstate $|B\rangle = (\varphi_a \sigma_{eg} + \sum_{\alpha=A/B,j} f_{\alpha,j} b_{\alpha,j}^\dagger)|g\rangle|0\rangle$ remains chiral, with the amplitudes $\varphi_a$ and $f_{\alpha,j}$ of the atomic and photonic components derived in Appendix B. This is best exemplified with $\gamma_a = \gamma_b$, i.e., $\Delta' = -i\gamma_b/2$. Here, regardless of the sublattice to which the QE is coupled, a BS with $\omega_{\mathrm{BS}} = -i\gamma_b/2$ always appears, but its wavefunction differs for the $A$-BS and $B$-BS. Specifically, for the $A$-BS, right-side localization occurs when $J_1 > J_2 - \gamma_b/2$, with the photonic amplitude $f_{A,j} = 0$, $f_{B,j<0} = 0$, and

$$f_{B,j>0} = \frac{\Omega\varphi_a}{J_2}\left(-\frac{J_2}{J_1 + \frac{\gamma_b}{2}}\right)^{j+1}, \quad J_1 > J_2 - \gamma_b/2.$$

In contrast, for the $B$-BS, right-side localization occurs if $J_1 < J_2 + \gamma_b/2$, with $f_{B,j} = 0$, $f_{A,j<0} = 0$, and

$$f_{A,j>0} = \frac{\Omega\varphi_a}{J_1 - \frac{\gamma_b}{2}}\left(-\frac{J_1 - \frac{\gamma_b}{2}}{J_2}\right)^{j}, \quad J_1 < J_2 + \gamma_b/2.$$

Figure 2(a) summarizes the configurations of photon BSs across phases (1)–(3) of the bath, revealing their direct correspondence to the underlying non-Hermitian band topology.

On the other hand, the BSs exhibit two distinct features from closed baths: (1) The $A$-BS and $B$-BS switch chirality at separate critical points: the $A$-BS at $J_1 = J_2 - \gamma_b/2$ and the $B$-BS at $J_1 = J_2 + \gamma_b/2$. (2) BSs in the point gap align homodirectionally irrespective of their sublattice association. This is in stark contrast to the BSs in the line gap, which orient either toward or away from each other, much like those in a closed bath.

We therefore focus on the unique point-gap regime, $J_2 - \gamma_b/2 < J_1 < J_2 + \gamma_b/2$, to investigate the interaction between QEs (analysis in line gap can be found in Appendix B 2). Since photon BSs exhibit homodirectional, right-side localization, excitations on downstream QEs could not hop upstream via the BS, thus inhibiting backward interactions [cf. Fig. 1(c)]. For two QEs, one in sublattice $A$ at unit cell $j_1 = 0$ and the second in sublattice $B$ at $j_2 = d > 0$, standard analysis within the single-pole approximation $\omega \approx \omega_{\mathrm{BS}}$ yields an effective interaction:

$$H_{\mathrm{int}} = \Sigma_d^{AB} \sigma_{eg}^1 \sigma_{ge}^2 + \Sigma_d^{BA} \sigma_{eg}^2 \sigma_{ge}^1, \tag{9}$$

where $\Sigma_d^{AB}$ and $\Sigma_d^{BA}$ represent the backward and forward interaction strengths, respectively. As detailed in Appendix B 2, for the example $\Delta' = -i\gamma_b/2$, one finds

$$\Sigma_d^{AB} = \begin{cases} 0, & J_1 < J_2 + \gamma_b/2, \\ \frac{\Omega^2}{J_2}\left(-\frac{J_2}{J_1 - \frac{\gamma_b}{2}}\right)^{d+1}, & J_1 > J_2 + \gamma_b/2, \end{cases} \tag{10}$$

suggesting that backward interactions should be absent.

However, the exact two-QE dynamics [Fig. 2(c)] do not align with this expectation, revealing an unexpected phenomenon: a downstream QE can indeed interact with an upstream one. In Fig. 2(c), the second QE at $d > 0$ is initially excited with a single excitation while the first QE starts in the vacuum state. Using the Green function approach [Appendix B], we numerically compute the probability amplitude $P_{1(2)}$ for finding the two QEs, respectively, in the excited state. Counterintuitively, excitation of the upstream QE is observed. This is especially prominent for $\gamma_a \ll \gamma_b$ (solid curves), when $\Delta'$ is near the top edge of the photonic point gap. Remarkably, we observe strong oscillations between the two QEs even for sufficiently large separations ($d \gtrsim 100$; see red curve). This oscillatory behavior indicates a bidirectional exchange of trapped, virtual photons and thus the presence of a finite backward interaction—contrary to the expectation from Eq. (9).

If photon BSs cannot contribute, then what, exactly, is the nature of this anomalous interaction?

## IV. MIRAGE BATH ON MULTILAYERED RS

As we will demonstrate, the aforementioned anomalous interaction is mediated by virtual photons from a "*mirage bath,*" rather than the physical bath. Remarkably, this mirage bath generates exactly identical emitter dynamics to the physical bath, yet exhibits strikingly different topological phases. We refer to it as the mirage bath because it unfolds on a higher RS layer of the complex frequency plane [Fig. 1(d)], serving as a dual to the physical bath.

We begin by mathematically demonstrating the appearance of the mirage bath through analytic continuation. The key insight is that the dynamics of QEs are mathematically determined by a contour integral in the complex frequency plane. The contour $C$ is confined to region I [see, e.g., Eq. (8)], which cannot be contracted across the branch loop (blue circle). However, by performing an analytic continuation, we can unfold the complex plane into a multisheeted RS, allowing the physical self-energy in region I of the first sheet to transition smoothly into region II on the second sheet (purple area). For instance, for a single QE, we can rewrite Eq. (8) as [Appendix C]

$$G_0(t) = \oint_{C\in\mathrm{I}} \frac{d\omega}{2\pi} G_0(\omega) e^{-i\omega t} \equiv \oint_{C'\in\mathrm{II}} \frac{d\omega}{2\pi} G_f(\omega) e^{-i\omega t}. \tag{11}$$

Here, $C'$ is a contour in region II of the second RS sheet, and the Green function is defined by

$$G_f(\omega) = \frac{1}{\omega - \Delta' - \Sigma_0^f(\omega)}, \tag{12}$$

where the self-energy $\Sigma_0^f(\omega)$ is the *analytic continuation* of the physical self-energy in Eq. (7) from region I to II:

$$\Sigma_0^f(\omega) = \frac{\Omega^2\left(\omega + \frac{i\gamma_b}{2}\right)\mathrm{sign}(|z_-| - |z_+|)}{\sqrt{\left[\left(\omega + i\frac{\gamma_b}{2}\right)^2 - \sigma_1\right]^2 - J_2^2\left(4J_1^2 - \gamma_b^2\right)}},$$

$$\omega \in \mathrm{I} + \mathrm{II}, \tag{13}$$

where $\sigma_1 = J_1^2 + J_2^2 - \gamma_b^2/4$.

Unlike $G_0(\omega)$ of the physical bath [Eq. (6)], which takes distinct expressions for $\omega$ outside and inside its branch loop, $G_f(\omega)$ [Eq. (12)] contains a unified self-energy $\Sigma_0^f(\omega)$ across the entire frequency plane, thus exhibiting only branch linecuts (with branch points) at $\omega = \omega_b^f(k)$ [red line in Fig. 1(d)]. We interpret these linecuts as representing the continuous spectrum of a mirage bath and interpret $G_f(\omega)$ as corresponding to an emitter coupled to this mirage bath with coupling rate $\Omega$. The explicit forms of $\omega_\mathrm{b}^f(k)$ can be obtained by observing that $\Sigma_0^f$ and its closed-bath counterpart [Eq. (A12)] share a similar functional form under the substitution: $\omega \to \omega + i\gamma_b/2$ and $J_1^2 \to J_1^2 - \gamma_b^2/4$, indicating a dissipation-renormalized coupling $\tilde{J}_1^2 = J_1^2 - \gamma_b^2/4$. Applying this transformation to $\omega_b(k) = \pm\sqrt{(J_1 + J_2\cos k)^2 + J_2^2\sin^2 k}$ of the Hermitian SSH model, we obtain

$$\omega_\mathrm{b}^f(k) = -i\frac{\gamma_b}{2} \pm \sqrt{(\tilde{J}_1 + J_2\cos k)^2 + J_2^2\sin^2 k}. \tag{14}$$

A more rigorous approach can be found in Appendix C.

We emphasize that the above analytic continuation is feasible because a branch loop, unlike branch cuts, does not possess branch points. This crucial feature allows us to mathematically replace the original contour integration in the frequency plane with branch cuts and poles on the higher RS sheet. Physically, this means that the real-time dynamics of QEs driven by the physical and mirage baths are effectively identical, as confirmed numerically in Fig. 3(a). (The case with a chain of QEs can be found in Appendix C).

Despite identical dynamics, the phase diagrams of the physical and mirage baths are drastically different [Fig. 3(b)]. The spectrum (14) now contains only line gaps, which close at $J_1 = \sqrt{J_2^2 + \gamma_b^2/4}$, corresponding to $\tilde{J}_1^2 = J_2^2$. Without point gaps, the mirage bath is topologically characterized in a conventional manner by the Zak phase, which is nontrivial for $J_1 < \sqrt{J_2^2 + \gamma_b^2/4}$ and trivial for $J_1 > \sqrt{J_2^2 + \gamma_b^2/4}$.

This distinction leads to a stark contrast in the regime $\sqrt{J_2^2 + \gamma_b^2/4} < J_1 < J_2 + \gamma_b/2$ [red region in Fig. 3(b)]: the mirage bath is in a trivial line-gapped phase, whereas the physical bath resides in a point-gapped phase. Nevertheless, both baths drive identical dynamics for emitters.

We now show that the anomalous point-gap interaction between QEs actually inherits the topology of the mirage bath, rather than the physical one. Intuitive insight can be obtained from the perspective of an observer on the second sheet of the RS. There, a QE with $\Delta' = -i\gamma_a/2$ is situated in the *line gap* of the mirage bath [Fig. 1(d)], which, for $\sqrt{J_2^2 + \gamma_b^2/4} < J_1 < J_2 + \gamma_b/2$, is in the trivial phase. Consequently, the $A$-BS and $B$-BS on the second RS sheet are expected to orient toward each other as in closed baths, allowing QEs [Fig. 2(c) inset] to interact. Thus, the cascaded topological inheritance linking atomic interaction and photonic bands, broken on the first RS sheet, is restored on the second layer.

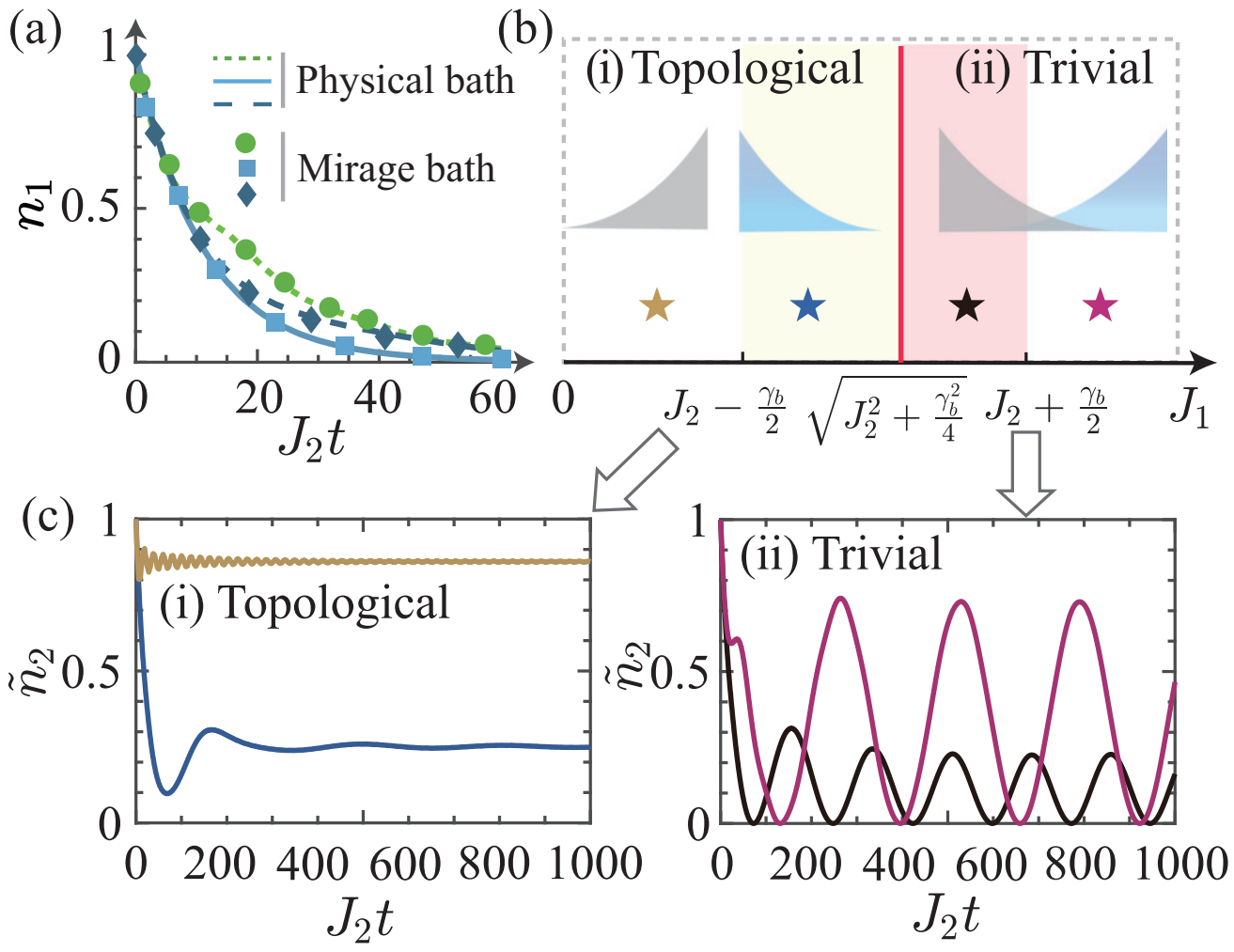

FIG. 3. Point-gap interaction reflects the mirage bath's topology. (a) Equivalent QE dynamics generated by the physical and mirage baths. The same parameters are used as in Fig. 2(b). (b) Topological phase diagram of the mirage bath. The colored band denotes where the mirage and the physical baths have different topologies. (c) Renormalized emission dynamics, $\tilde{n}_2(t) = e^{\gamma_b t} n_2(t)$, of the downstream QE reveal topology of the mirage bath. The left panel shows dynamics for $J_1/J_2 = 0.7, 0.98$ (brown, blue). The right panel shows dynamics for $J_1/J_2 = 1.02, 1.1$ (black, red). For other parameters, $\gamma_a/J_2 = \gamma_b/J_2 = 0.05$, $\Omega/J_2 = 0.2$, and $\Delta/J_2 = 0$, $d = 10$.

Guided by this insight, we calculate the photon BS and the interaction strength in the regime $\sqrt{J_2^2 + \gamma_b^2/4} < J_1 < J_2 + \gamma_b/2$ using the *mirage bath*. The energy of a single-QE BS is found from Eq. (12), $G_f^{-1}(\omega_\mathrm{BS}) = 0$, and the corresponding wavefunction can be derived similarly to previous methods [Appendix C]. The expressions simplify for $\Delta' = -i\gamma_b/2$, where the energy of a BS is $\omega_\mathrm{BS} = -i\gamma_b/2$, independent of sublattices. For $\sqrt{J_2^2 + \gamma_b^2/4} < J_1 < J_2 + \gamma_b/2$, the photon $A$-BS is localized to the right, with $f_{A,j} = 0$, $f_{B,j<0} = 0$, and

$$f_{B,j\geqslant 0} = (-1)^{j+1}\frac{\Omega\varphi_a}{\sqrt{J_1^2 - \gamma_b^2/4}}\left(\frac{J_2}{\sqrt{J_1^2 - \gamma_b^2/4}}\right)^j. \tag{15}$$

In contrast, the photon $B$-BS localizes to the left, with $f_{B,j} = 0$, $f_{A,j>0} = 0$, and

$$f_{A,j\leqslant 0} = (-1)^{|j|+1}\frac{\Omega\varphi_a}{\sqrt{J_1^2 - \gamma_b^2/4}}\left(\frac{J_2}{\sqrt{J_1^2 - \gamma_b^2/4}}\right)^{|j|}. \tag{16}$$

The BS shapes in different regimes of the mirage bath are schematically shown in Fig. 3(b). In this configuration, QEs on different sublattices can interact via the photon BS. Within the single-pole approximation $\omega \approx \omega_\mathrm{BS}$, the interaction Hamiltonian of two QEs takes the form

$$H_{\text{int}} = \Sigma^{AB}_{d,f}\sigma^1_{eg}\sigma^2_{ge} + \Sigma^{BA}_{d,f}\sigma^2_{eg}\sigma^1_{ge}, \tag{17}$$

where the backward interaction strength is

$$\Sigma^{AB}_{d,f} = \begin{cases} -\frac{\Omega^2}{J_1-\frac{\gamma_b}{2}}\left(\frac{-J_2}{J_1-\frac{\gamma_b}{2}}\right)^d, & J_1 > \sqrt{J_2^2+\gamma_b^2/4}, \\ 0, & J_1 < \sqrt{J_2^2+\gamma_b^2/4}. \end{cases} \tag{18}$$

We refer to Appendix C for detailed calculations of both backward and forward interactions. Equation (18) indicates the presence of backward interaction when $\sqrt{J_2^2+\gamma_b^2/4} < J_1 < J_2 + \gamma_b/2$, which is facilitated by the mirage bath.

This prediction is numerically confirmed in Fig. 3(c). There, we plot the renormalized population dynamics $\tilde{n}_2(t) = e^{\gamma_b t}n_2(t)$ of the second QE, initially excited, taking the real-time dynamics of $n_2(t)$ with $\gamma_a = \gamma_b$ [cf. Fig. 2(c)]. The dynamics clearly distinguish between the two distinct regimes relative to the phase boundary $J_1 = \sqrt{J_2^2+\gamma_b^2/4}$ of the mirage bath. In the trivial regime ($J_1 > \sqrt{J_2^2+\gamma_b^2/4}$), shown in the right panel of Fig. 3(c), the dynamics show persistent, slow oscillations with a frequency $\propto \Omega^2/J_1 \ll 1$, as expected from Eq. (18). In contrast, the topological phase (left panel) shows no oscillations at the timescale $\sim J_1/\Omega^2$, highlighting the stark difference in dynamics. Note that fast oscillations (brown curve) appear in the trivial regime, with frequency $\sim|J_2 - \tilde{J}_1|$ determined by the gap of the mirage bath.

Using the above-described bath duality, we can also explain the observation in Fig. 2(c) where oscillation amplitude increases as the effective detuning $\Delta' = -i\gamma_a/2 \ll -i\gamma_b/2$ approaches the upper edge of the point gap. From the perspective of the mirage bath on the second RS sheet, $\Delta'$ in this case is actually the farthest from the mirage photonic band edge [inset of Fig. 2(c)], so that the resulting BS is predominantly an atomic excitation with a small photonic component, leading to enhanced transfer probability between QEs. While anomalous interactions can in principle be analyzed through the physical bath, insight into their topological nature may be obscured.

Thus, we conclude that the point-gap interaction between QEs occurs via the exchange of virtual photons of the mirage bath, inheriting its topology rather than that of the physical bath. Emerging on a different RS layer of the complex frequency plane, the mirage bath acts as a dual to the actual bath in the sense that they produce same emitter dynamics but exhibit distinct properties. This multilayered nature of the interaction is rooted in the point-gap topology of the bath's non-Hermitian spectrum, fundamentally distinguishing it from traditional interactions mediated by a closed bath.

## V. MIRAGE BATH FOR ARBITRARY 1D BATH WITH POINT-GAP TOPOLOGY

In general, any PBC bath with a complex spectrum featuring point gaps has a corresponding mirage on a different sheet of the RS in the complex frequency plane. Intriguingly, however, its excitation spectrum matches the bulk spectrum of the physical bath with OBCs, even though the latter exhibits skin modes. This spectral equivalence is not coincidental: in the thermodynamic limit, a photon emitted by a single QE takes an infinite amount of time to return; consequently, the emitter dynamics are not affected by the bath's boundary conditions. This boundary insensitivity of emitter dynamics has been noticed in previous works [60,75]. We refer the readers to Appendix E for further comparisons between the mirage bath and the physical bath with OBC.

Yet, we emphasize that the mirage bath is an intrinsic feature of any physical bath with spectral topology, functioning as its dual, independent of the presence of QEs. Below we provide a general definition for the mirage bath (Sec. V A). Then, we systematically derive its spectrum and density of state (DOS) (Sec. V B), showing its spectral equivalence with the physical bath under OBC. Most importantly, we show that the mirage bath is not only a mathematical concept. We explicitly demonstrate that the real-time dynamics reflects the spectral properties of the mirage bath (Sec. V D).

### A. Definition of mirage baths

Let us first provide a general construction and definition for a mirage bath. Consider a generic $N_s$-band dissipative bath associated with the effective lattice Hamiltonian $H_b' = \sum_{ij} c_i^\dagger \mathcal{H}_{ij} c_j$ under PBC, where $\mathcal{H} \neq \mathcal{H}^\dagger$ is non-Hermitian. The corresponding Bloch Hamiltonian $H_b'(k)$ exhibits the spectrum $\omega_b'(k)$ that features point gaps in the complex frequency plane.

The physical quantity of interest is the two-point Green function for times $t > 0$, i.e.,

$$G_{b,ij}(t) = -i\langle[c_i(t), c_j^\dagger(0)]_\mp\rangle = \int_{-\infty}^{\infty}\frac{d\omega}{2\pi}\frac{e^{-i\omega t}}{\omega - \mathcal{H}_{ij} + i0^+}, \tag{19}$$

where $[\cdot,\cdot]_\mp$ denotes the commutator (for bosons) or anticommutator (for fermions). In the thermodynamic limit, the function becomes $G_{b,ij}(t) = \int_{-\infty}^{\infty}\frac{d\omega}{2\pi}e^{-i\omega t}\int_{-\pi}^{\pi}\frac{dk}{2\pi}\frac{e^{ik(i-j)}}{\omega - H_b'(k)+i0^+}$. Introducing $z = e^{ik}$, Eq. (19) becomes integrals along contours in both the complex $\omega$ plane and $z$ plane, i.e.,

$$G_{b,ij}(t) = \int_C \frac{d\omega}{2\pi}e^{-i\omega t}G_{b,ij}(\omega) \tag{20}$$

$$= \int_C \frac{d\omega}{2\pi}e^{-i\omega t}\oint_{|z|=1}\frac{dz}{2\pi i z}\frac{z^{i-j}}{\omega - H_b'(z)}, \tag{21}$$

where the contour $C$ is outside the branch loop we describe below. For convenience, we focus exclusively on the case $i \geqslant j$ (for the case $i < j$, a similar analysis applies by redefining $z = e^{-ik}$).

Let us first outline the key properties of $G_{b,ij}(\omega)$ as an integral in the $z$ plane. In Eq. (21), the poles $\bar{z}_s$ of $[\omega - H_b'(z)]^{-1}$ are determined by the condition

$$\lambda(\omega, \bar{z}_s) \equiv \det[\omega - H_b'(\bar{z}_s)] = 0, \tag{22}$$

yielding $m$ solutions sorted by increasing modulus $|\bar{z}_1| \leqslant \cdots \leqslant |\bar{z}_m|$. We summarize below three properties:

(1) For $z$ on the unit circle in the $z$ plane, Eq. (22) maps to the curves $\omega = \omega_b'(k)$ in the $\omega$ plane, defining the branch cuts of $G_{b,ij}(\omega)$. When $\omega_b'(k)$ exhibits point gaps, these cuts form branch loops without branch points [cf. Fig. 1(d)].

(2) For a given $\omega$, the $z$ integral in Eq. (21) is determined by poles inside $|z| = 1$, with their number $n_i$ depending on $\omega$ [cf. Fig. 5(a)]. This contrasts markedly with the closed baths, where $n_i$ is a constant. Consequently, the first RS layer of the complex frequency plane is partitioned into distinct regions [cf. Fig. 5(b)], separated by branch loops across which $n_i$ changes. In each region, $G_{b,ij}(\omega)$ takes a distinct analytic form (when unfolding the RS, each frequency region corresponds to an RS sheet).

(3) Let I denote frequency region outside all branch loops. Here, $[\omega - H_b'(z)]^{-1}$ has $n_w$ poles inside the unit circle in the $z$ plane, giving

$$G_{b,ij}(\omega) = \sum_{s=1}^{n_w} \mathrm{Res}(\bar{z}_s) \equiv \mathcal{F}(\omega), \quad \omega \in \mathrm{I}, \tag{23}$$

where the sum includes only residues from these $n_w$ poles. These $n_w$ poles represent "dynamical observable" poles in the real-time dynamics we explain below.

Let us return to the time domain. The correlation dynamics is governed by $G_{b,ij}(t) = \int_C \frac{d\omega}{2\pi} e^{-i\omega t} G_{b,ij}(\omega)$, where the contour $C$ must lie outside branch loops. The causality condition restricts the branch loops to the lower half of $\omega$ plane, ensuring $C$ never intersects them [cf. Fig. 1(d)]. This leads to a key observation: only region I contributes to dynamics, whereas the functional forms of $G_{b,ij}(\omega)$ inside loops are irrelevant. This observation permits an analytic continuation of Eq. (23) to the *entire* complex $\omega$ plane (denoted by $\mathbb{C}$),

$$G^f_{b,ij}(\omega) = \mathcal{F}(\omega), \quad \omega \in \mathbb{C}. \tag{24}$$

The $G^f_{b,ij}(\omega)$ is characterized by two properties:

(1) Unlike its parent function $G_{b,ij}(\omega)$, which is piecewise defined, $G^f_{b,ij}(\omega)$ has a unified expression across the entire $\omega$ plane, so that its discontinuities are restricted to *branch lines* with branch points.

(2) Despite spectrally distinct in frequency domain, $G^f_{b,ij}(\omega)$ and $G_{b,ij}(\omega)$ produce identical dynamics in time domain, i.e.,

$$G_{b,ij}(t) = \int_C \frac{d\omega}{2\pi} e^{-i\omega t} G_{b,ij}(\omega) = \int_{C'} \frac{d\omega}{2\pi} e^{-i\omega t} G^f_{b,ij}(\omega). \tag{25}$$

This is ensured by analytic continuation that preserves the $n_w$ poles in Eq. (23) that govern the time evolution.

The mirage bath is defined inversely via the analytic structures of $G^f_{b,ij}(\omega)$: its linecuts $\omega = \omega^f_b$ define the mirage-bath's spectrum $\omega^f_b$. Analogous to the original bath, we formally define a mirage-bath's Hamiltonian $H^f_b(k)$ through $G^f_{b,ij}(\omega) = \int \frac{dk}{2\pi} \frac{e^{ik(i-j)}}{\omega - H^f_b(k) + i0^+}$. Note that our construction can be easily extended to when emitters are present. There, the self-energy matrix is related to the bath's Green function via $\Sigma_{ij}(\omega) = \Omega^2 G_{b,ij}(\omega)$. The mirage BSs are thus defined from the poles of the continued emitter Green function.

### B. Mirage-bath's spectrum and DOS

We now present a general scheme to calculate the mirage-bath's spectrum. Since $G^f_{b,ij}(\omega)$ preserves the analytic form in Eq. (23) in the entire $\omega$ plane, when $\omega$ is varied across its branch line, the number of poles in the corresponding $z$ integral must maintain at $n_w$. This indicates the condition for the line cut can only be

$$\left|\bar{z}_{n_w}\left(\omega^f_b\right)\right| = \left|\bar{z}_{n_w+1}\left(\omega^f_b\right)\right|. \tag{26}$$

Parametrizing $\bar{z}_{n_w+1} = \bar{z}_{n_w} e^{i\theta}$ with $\theta \in (-\pi, \pi]$, we find $\omega^f_b$ by solving simultaneous solutions to $\lambda(\omega, \bar{z}_w) = 0$ and $\lambda(\omega, \bar{z}_w e^{i\theta}) = 0$. As detailed in Appendix H, this can be achieved using standard Sylvester's elimination method to eliminate $\bar{z}$, which gives

$$\det(R) \equiv f\left(\omega^f_b, \theta\right) = 0, \tag{27}$$

with a $2m$-dimensional Sylvester matrix $R$. Each solution of $\omega^f_b$ for $\theta \in [-\pi, \pi)$ forms the mirage-bath spectrum [cf. white curves in Fig. 5(c)].

Analogous to the DOS characterizing Hermitian band structures, we can obtain the DOS associated with the mirage-bath's complex spectrum, defined by

$$D\left(\omega^f_b\right) = \int_{-\pi}^{\pi} \frac{d\theta}{2\pi} \delta\left[\omega^f_b - \omega^f_b(\theta)\right] = \frac{1}{2\pi}\left|\frac{1}{\partial_\theta \omega^f_b(\theta)}\right|. \tag{28}$$

Using the identity $\partial_\theta(\det R) = \mathrm{Tr}(R^* \partial_\theta R)$, with $R^*$ the adjoint matrix of $R$, we obtain $\partial_\theta R = R_1 \partial_\theta \omega^f_b + R_2$ with the coefficient matrices $R_1$ and $R_2$. Consequently, the DOS is derived as

$$D\left(\omega^f_b\right) = \frac{1}{2\pi}\left|\frac{\mathrm{Tr}(R^* R_1)}{\mathrm{Tr}(R^* R_2)}\right|. \tag{29}$$

### C. Spectral equivalence between mirage bath and physical bath under OBC

Two remarks are in order. First, Eq. (26), which defines the branch linecut (and thus the spectrum) of mirage bath (PBC), is precisely the condition determining the spectrum in an OBC system in the thermodynamic limit [64]. This spectral equivalence arises because the Green function between two finitely separated sites, $i - j$, becomes insensitive to boundary conditions in the thermodynamic limit. Consequently, PBC and OBC systems exhibit identical bulk time evolutions. Technically, it means that analytic continuation of the Green's function [e.g., $G_{b,ij}(\omega)$] from outside the branch loops into their entire interior must correspond to the OBC spectrum, as established by Eq. (26). Thus, dynamical equivalence in the bulk enforces spectral equivalence between the mirage bath (PBC) and physical baths under OBC.

Second, the observable dynamics preselects a fundamental arrow of analytic continuation—from the exterior into the branch loop interior—as the physically meaningful one. While there exist many such continuations [cf. Fig. 5(h)], only the one producing a unified function across the entire $\omega$ plane uniquely defines the mirage bath. For instance, Fig. 5(h) illustrates a continuation from region I to encompass region II, yielding branch loops instead of open cuts. Physically, it means that, for instance, any BSs spectrally inside these loops remain hidden from dynamics. Thus, only the mirage bath always leaves observable manifestations in real-time dynamics and is spectrally equivalent to the OBC.

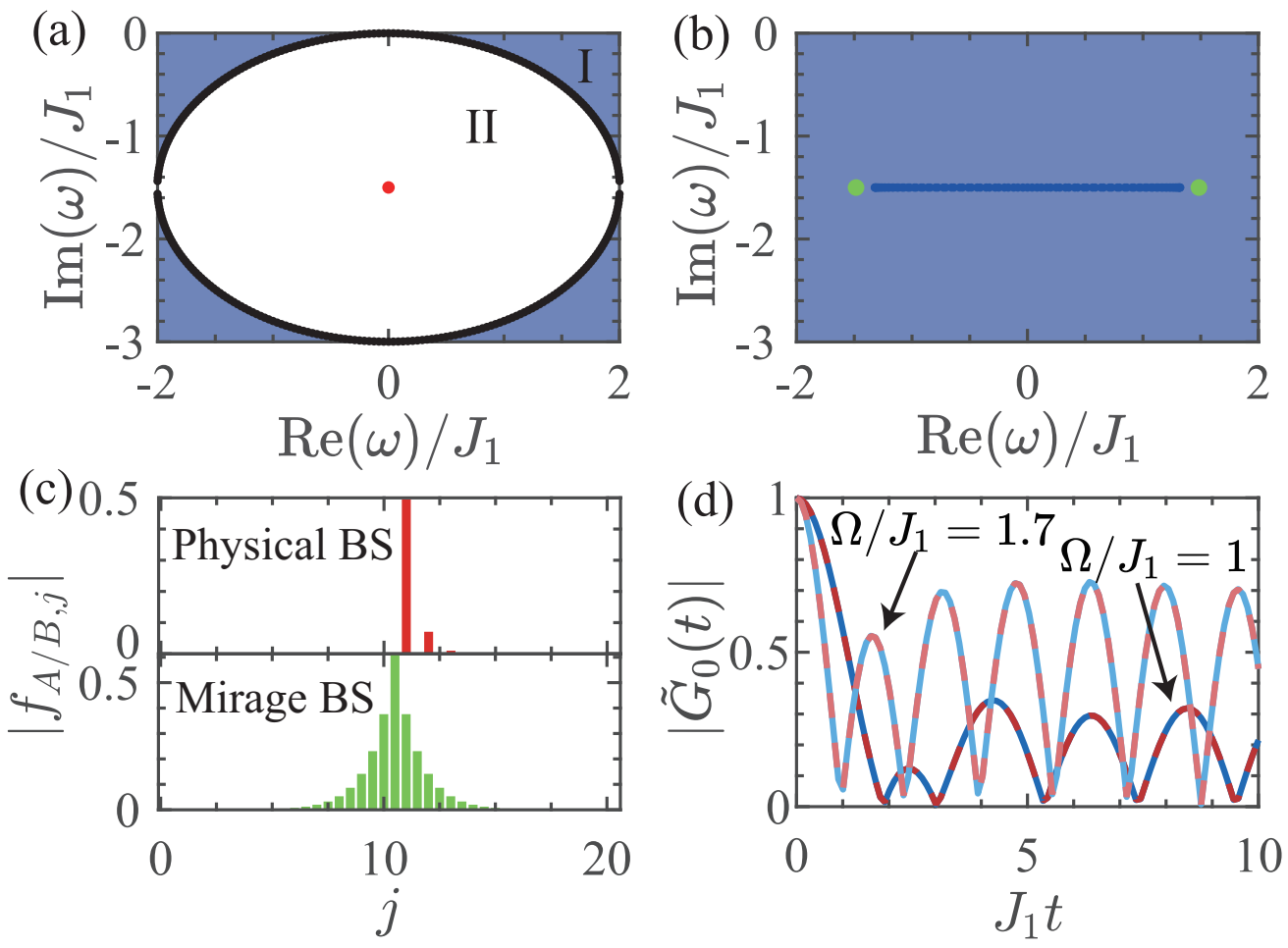

FIG. 4. Single-QE dynamics in a Hantao-Nelson bath: (a) Frequency plane partitioned into two regions by the branch loop. The self-energy $\Sigma_0(\omega)$ of the physical bath takes distinct expressions (see main text) outside (I) and inside (II) the loop (black curve). The red dot denotes the physical BS. (b) Mirage bath via analytic continuation. The blue line denotes the branch cut representing the mirage spectrum $\omega_b^f(k)$; the green dots denote the mirage BS (see main text). (c) Physical vs mirage BS: photonic spatial distribution. The energy of the BS associated with the physical (mirage) bath is $\omega_{\rm BS} = -1.5i$ ($\omega_{\rm BS}^f = \pm 1.48 - 1.5i$). (d) Dynamics $|\tilde{G}_0(t)| = |e^{\gamma_a t/2} G_0(t)|$ of an excited QE, for $\Omega/J_1 = 1, 1.7$, respectively. In panels (a)–(d), $\gamma_b/J_1 = 1.5$, $\Delta'/J_1 = -1.5i$, $\gamma_a/J_1 = 3$, and in panels (a)–(c) $\Omega/J_1 = 1$.

### D. Illustrative examples

In this section, we consider emitters couple to three different types of baths, focusing on where emitter frequencies $\Delta'$ fall within the point gaps of the physical bath. We shall focus on the key results, delegating calculation details in Appendix J.

#### *1. Hatano-Nelson bath*

We first consider a single QE coupled to a Hatano-Nelson (HN) bath, governed by the master equation (4) with $H_{\rm b} = \sum_j J_1 b_j^\dagger b_j + \mathrm{H.c.}$ and the jump operator $l_j = -ib_j + b_{j+1}$. The resulting effective bath Hamiltonian $H_b' = \sum_{j=1}^{N_b} [(J_1 + \gamma_b/2) b_j^\dagger b_{j+1} + (J_1 - \gamma_b/2) b_{j+1}^\dagger b_j - i\gamma_b b_j^\dagger b_j]$ describes a Hatano-Nelson model (PBC). Its spectrum $\omega_b'(k) = 2J_1 \cos(k) - i\gamma_b[1 - \sin(k)]$ forms a point gap.

As first shown in Ref. [58], when $\Delta'$ is in the point gap, a dynamically hidden BS emerges. The self-energy $\Sigma_0(\omega) = \Omega^2 \int dk/(2\pi)[\omega - \omega_b'(k)]^{-1}$ exhibits a branch loop [Fig. 4(a)]: $\Sigma_0(\omega) = \frac{\Omega^2 \mathrm{sign}(|z_-| - |z_+|)}{\sqrt{(\omega + i\frac{\gamma_b}{2})^2 - 4J_1^2 + \gamma_b^2}}$ outside [58], while $\Sigma_0(\omega) = 0$ inside. The physical BS in the branch loop (red dot) has the fixed energy $\omega_{\rm BS} = \Delta'$ [its wavefunction shown in Fig. 4(c)], independent of coupling rate $\Omega$. Yet, dynamics in Fig. 4(d) reveals $\Omega$-dependent oscillations, which cannot be explained by $\Omega$-independent BSs. This is in contrast to its Hermitian-bath counterpart, where such oscillations originate from photon BSs.

Instead, these oscillations probe the mirage BSs on the second RS layer. Analytic continuation yields $\Sigma_0^f(\omega) = \frac{\Omega^2 \mathrm{sign}(|z_-| - |z_+|)}{\sqrt{(\omega + i\frac{\gamma_b}{2})^2 - 4J_1^2 + \gamma_b^2}}$ in the entire $\omega$ plane [Fig. 4(b)]. Since $\Sigma_0^f(\omega)$ is formally analogous to its Hermitian counterpart under the substitution: $\omega \to \omega + i\gamma_b/2$ and $J_1^2 \to J_1^2 - \gamma_b^2/4$, we obtain the mirage spectrum $\omega_b^f(k) = -i\gamma_b/2 + 2\tilde{J}_1 \cos(k)$, coinciding with that of the physical bath under OBC. Unlike their physical counterpart, the mirage BS [solutions of $\omega_{\rm BS}^f - \Delta' - \Sigma_0^f(\omega_{\rm BS}^f) = 0$] depends strongly on $\Omega$. For instance, for $\Omega/J = 1$, while only one physical BS exists in the point gap at strong couplings, two mirage ones with $\omega_B^f = \pm 1.48 - 1.5i$ appear [green dots in Fig. 4(b)], with distinct shapes from the physical one [Fig. 4(b)]. Indeed, oscillations in Fig. 4(d) ($\Omega/J = 1$) align with $\tilde{G}_a(t) \propto \cos(1.48t)$ at long times.

#### *2. Trefoil Bath*

The preceding analysis extends naturally to systems with more intricate spectral topology (Fig. 5), such as when the energy bands braid multiple times, creating nontrivial knot structures that have recently attracted significant interests [38]. Consider Eq. (4) with $H_{\rm b} = \sum_j (J_1 b_{A,j}^\dagger b_{B,j} + J_2 b_{A,j}^\dagger b_{A,j+1} + J_3 b_{A,j}^\dagger b_{A,j+3} + \mathrm{H.c.})$ and $l_j = -ib_{A,j} + b_{A,j+3}$. The effective bath Hamiltonian $H_{\rm b}' = \sum_j (J_1 b_{A,j}^\dagger b_{B,j} + J_2 b_{A,j}^\dagger b_{A,j+1} + \mathrm{H.c.}) + \sum_j [(J_3 + \gamma_b/2) b_{A,j}^\dagger b_{A,j+3} + (J_3 - \gamma_b/2) b_{A,j+3}^\dagger b_{A,j} - i\gamma_b b_{A,j}^\dagger b_{A,j}]$ realizes a Trefoil model [38] with a spectrum forming a trefoil knot in the complex frequency plane.

Figure 5(a) displays the $\omega$-dependent pole structures of $\Sigma_0(\omega)$ in the $z$ plane, partitioning the frequency plane into four distinct regions [Fig. 5(b)], where $\Sigma_0(\omega)$ takes different expressions. Still, only region I—and thus $n_w = 3$ poles within $|z| = 1$ in $z$ plane—is dynamically relevant, where $\Sigma_0(\omega) = \sum_{s=1}^3 \mathrm{Res}(z_s)$. Analytical continuation yields $\Sigma_0(\omega) = \sum_{s=1}^3 \mathrm{Res}(z_s)$ in the entire $\omega$ plane. Its branch linecut [determined by Eq. (26)] defines the mirage spectrum [Fig 5(c)]. The calculated DOS in Fig. 5(d) reveals linecuts terminating at branch points, where DOS can be discontinuous. The equivalence between the mirage spectrum and the OBC spectrum of the physical bath is validated in Fig. 5(e). Figure 5(f) shows the dynamical equivalence of dual baths. Yet, only the mirage BSs are dynamically observable. Comparisons in Fig. 5(g) explicitly show that the dynamics at intermediate and long times is captured by $G_0(t) \approx \sum_j Z_j e^{-i\omega_{\rm f,BS}^j t}$, where $\omega_{\rm f,BS}^j$ is a mirage BS' energy and $Z_j$ is the corresponding residue.

#### *3. Nonreciprocal SSH bath with NNN coupling*

As previously shown in Fig. 2 for the nonreciprocal SSH bath, for a single emitter, the photon BSs emerging in the complex band gap always faithfully inherit the underlying topology. However, if $\Delta'$ are in the point gap, the BSs therein have no apparent contribution to atomic interactions, as they are dynamically unobservable. Instead, the point-gap interaction is explicitly mediated through the mirage BSs, thus inheriting the mirage-bath's topology. Here, we generalize this analysis to an extended nonreciprocal SSH bath with next-to-nearest-neighbor (NNN) couplings. Consider Eq. (4) with $H_\text{b} = \sum_j (J_1 b_{A,j}^\dagger b_{B,j} + J_2 b_{A,j+1}^\dagger b_{B,j} + J_3 b_{A,j}^\dagger b_{B,j+1} + \text{H.c.})$, yielding $H_\text{b}' = \sum_j [(J_1 + \gamma_b/2) b_{A,j}^\dagger b_{B,j} + (J_1 - \gamma_b/2) b_{B,j}^\dagger b_{A,j} - i\gamma_b/2 b_{A,j}^\dagger b_{A,j} - i\gamma_b/2 b_{B,j}^\dagger b_{B,j}] + \sum_j (J_2 b_{A,j+1}^\dagger b_{B,j} + J_3 b_{A,j}^\dagger b_{B,j+1} + \text{H.c.})$ that represents an extended dissipative SSH bath with NNN coupling; its topological phase diagram is depicted in Fig. 6(c). In this case, the frequency plane is partitioned into two regions, in which self-energy exhibits different, nontrivial analytic forms [Fig. 6(a)]. Following schemes in Sec. V B, we obtain the mirage spectrum [Fig. 6(b)] and phase diagram [Fig. 6(c)], which are radically different from the original bath's. Yet, the correlation dynamics of two QEs in the point gap probes the mirage-bath's topology [Fig. 6(d)], consistent with our analysis.

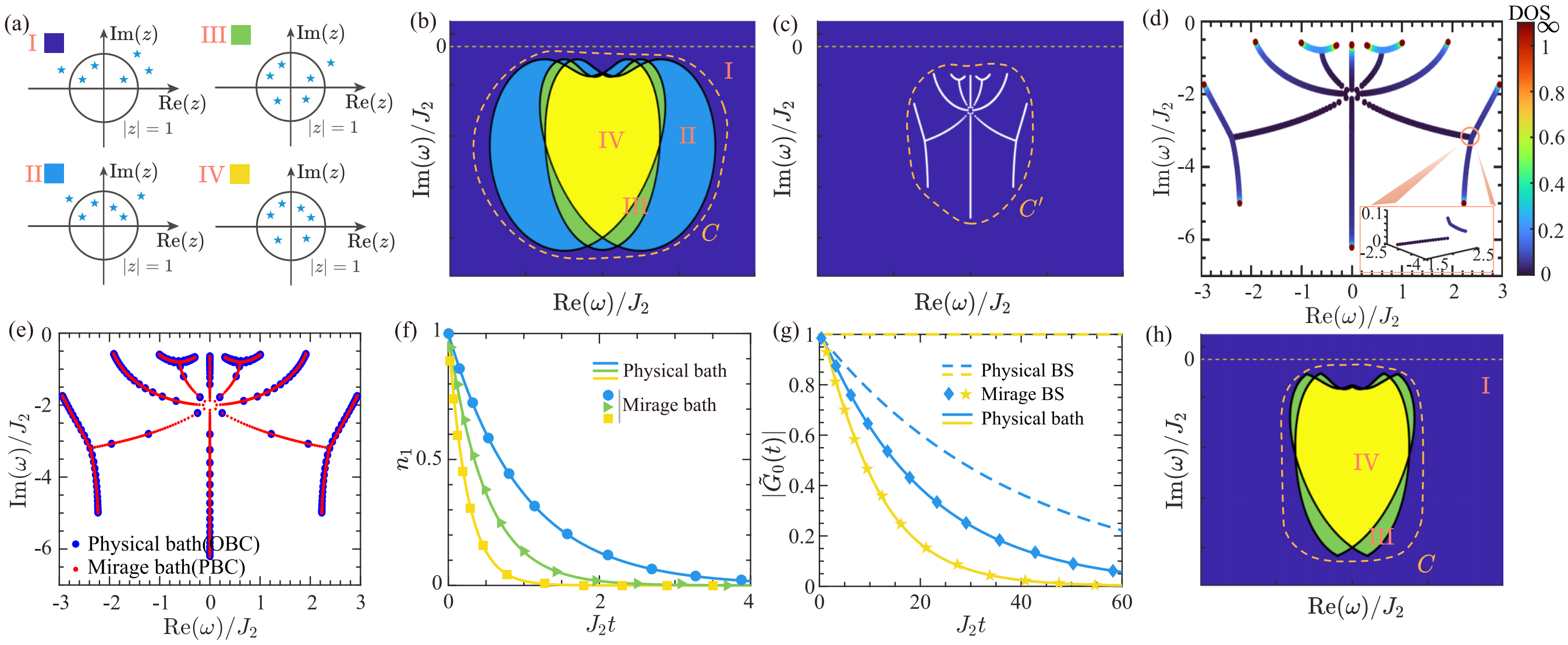

FIG. 5. Single-QE dynamics in a trefoil bath. (a) $\omega$-dependent pole structure in the $z$ plane. (b) Frequency plane partitioned into multiple regions by the branch loop (black curve). In each region, $\Sigma_0(\omega)$ has a distinct analytic form (see detailed calculations in Appendix J). (c) Mirage bath via analytic continuation. The branch linecuts of $\Sigma_0^f(\omega)$ (white curves) are obtained from Eq. (27), which defines the mirage spectrum. (d) DOS of the mirage bath, obtained from Eq. (29). (e) Comparison between the mirage spectrum and the OBC spectrum of the physical bath. The OBC spectrum is obtained from diagonalizing the physical bath Hamiltonian (J10) with a size $N_b = 100$ under OBC. (f) Population dynamics $n_1(t) = |G_0(t)|^2$ obtained from the physical vs mirage baths for $\Omega/J_2 = 0.2$ and $\Delta'/J_2 = 3.5 - 0.5i$, $3 - i$, and $2 - 2i$ (blue, green, and yellow). (g) Comparison between $|\tilde{G}_0(t)|$ (solid curves) and its pole (BS) approximation when $\Omega/J_2 = 0.5$. We choose $\Delta'/J_2 = 1.2 - 0.01i, 2 - 0.3i$ (blue, yellow) lying in region II and IV in panel (b), respectively. Here, only a single pole (BS) with the largest residue is kept in the approximation [e.g., dashed curves denote the single-pole approximation to $G_0(\omega)$ of the physical bath]. For both mirage and physical baths, the pole with the largest residue is chosen in making the single-pole approximation to $G_0(t)$. (h) Illustration of different analytic continuation. For other parameters, in panels (d)–(g) $J_1/J_2 = 2$, $J_3/J_2 = 1.7$, and $\gamma_b/J_2 = 4$.

## VI. TOPOLOGICAL IMPLICATION: DYNAMICAL EQUIVALENCE UNIFIES NON-HERMITIAN PBC AND OBC TOPOLOGIES

Why does a dissipative topological bath with point gaps generate a mirage bath with identical dynamics yet distinct topology? As we show, this topological mismatch is fundamental rather than incidental: it emerges necessarily from the requirement of dynamical equivalence. This equivalence acts as a filter, preserving only those Green's function poles that govern observable dynamics while excluding dynamically unobservable ones, thus inducing a topological change. As a remarkable consequence, dynamical equivalence provides a unifying principle to determine the topological phase diagram of mirage baths (and thus that of OBC) directly from the original bath's properties through analytic continuation.

As is well known, when PBC bands exhibit point-gap winding, skin effects generically arise in the system with OBC, causing the conventional bulk-edge correspondence and Bloch theorem to break down. To determine the OBC topology—and physical properties in general—one typically uses non-Bloch band theory [39,42,44,52,64], which involves the construction of a generalized Brillouin zone (GBZ). Below we present a tantalizing alternative to study OBC topology bypassing GBZ construction.

We illustrate the essential ideas through specific examples, while detailed calculations can be found in Appendix I. Consider first the nonreciprocal SSH bath (PBC) with sublattice symmetry studied previously. Its topological property is characterized by the index [76],

$$\begin{aligned}
\nu &= \lim_{\omega\to\omega_\text{EP}} \nu(\omega) \\
&= \lim_{\omega\to\omega_\text{EP}} \int_{-\pi}^{\pi} \frac{dk}{4\pi i} \operatorname{tr}[\sigma_z G_b(k,\omega)^{-1} \partial_k G_b(k,\omega)] \\
&= \lim_{\omega\to\omega_\text{EP}} \oint_{|z|=1} \frac{dz}{4\pi i} \operatorname{tr}[\sigma_z G_b(z,\omega)^{-1} \partial_z G_b(z,\omega)]. \qquad (30)
\end{aligned}$$

Here, $G_b(z,\omega) = [\omega - H_b'(z)]^{-1}$ is the bath's Green function that exhibits a total of two poles ($|\bar{z}_1| < |\bar{z}_2|$) and $\omega_\text{EP} = -i\gamma_b/2$ marks the EP of the bath's spectrum. Unlike the

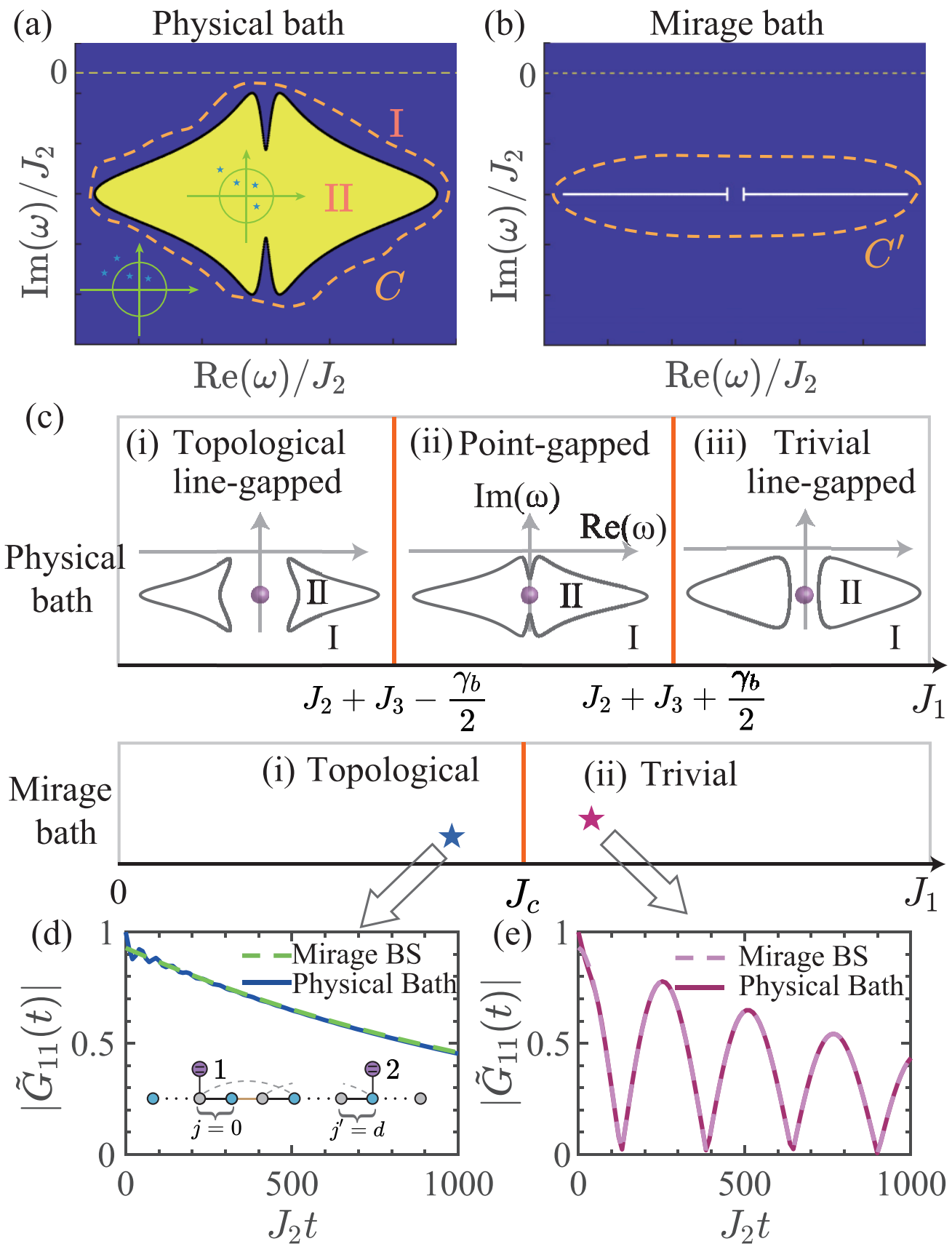

FIG. 6. Two-QE dynamics in an extended nonreciprocal SSH bath with NNN coupling. (a) Partition of the frequency plane according to the corresponding pole structures in the $z$ plane. (b) Mirage bath by analytic continuation. The resulting branch linecut (white curve) defines the mirage spectrum. (c) Phase diagrams of the physical vs mirage bath. (d), (e) Emission dynamics, $\tilde{G}_{11}(t) = -ie^{\gamma_a t/2}\langle g|\sigma^1_{ge}(t)\sigma^1_{eg}(0)|g\rangle$. The upstream (downstream) QE is coupled to sublattice $A$ ($B$). Initially, the upstream QE is excited (vacuum). Results are shown for their distance $d = 0$, and (d) $J_1/J_2 = 1.3$ and (e) $J_1/J_2 = 1.48$. The dashed line indicates the contribution of the mirage two-emitter BSs with the large residues to the dynamics. For other parameters, $J_3/J_2 = 0.39$, $\gamma_b/J_2 = 0.2$, $\Omega/J_2 = 0.1$, and $\Delta'/J_2 = -0.09i$.

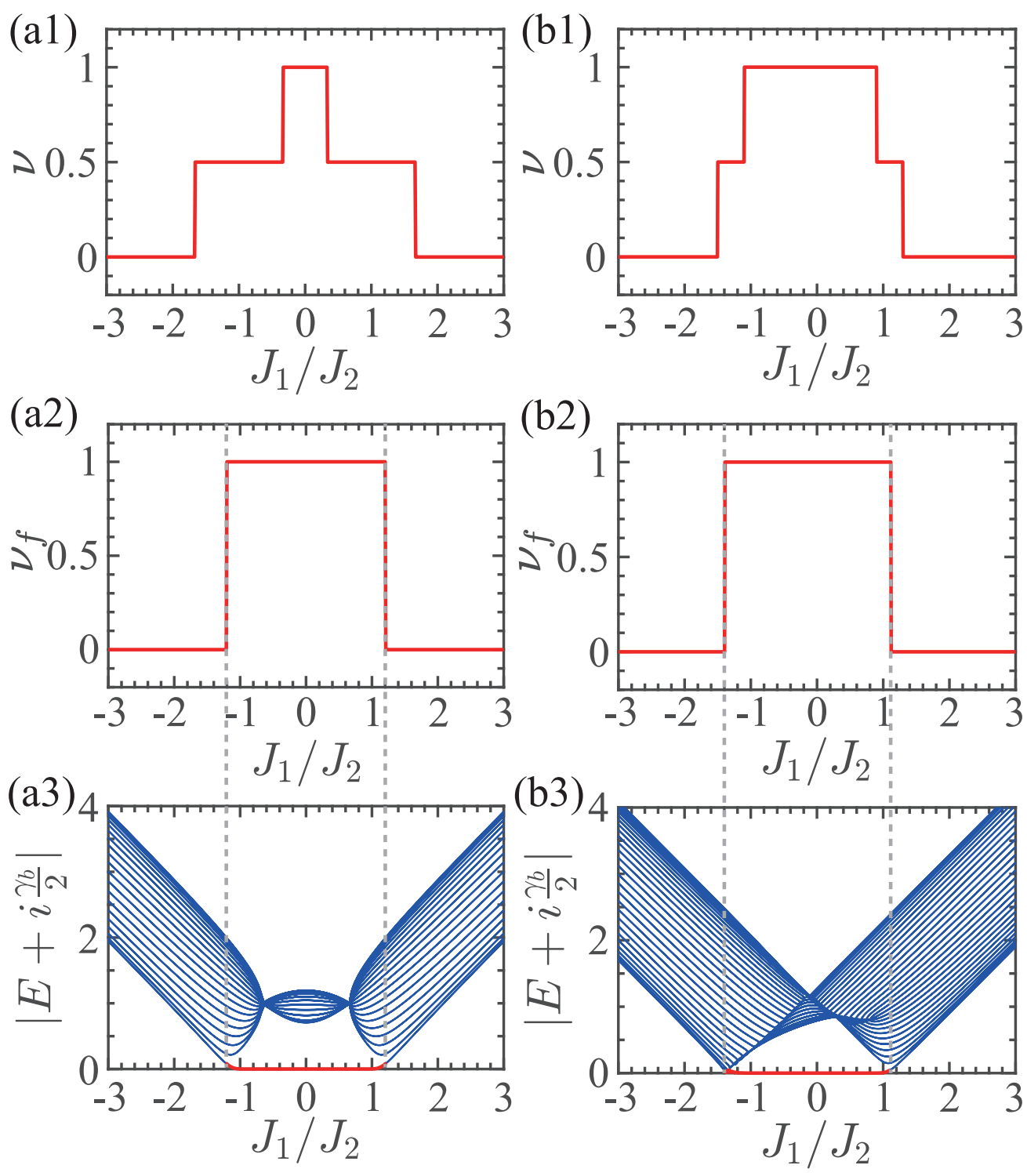

FIG. 7. Topological phase diagram of the mirage bath obtained from analytic continuation. The physical bath (PBC) is described by (a1)–(a3) nonreciprocal SSH model (I1) and (b1)–(b3) extended nonreciprocal SSH model with long-range couplings (see text). (a1), (b1) topological index $\nu$ of the physical bath [Eq. (30)] as a function of $J_1/J_2$. (a2), (b2) topological index $\nu_f$ of the mirage bath [Eq. (31)] as a function of $J_1/J_2$. (a3), (b3) Spectrum of the physical bath under OBC as a function of $J_1/J_2$. Results are obtained from diagonalization of the physical bath Hamiltonian with size (a3) $N_b = 20$ and (b3) $N_b = 30$ under OBC. The zero-mode line is shown in red. In panels (a1)–(a3), $\gamma_b/J_2 = 4/3$. In panels (b1)–(b3), $J_3/J_2 = 0.2$, $J_4/J_2 = 0.1$, and $\gamma_b/J_2 = 0.4$.

real-time dynamics $G_{b,ij}(t)$, which involves integrating over $\omega$, the topological indices in Eq. (30) are evaluated at a fixed frequency $\omega = \omega_{\rm EP}$. As detailed in Appendix I, for $\omega_{\rm EP}$ in the line gap [cf. (i) and (iii) in Fig. 2(a)], only $n_w = 1$ pole ($\bar{z}_1$) of $G_b(z,\omega)$ lies inside $|z| = 1$, yielding quantized index $\nu = 1(0)$. In contrast, when $\omega_{\rm EP}$ enters the point gap (i.e., inside branch loop), $n_w = 2$ poles of $G_b(z,\omega)$ are enclosed, leading to $\nu = 0.5$. As discussed in Ref. [77], this corresponds to the existence of one zero-mode edge state. As shown in Fig. 7(a1), the topological index $\nu$ reproduces the phase diagram of the nonreciprocal SSH bath.

The corresponding mirage bath is constructed to reproduce the original bath's observable dynamics through analytic continuation, so that $G^f_b(\omega)$ preserves the analytic forms of $G_b(\omega)$ outside its branch loop, where only one pole of $G_b(z,\omega)$ is inside $|z| = 1$. This key property allows us to obtain the mirage-bath's topological index $\nu_f$ via continuation:

$$\nu_f = \lim_{\omega\to\omega_{\rm EP}} \nu^f(\omega)$$

$$= \lim_{\omega\to\omega_{\rm EP}} \oint_{C_z} \frac{dz}{4\pi i}{\rm tr}[\sigma_z G_b(z,\omega)^{-1}\partial_z G_b(z,\omega)], \tag{31}$$

where the deformed contour $C_z$ must enclose the origin ($z = 0$) and pole $z_1$ of $G_b(\omega)$ (the shape of $C_z$ not important). Subsequent calculation follows identical residue analysis to Eq. (30) for $\omega_{\rm EP}$ outside the branch loop (where quantized $\nu$ is obtained). Figure 7(a2) shows $\nu_f = 1(0)$, distinguishing topologically nontrivial (trivial) phases of the mirage bath. As shown in Fig. 7(a3), $\nu_f = 1(0)$ directly diagnoses the presence (absence) of edge modes in the nonreciprocal SSH model under OBC.

Our above analysis can be easily generalized to more complicated cases, such as the extended nonreciprocal SSH bath with the long-range couplings described by $H'_b = \sum_j[(J_1 + \gamma_b/2)b^\dagger_{A,j}b_{B,j} + (J_1 - \gamma_b/2)b^\dagger_{B,j}b_{A,j}] + \sum_j(J_2 b^\dagger_{A,j+1}b_{B,j} + J_3 b^\dagger_{A,j}b_{B,j+1} + J_4 b^\dagger_{A,j}b_{B,j+2} + {\rm H.c.}) - i\gamma_b/2\sum_j(b^\dagger_{A,j}b_{A,j} +$ $b^{\dagger}_{B,j} b_{B,j}$). In this case, $G_b(z, \omega)$ exhibits a total of six poles, $|\bar{z}_1| \leqslant \cdots \leqslant |\bar{z}_6|$, but only three poles ($\bar{z}_{1,2,3}$) govern the dynamics [cf. Fig. 5(a)]. Figure 7(b1) presents the topological index $\nu$ of the original bath calculated from Eq. (30). The topological index of the corresponding mirage bath is obtained from the analytic continuation [Eq. (31)] that preserves three dynamically relevant poles ($\bar{z}_{1,2,3}$) inside the contour. Consequently, the resulting $\nu_f = 1(0)$ is necessarily quantized, yielding the phase diagram in Fig. 7(b2), with direct correspondence to the presence or absence of edge modes in the physical OBC bath [Fig. 7(b3)].

To conclude, the topological mismatch between the dual baths is a fundamental result of point-gap topology and dynamical equivalence. In the light that the mirage bath and the physical OBC bath are spectrally and topologically equivalent, the dynamical equivalence naturally bridges non-Hermitian PBC and OBC topologies—through a pole-filtering process. Such topological unification allows the intuition and results from topological matter and field theory to be directly transferred to study the bulk properties under OBC—via the mirage bath, facilitating explorations of non-Hermitian topology in genuine quantum regimes.

### A. Dual topological detection on a single platform

Another intriguing implication of the mirage bath is the possibility to probe the topology of a non-Hermitian system with different boundary conditions—all within a single setup. Specifically, coupling a single QE to a PBC bath, while its dynamics is insensitive to the bath's topology, the spatial profiles of the BSs within the band gap—be it a line gap or point gap—faithfully reflect it. The latter can be detected, e.g., via the state engineering technique [62]. In the presence of multiple emitters, the photon BS mediates emitter-emitter interactions, allowing the multi-QE dynamics to probe the BS's properties. However, the nature of such BS mediator depends crucially on whether the emitters are in a line gap or point gap. Notably, if emitters are in a point gap, their dynamics probe the mirage-bath's topology rather than the physical one [Figs. 3(c), 6(d), and 6(e)]. Consequently, two-emitter dynamics provide a sensitive probe to OBC topology, despite the bath actually being under PBCs. This unified capability promises a versatile and efficient method for detecting non-Hermitian topological phases in open quantum systems, opening possibilities for state-of-the-art experiments using synthetic quantum materials with controllable dissipations [47,59,62,63,65–68].

## VII. SYSTEMATIC FRAMEWORK FOR STUDYING FEW-BODY AND MANY-BODY QUANTUM DYNAMICS

Beyond its conceptual interest, the mirage bath provides a highly efficient method for studying few-body and many-body physics in open quantum systems with non-Hermitian topology. This has been highly desired but remains challenging due to the general difficulty to treat long-time dynamics of open quantum systems. For instance, beyond the single-excitation sector, already the spectrum of the dissipative SSH bath with two excitations becomes highly complex, exhibiting dense regions in the frequency plane [Fig. 8(a)]. In contrast, the much simpler analytic structure of the mirage bath [Fig. 8(b)] affords significant advantages in studying, for instance, multiphoton process and quantum correlated statistics of photons.

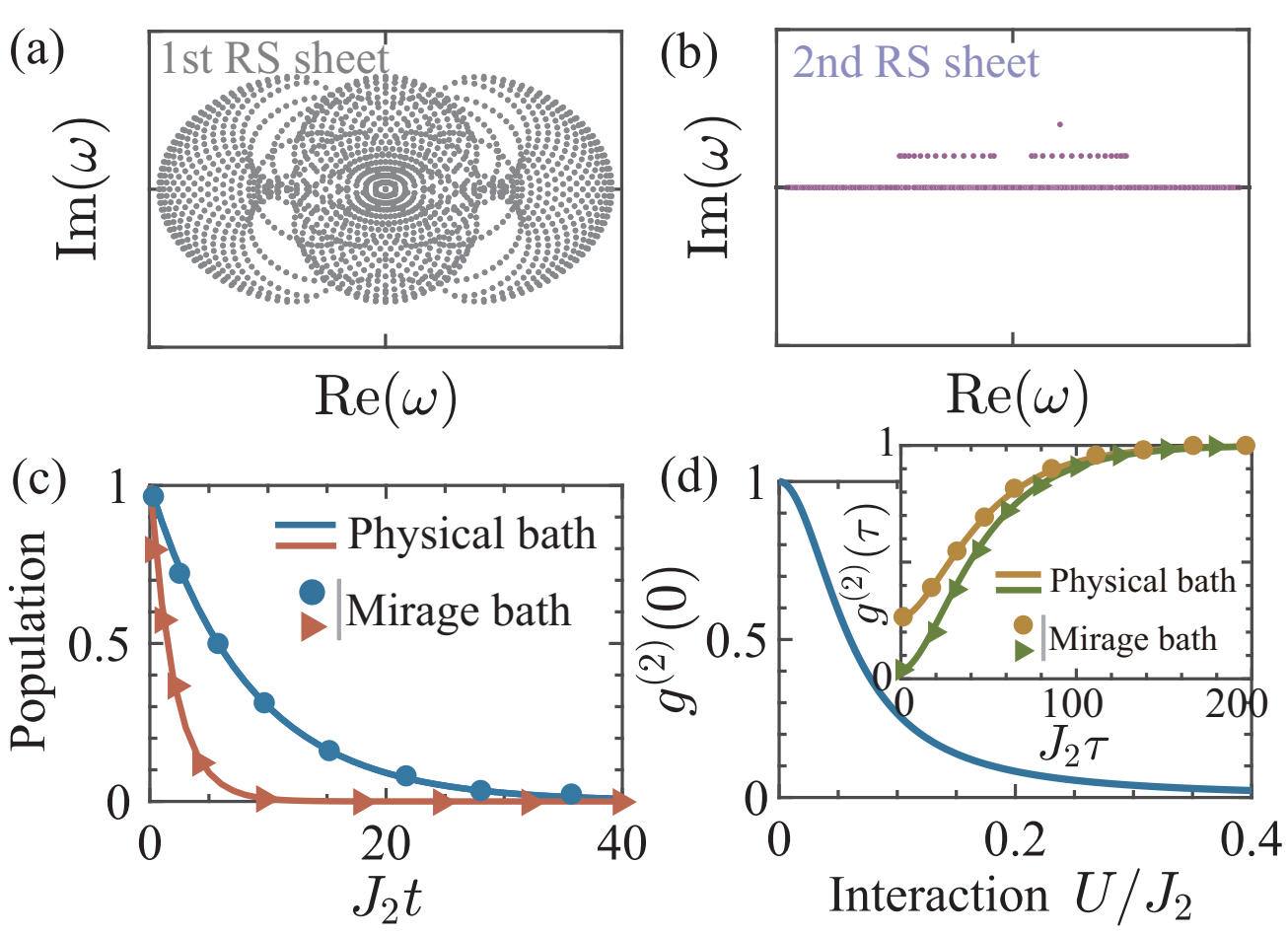

FIG. 8. Second-order correlation of a driven nonlinear QE with multiple excitations. The two-excitation spectrum for (a) physical and (b) mirage photon baths. (c) Spontaneous emission dynamics of two excitations in the absence of driving ($\epsilon = 0$), computed via the physical and mirage baths, respectively. The blue color denotes where $J_1/J_2 = 1.01$, $\Omega/J_2 = 0.01$, $\gamma_a/J_2 = 0.06$, $\gamma_b/J_2 = 0.1$, and $\Delta/J_2 = 0$, $U/J_2 = 0.1$. The red color denotes where $J_1/J_2 = 1.1$, $\Omega/J_2 = 0.2$, $\gamma_a/J_2 = 0.2$, $\gamma_b/J_2 = 1$, $\Delta/J_2 = 0$, and $U/J_2 = 0.4$. (d) $g^{(2)}(0)$ of a weakly driven emitter ($\epsilon \neq 0$) as a function of interaction strength $U/J_2$. The inset shows $g^{(2)}(\tau)$ for $U/J_2 = 0.1, 0.3$ (brown, green) computed using the physical and mirage photon baths, respectively. Other parameters are the same as the blue line in Fig. 8(c).

To demonstrate this, we extend beyond single excitation and study the generation of quantum light from a driven nonlinear emitter, where the quantum nature of light is characterized by the second-order correlations. Consider specifically a nonlinear emitter $a$ with multiple bosonic excitations, such as a cavity with Kerr interactions, coupled to a dissipative photonic SSH bath. The corresponding master equation becomes $\dot{\rho} = -i[\tilde{H}_a + H_b + H_{ab}, \rho] + (\gamma_b/2)\sum_j \mathcal{D}_{\rm b}[l_j]\rho + (\gamma_a/2)\sum_m \mathcal{D}_a[a]\rho$, with $H_b$, $H_{ab}$, and $\mathcal{D}_{\rm b}[l_j]$ following the same form as in Eq. (4). The emitter Hamiltonian is

$$\tilde{H}_a = \Delta a^{\dagger} a + \frac{U}{2} a^{\dagger} a^{\dagger} a a + \epsilon (a^{\dagger} e^{-i\omega_d t} + \text{H.c.}), \tag{32}$$

describing a nonlinear QE with local interaction strength $U$, driven by an external pump of strength $\epsilon$ and frequency $\omega_d$. In the hardcore boson limit $U \to \infty$, Eq. (32) reduces to the case of two-level QEs. As shown in Fig. 8(c) for $\epsilon = 0$, in the two-excitation sector, the mirage bath produces the same emitter dynamics, $D(t) = -i\langle 0|a^2(t) a^{\dagger 2}(0)|0\rangle$, as the original bath (see details in Appendix D).

Next, we study the steady-state second-order correlation function for a driven emitter:

$$g^{(2)}(\tau) = \frac{1}{n^2} \text{Tr}[a^{\dagger} a^{\dagger}(\tau) a(\tau) a \rho_{\rm ss}], \tag{33}$$

where $n = \text{Tr}(a^{\dagger} a \rho_{\rm ss})$ is the first-order correlation function and $\rho_{\rm ss}$ is the steady state of the master equation with $\epsilon \neq 0$. The quantum nature of the photon statistics is characterized by $g^{(2)}(0) < 1$. Using the mirage bath, one has $g^{(2)}(0) = |1 - U\Pi_f(2\omega_d)|^{-2}$ [61], Appendix D, where $\Pi_f(\tau) = i\int \frac{d\omega'}{2\pi} G_f(\omega_d + \omega')G_f(\omega_d - \omega')e^{-i\omega'\tau}$ is associated with the mirage bath and $\Pi_f(2\omega_d)$ corresponds to $\Pi_f(0)$. By using the mirage Green function, we can then apply the well-established techniques, such as the Lehmann spectral representation, to simplify computations. Figure 8(d) shows $g^{(2)}(0)$ for various nonlinear interaction strength $U$, indicating the occurrence of photon antibunching. In the inset, we compare $g^{(2)}(\tau)$ obtained from the Green functions associated with the original and mirage baths, respectively. A perfect agreement is found. This example highlights the significant advantage of the mirage bath in accessing quantum correlated dynamics in open quantum systems with non-Hermitian topological properties.

The above dynamical equivalence of $g^{(2)}(\tau)$ arises because although the dual baths exhibit distinct spectral structures in the frequency domain—branch linecuts on higher RS versus loops on the first RS—they yield identical single-particle Green function in the *time* domain [see Eq. (11)]. This analytic continuation approach extends naturally to study many-body dynamics governed by master equations, where the time-dependent density matrix evolves an incoherent superposition across different excitation sectors. Transitions between these sectors are induced by jump operators, analogous to the Keldysh terms in nonequilibrium Green function theory, while intrasector dynamics are governed by multipoint retarded (advanced) Green functions typically expressed as integrals of single-particle Green functions in the time domain. As the mirage bath and the original physical bath yield identical single-particle Green functions in the time domain, this equivalence allows one to replace the original bath with its mirage to study many-body correlation dynamics.

## VIII. CONCLUSION

In conclusion, we unveil a universal mechanism for quantum interactions in open quantum systems with non-Hermitian topological properties: interactions between particles are mediated by virtual excitations of a “mirage bath,” inheriting its topology rather than the physical bath. Residing on a distinct RS layer in the complex-energy plane, the mirage bath acts as a dual to the physical bath, generating identical dynamics. Intriguingly, the mirage-bath’s spectrum and topology are equivalent to that of the physical bath under OBC. Such duality between the mirage and physical baths enables a mechanism for nonlocal interactions and quantum correlations across the multilayered RS, with no analogs in conventional closed-bath settings.

Our results offer profound insights into non-Hermitian topology and open quantum many-body physics. In particular, we show how the topological mismatch arises necessarily from dynamical equivalence that requires to discard dynamically “unobservable” poles. Consequently, preserving identical dynamics unifies non-Hermitian topologies under different boundaries, without explicit construction of GBZ. Finally, our findings have direct experimental implications, such as probing topological properties under different boundary conditions within a single experimental setup.

## ACKNOWLEDGMENTS

This research is funded by the National Natural Science Foundation of China (Grants No. 12374246, No. 12135018, and No. 12047503) and the National Key Research and Development Program of China (No. 2022YFA1404003, No. 2022YFA1203903, and No. 2021YFA0718304). Y.H. acknowledges support by Beijing National Laboratory for Condensed Matter Physics (No. 2023BNLCMPKF001).

## DATA AVAILABILITY

The data that support the findings of this article are not publicly available. The data are available from the authors upon reasonable request.

## APPENDIX A: EMITTERS IN A CLOSED SSH BATH

Following Ref. [21], in this section we review the dynamics of two-level QEs coupled to a closed photonic SSH bath. Specifically, in Appendix A 1 we analyze the single QE and derive the wavefunctions of the bound state (BS) in the topological band gap, and in Appendix A 2, we analyze two QEs and derive their interaction strength.

The total system of $N_a$ QEs coupled to the SSH bath is described by the Hamiltonian

$$H_0 = H_a + H_{ab} + H_b. \tag{A1}$$

Here, Hamiltonian $H_b$ describes a closed SSH bath under PBC, i.e.,

$$H_b = \sum_j (J_1 b^\dagger_{A,j} b_{B,j} + J_2 b^\dagger_{B,j} b_{A,j+1} + \text{H.c.}), \tag{A2}$$

where $b^\dagger_{A/B,j}(b_{A/B,j})$ is the creation (annihilation) operator of a photonic mode at sublattice $A/B$ in unit cell $j$. In the thermodynamic limit, the bath exhibits two energy bands

$$\omega_b(k) = \pm\sqrt{(J_1 + J_2\cos k)^2 + J_2^2\sin^2 k}, \tag{A3}$$

where $k \in [-\pi, \pi)$ is the quasimomentum. The emitter Hamiltonian $H_a$ can be written using the language of hardcore bosons, reading

$$H_a = \Delta \sum_{m=1}^{N_a} a^\dagger_m a_m, \tag{A4}$$

where $a_m (m = 1, \ldots, N_a)$ is the annihilation operator of hardcore bosons and $\Delta$ is the transition frequency of the QE with respect to the central frequency $\omega_r = 0$ of the bath. Assuming the $m$th QE is locally coupled to the sublattice site $\alpha_{j_m} \in \{b_{A,j_m}, b_{B,j_m}\}$ in the unit cell $j_m$, the coupling Hamiltonian $H_{ab}$ is written as

$$H_{ab} = \Omega \sum_m \left(\alpha^\dagger_{j_m} a_m + \text{H.c.}\right), \tag{A5}$$

where $\Omega$ is the local coupling rate.

### 1. Single emitter

#### *a. Dynamics*

We first consider a single excited QE ($N_a = 1$) coupled to $j = 0$ of the bath and exactly calculate its spontaneous emission dynamics, $G_0(t) = -i\langle 0|a_1(t)a_1^{\dagger}(0)|0\rangle$, using Green's function [6,61].

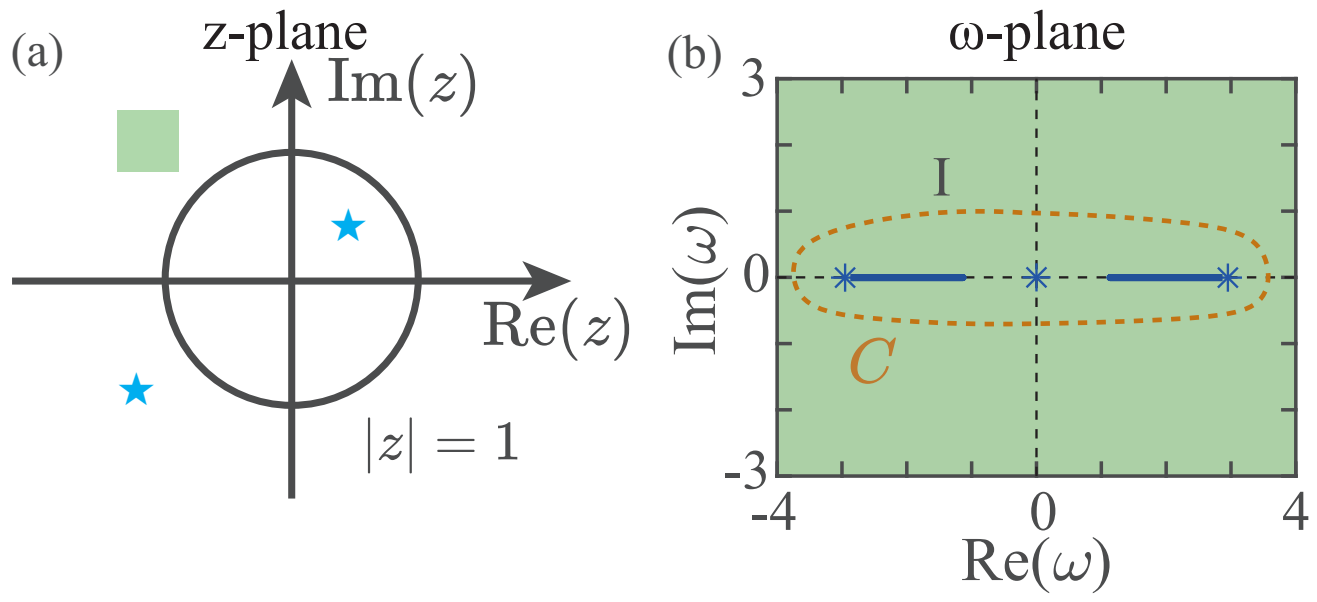

FIG. 9. Analytic structure of the self-energy and Green function for the closed bath. (a) Distribution of the poles of the self-energy in $z \equiv e^{ik}$ plane. For $\omega \neq \omega_b(k)$ [green region in panel (b)], one of the two poles (pentagram) of the self-energy [see Eq. (A8)] is inside the unit circle, while the other is outside. (b) Analytic structure of the emitter's Green function $G_0(\omega)$ [see Eq. (A7)] in the complex frequency plane. The blue lines depict the branch cut $\omega = \omega_b(k)$ [see Eq. (A3)]. The asterisk depicts the pole, representing the BS. For the SSH bath, there exist three poles: two poles are near the band edges and one is within the gap between two bands.

Since $H_0$ conserves the total number of excitations, the emitter dynamics can be calculated from the Fourier transform

$$G_0(t) = \int \frac{d\omega}{2\pi} e^{-i\omega t} G_0(\omega), \tag{A6}$$

where we have

$$G_0(\omega) = \frac{1}{\omega - \Delta - \Sigma_0(\omega)}. \tag{A7}$$

Here, $\Sigma_0(\omega)$ is the self-energy associated with the SSH bath, reading

$$\begin{aligned}\Sigma_0(\omega) &= \Omega^2 \int_{-\pi}^{\pi} \frac{dk}{2\pi} \frac{\omega}{\omega^2 - |\omega_b(k)|^2} \\ &= \Omega^2 \oint_{|z|=1} \frac{dz}{2\pi i z} \frac{\omega}{\omega^2 - (J_1 + J_2 z)(J_1 + J_2 z^{-1})}.\end{aligned} \tag{A8}$$

In the second line, we have transformed to the variable $z = e^{ik}$ and expressed the self-energy as the contour integral along a unit circle in the $z$ plane. The self-energy (A8) can then be explicitly calculated using the residue theorem. The poles are found from the two roots $z_{\pm}$ of the equation

$$-J_1 J_2 z^2 + \left(\omega^2 - J_1^2 - J_2^2\right) z - J_1 J_2 = 0, \tag{A9}$$

which yields

$$z_{\pm}(\omega) = -\frac{J_1^2 + J_2^2 - \omega^2 \pm \Lambda(\omega)}{2J_1 J_2}, \tag{A10}$$

with the notation

$$\Lambda(\omega) = \sqrt{\left(\omega^2 - J_1^2 - J_2^2\right)^2 - 4J_1^2 J_2^2}. \tag{A11}$$

Since for the closed bath $z_+ z_- = 1$ is satisfied, only one of the two roots is necessarily within the unit circle [see Fig. 9(a)]. Therefore, using the residue theorem and Eq. (A10), Eq. (A8) is calculated as

$$\begin{aligned}\Sigma_0(\omega) &= -\frac{\Omega^2 \omega}{J_1 J_2} \oint_{|z|=1} \frac{dz}{2\pi i} \frac{1}{(z - z_+)(z - z_-)}, \\ &= \frac{\Omega^2 \omega}{\Lambda(\omega)} \mathrm{sign}(|z_-| - |z_+|).\end{aligned} \tag{A12}$$

Finally, substituting Eq. (A12) into Eq. (A7), Eq. (A6) can be calculated. According to the Lehmann spectral representation, the emitter dynamics in Eq. (A6) is fully determined by the analytic structure, i.e., poles and branch cut, of the Green function $G_0(\omega)$ in the complex $\omega$ plane. As shown in Fig. 9(b), there are three isolated poles of the Green function [i.e., $G_0^{-1}(\omega) = 0$] corresponding to the energies of the single-QE BSs, and the branch cuts represent the continuum of the bath.

#### b. Chiral BS in the topological band gap

For a single QE, when its transition frequency lies in the gap between the two bands of the SSH photonic bath, a chiral BS emerges in the gap. In this subsection, we derive the energy and wavefunction of the BS across different phases of the bath.

The energy of the BSs is given by the pole equation in Eq. (A7), i.e.,

$$\omega_{\mathrm{BS}} - \Delta - \Sigma_0(\omega_{\mathrm{BS}}) = 0. \tag{A13}$$

For $\Delta = 0$, $\omega_{\mathrm{BS}}$ can be analytically found: substituting Eq. (A12) into Eq. (A13), there always exists a solution $\omega_{\mathrm{BS}} = 0$, irrespective of $J_1/J_2$.

The wavefunction of the BS in real space is written as

$$|B\rangle = \left(\varphi_a a^{\dagger} + \sum_{\alpha = A/B, j} f_{\alpha,j} b_{\alpha,j}^{\dagger}\right)|0\rangle, \tag{A14}$$

where $\varphi_a$ and $f_{\alpha,j}$ are the amplitudes of atomic and photonic components. Below, we calculate the photonic wavefunction $f_{\alpha,j}$ for the $A$-BS ($B$-BS), which depends on the sublattice $A$ ($B$) to which the emitter is coupled.

(1) $A$-BS: for the emitter coupled to the $A$ sublattice at $j = 0$, $f_{A,j}$ is given by [21]

$$\begin{aligned}f_{A,j} &= \Omega \varphi_a \int_{-\pi}^{\pi} \frac{dk}{2\pi} \frac{\omega_{\mathrm{BS}} e^{ikj}}{\omega_{\mathrm{BS}}^2 - |\omega_b(k)|^2} \\ &= -\frac{\Omega \varphi_a}{J_1 J_2} \oint_{|z|=1} \frac{dz}{2\pi i} \frac{\omega_{\mathrm{BS}} z^j}{(z - z_+)(z - z_-)},\end{aligned} \tag{A15}$$

where $z_{\pm}$ is given by Eq. (A10) and $\varphi_a$ can be determined from the normalization conditions. Applying the residue theorem and using $\Lambda(\omega)$ in Eq. (A11), we obtain

$$f_{A,j} = \begin{cases} \frac{\Omega \varphi_a \omega_{\mathrm{BS}} [z_+^j \Theta_+(z_+) - z_-^j \Theta_+(z_-)]}{\Lambda(\omega_{\mathrm{BS}})}, & j \geqslant 0, \\ \frac{\Omega \varphi_a \omega_{\mathrm{BS}} [z_-^j \Theta_-(z_-) - z_+^j \Theta_-(z_+)]}{\Lambda(\omega_{\mathrm{BS}})}, & j < 0, \end{cases} \tag{A16}$$

where we have introduced the function

$$\Theta_{\pm}(z) = \Theta(\pm 1 \mp |z|) \tag{A17}$$

in terms of Heaviside's step function $\Theta(z)$.

TABLE I. Analytical expressions of the photonic distribution $f_{A/B,j}$ of the single-QE BS in the closed bath. The emitter's transition frequency $\Delta = 0$ is within the band gap, and the emitter is coupled to sublattice $A/B$ at unit cell $j = 0$. The first (second) row presents the expressions when the bath is in the trivial (topological) phase.

| | $A$-BS | $B$-BS |
|---|---|---|
| $\lvert J_1\rvert > \lvert J_2\rvert$ | $f_{A,j} = 0$ | $f_{B,j} = 0$ |
| | $f_{B,j} = \begin{cases} -\frac{\Omega\varphi_a}{J_1}\left(-\frac{J_2}{J_1}\right)^{j}, & j \geqslant 0, \\ 0, & j < 0 \end{cases}$ | $f_{A,j} = \begin{cases} 0, & j > 0, \\ -\frac{\Omega\varphi_a}{J_1}\left(-\frac{J_2}{J_1}\right)^{\lvert j\rvert}, & j \leqslant 0 \end{cases}$ |
| $\lvert J_1\rvert < \lvert J_2\rvert$ | $f_{A,j} = 0$ | $f_{B,j} = 0$ |
| | $f_{B,j} = \begin{cases} 0, & j \geqslant 0, \\ \frac{\Omega\varphi_a}{J_1}\left(-\frac{J_1}{J_2}\right)^{\lvert j\rvert}, & j < 0 \end{cases}$ | $f_{A,j} = \begin{cases} \frac{\Omega\varphi_a}{J_1}\left(-\frac{J_1}{J_2}\right)^{j}, & j > 0, \\ 0, & j \leqslant 0 \end{cases}$ |

The values of $f_{B,j}$ can be obtained in a similar fashion, i.e.,

$$f_{B,j} = \Omega\varphi_a \int_{-\pi}^{\pi} \frac{dk}{2\pi} \frac{(J_1 + J_2 e^{ik})e^{ikj}}{\omega_{\rm BS}^2 - |\omega_b(k)|^2}$$

$$= \begin{cases} \frac{\Omega\varphi_a[F_j(z_+)\Theta_+(z_+) - F_j(z_-)\Theta_+(z_-)]}{\Lambda(\omega_{\rm BS})}, & j \geqslant 0, \\ \frac{\Omega\varphi_a[F_j(z_-)\Theta_-(z_-) - F_j(z_+)\Theta_-(z_+)]}{\Lambda(\omega_{\rm BS})}, & j < 0, \end{cases} \tag{A18}$$

where we have introduced the function

$$F_j(z) = (J_1 + J_2 z)z^j. \tag{A19}$$

(2) $B$-BS: when the emitter is coupled to the $B$ sublattice at $j = 0$, the amplitude of photonic component $f_{A,j}$ is derived as

$$f_{A,j} = \Omega\varphi_a \int_{-\pi}^{\pi} \frac{dk}{2\pi} \frac{(J_1 + J_2 e^{-ik})e^{ikj}}{\omega_{\rm BS}^2 - |\omega_b(k)|^2},$$

$$= -\frac{\Omega\varphi_a}{J_1 J_2} \oint_{|z|=1} \frac{dz}{2\pi i} \frac{(J_1 + J_2 z^{-1})z^j}{(z - z_+)(z - z_-)}. \tag{A20}$$

After applying the residue theorem and having in mind $z_+ z_- = 1$, we obtain

$$f_{A,j} = \begin{cases} \frac{\Omega\varphi_a[F_{-j}(z_-)\Theta_+(z_+) - F_{-j}(z_+)\Theta_+(z_-)]}{\Lambda(\omega_{\rm BS})}, & j > 0, \\ \frac{\Omega\varphi_a[F_{-j}(z_+)\Theta_-(z_-) - F_{-j}(z_-)\Theta_-(z_+)]}{\Lambda(\omega_{\rm BS})}, & j \leqslant 0, \end{cases} \tag{A21}$$

where $\Lambda(\omega)$, $\Theta_\pm(x)$, and $F_j(x)$ were defined in Eqs. (A11), (A17), and (A19), respectively.

Similarly, we can obtain $f_{B,j}$ as

$$f_{B,j} = \Omega\varphi_a \int_{-\pi}^{\pi} \frac{dk}{2\pi} \frac{\omega_{\rm BS} e^{ikj}}{\omega_{\rm BS}^2 - |\omega_b(k)|^2},$$

$$= \begin{cases} \frac{\Omega\varphi_a\omega_{\rm BS}[z_+^j\Theta_+(z_+) - z_-^j\Theta_+(z_-)]}{\Lambda(\omega_{\rm BS})}, & j > 0, \\ \frac{\Omega\varphi_a\omega_{\rm BS}[z_-^j\Theta_-(z_-) - z_+^j\Theta_-(z_+)]}{\Lambda(\omega_{\rm BS})}, & j \leqslant 0, \end{cases} \tag{A22}$$

When $\Delta = 0$, the BS has $\omega_{\rm BS} = 0$, and the corresponding expressions of $f_{A/B,j}$ in Eqs. (A16)–(A22) can be greatly simplified, as summarized In Table I. There, we see that the photon $A$-BS ($B$-BS) has $f_{A(B),j} = 0$ and $f_{B(A),j}$ is localized on the left or right sides of the QE, depending on the bath's topology. The shape of the $A$-BS and $B$-BS for $\Delta = 0$ and $\Delta \neq 0$ can be visualized in Figs. 10(a) and 10(b). There, we also compare the analytical results of $f_{A/B,j}$ with those obtained from the numerical diagonalization of $H_0$, for both $\Delta = 0$ and $\Delta \neq 0$, respectively. A perfect agreement is found.

### 2. Two emitters

In this subsection, we consider two QEs ($N_a = 2$). Without loss of generality, we assume the first QE $a_1$ is coupled to the sublattice $A$ at $j_1 = 0$, while the second QE downstream $a_2$ is coupled to sublattice $B$ on the right at $j_2 = d > 0$, both lying in the topological band gap. As explained in the main text, the chiral BS mediates an effective interaction between them, i.e., $H_{\rm int} = \Sigma_d^{AB} a_1^\dagger a_2 + \text{H.c.}$ Here, we derive $\Sigma_d^{AB}$, which leads to Eq. (2).

The dynamics of two QEs is determined by the two-emitters' Green function

$$G_d(\omega) = \frac{1}{\omega - \Delta - \Sigma_d(\omega)}, \tag{A23}$$

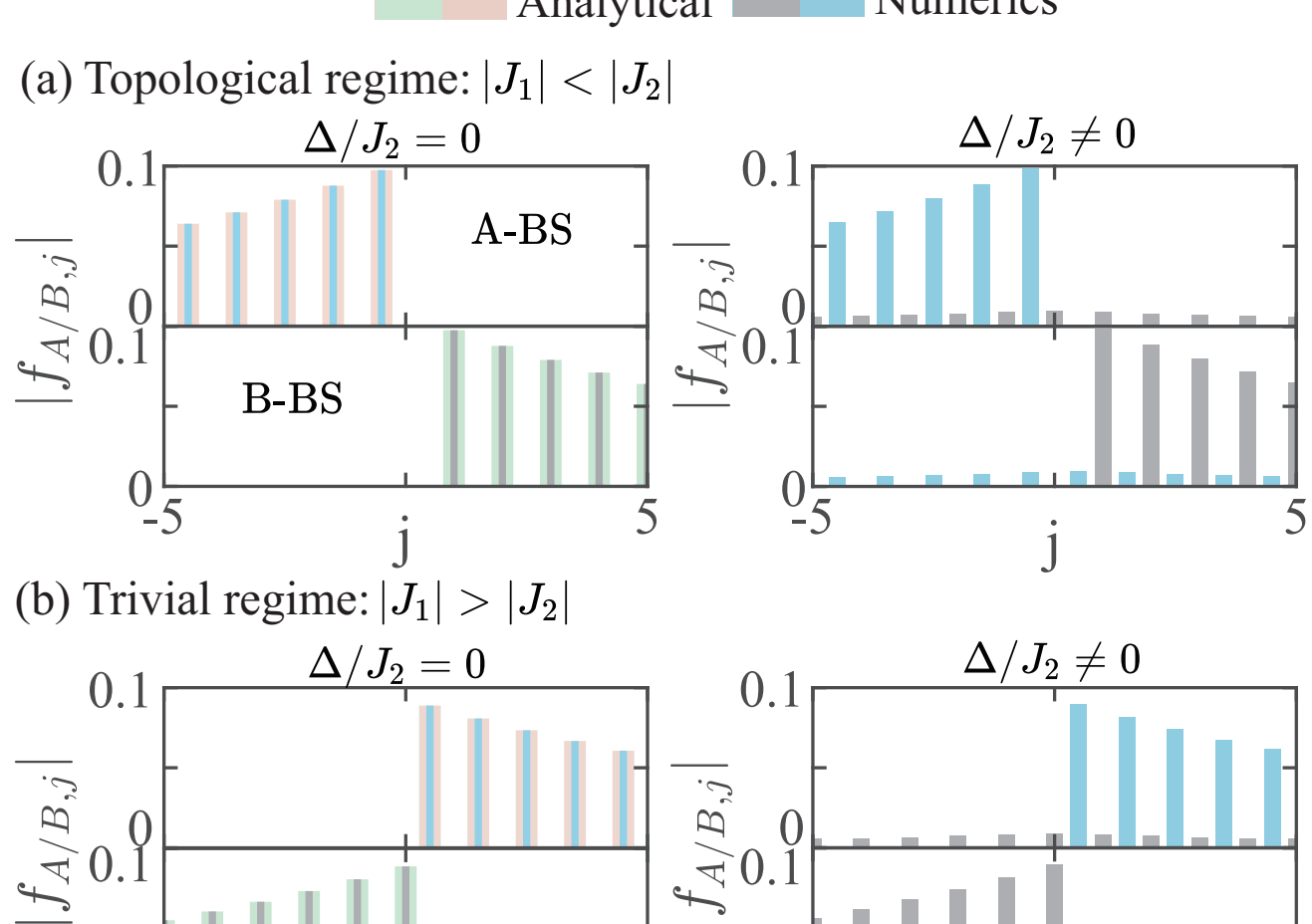

FIG. 10. Comparison between the analytic and numerical results for the photonic distribution $|f_{A/B,j}|$ of the single-emitter BS in the closed SSH bath. Results are shown for the bath with (a) $J_1/J_2 = 0.9$ in the topological phase and (b) $J_1/J_2 = 1.1$ in the trivial phase, when the emitter-bath coupling is $\Omega/J_2 = 0.1$. In both panels (a) and (b), we consider $\Delta = 0$ in the left panel and $\Delta/J_2 = 0.02$ in the right panel. Both $A$-BS and $B$-BS are shown. Analytical results are obtained from Table I. Numerical results are obtained through the numerical diagonalization of the total Hamiltonian (A1), taking $N_a = 1$ and the bath size $N_b = 500$.

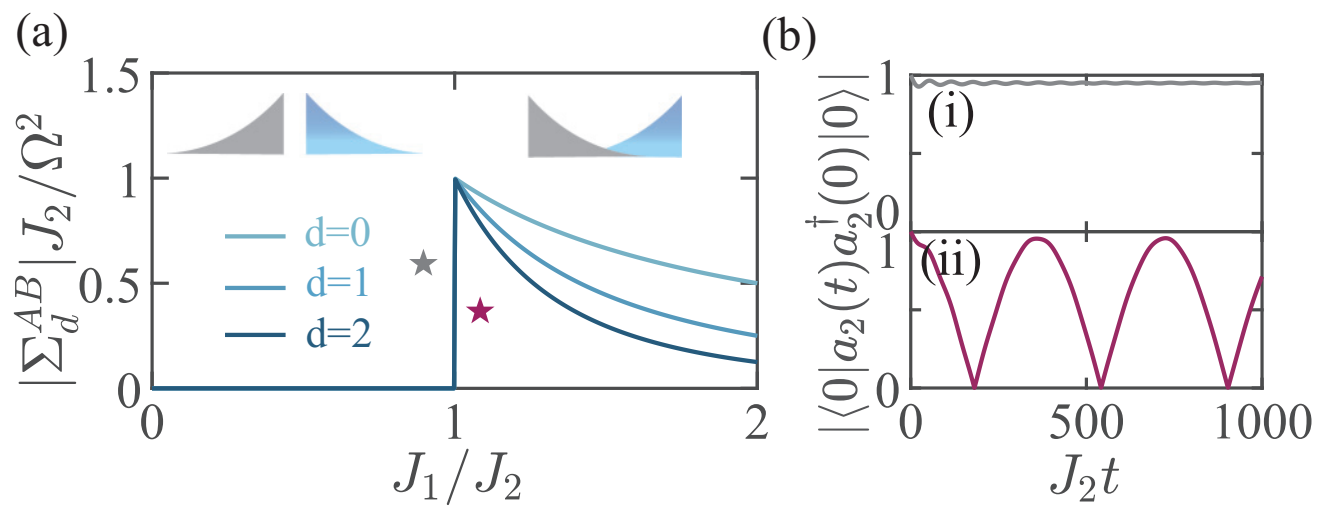

FIG. 11. Topological interaction mediated by the photon BS in the closed SSH bath. The first QE ($a_1$) is coupled to sublattice $A$ at $j_1 = 0$, and the second QE ($a_2$) is coupled to sublattice $B$ at $j_2 = d$. (a) Absolute value of the dipolar coupling $\Sigma_d^{AB}$ for $\Delta = 0$ and $\Omega/J_2 = 0.1$. Results are obtained from Eq. (A27) when $0 < J_1/J_2 < 1$ and Eq. (A26) when $J_1/J_2 > 1$. The insets show the shape of the photon $A$-BS (gray) and $B$-BS (blue) in the topological and the trivial phases, respectively. (b) Population dynamics $|\langle 0|a_2(t)a_2^\dagger(0)|0\rangle|$ for (i) $J_1/J_2 = 0.9$ in the topological regime and (ii) $J_1/J_2 = 1.1$ in the trivial regime. Initially at $t = 0$, the second QE $a_2$ is populated with one excitation. Shown are the results for $d = 0$, which are obtained from the Green function approach. Other parameters are the same as panel (a).

where $\Sigma_d(\omega)$ is the self-energy matrix with the elements

$$\Sigma_d(\omega) = \begin{pmatrix} \Sigma_0(\omega) & \Sigma_d^{AB}(\omega) \\ \Sigma_d^{BA}(\omega) & \Sigma_0(\omega) \end{pmatrix}. \tag{A24}$$

For the closed SSH bath, the off-diagonal elements satisfy $\Sigma_d^{AB} = (\Sigma_d^{BA})^*$; it can be calculated as

$$\Sigma_d^{AB}(\omega) = \Omega^2 \int \frac{dk}{2\pi} \frac{(J_1 + J_2 e^{-ik})e^{-idk}}{\omega^2 - |\omega_b(k)|^2}, \\ = \begin{cases} \frac{\Omega^2[F_d(z_+)\Theta_-(z_-) - F_d(z_-)\Theta_-(z_+)]}{\Lambda(\omega)}, & d \geqslant 0, \\ \frac{\Omega^2[F_d(z_-)\Theta_+(z_+) - F_d(z_+)\Theta_+(z_-)]}{\Lambda(\omega)}, & d < 0, \end{cases} \tag{A25}$$

where $z_\pm$ is given by Eq. (A10) and $\Theta_\pm(x)$ is defined in Eq. (A17).

For $\Delta = 0$ in the middle of the gap, where the BS has the energy $\omega_{\rm BS} = 0$, and a weak coupling $\Omega$, Eq. (A25) can be calculated using the single-pole approximation $\omega \approx 0$. After straightforward calculation as before, in the regime $|J_1| > |J_2|$, we find

$$\Sigma_d^{AB} = \begin{cases} -\frac{\Omega^2}{J_1}\left(-\frac{J_2}{J_1}\right)^d, & d \geqslant 0, \\ 0, & d < 0, \end{cases} \tag{A26}$$

and for $|J_1| < |J_2|$, we have

$$\Sigma_d^{AB} = \begin{cases} 0, & d \geqslant 0, \\ \frac{\Omega^2}{J_1}\left(-\frac{J_1}{J_2}\right)^{|d|}, & d < 0. \end{cases} \tag{A27}$$

Equations (A26) and (A27) show that the QE downstream can only interact with the first QE when the bath is in the trivial phase, consistent with the shape of the $A$-BS and $B$-BS in this case.

In Fig. 11(a), we illustrate $\Sigma_d^{AB}$ as a function of $J_1/J_2$ for $\Delta = 0$ and various $d$. In Fig. 11(b), we assume $a_2$ is initially populated with one excitation while $a_1$ is initially in the vacuum state. We calculate the population dynamics on the second QE for $J_1/J_2 < 1$, which reveal distinct behaviors in the topological phase [panel (i)] and trivial phase [panel (ii)]. In the topological phase, the probability to find the second QE in the excited state remains almost unchanged, indicating the absence of interaction between the first QE. In contrast, oscillation is observed in the trivial phase, indicating the interaction of QEs through the exchange of localized photons.

We note that in the above we have focused on the case where $a_1$ and $a_2$ couple to different sublattices. It can be shown that two QEs coupled to the same sublattice exhibit no interactions when $\Delta = 0$. Specifically, the interaction strength between two QEs on the sublattice $A$ ($B$) can calculated straightforwardly as

$$\Sigma_d^{AA/BB}(\omega) = \Omega^2 \int \frac{dk}{2\pi} \frac{\omega e^{idk}}{\omega^2 - |\omega_b(k)|^2}, \\ = \begin{cases} \frac{\Omega^2 \omega[z_+^d \Theta_-(z_+) - z_-^d \Theta_-(z_-)]}{\Lambda(\omega)}, & d \geqslant 0, \\ \frac{\Omega^2 \omega[z_-^d \Theta_+(z_-) - z_+^d \Theta_+(z_+)]}{\Lambda(\omega)}, & d < 0. \end{cases} \tag{A28}$$

For $\Delta = 0$ and under the single-pole approximation with $\omega \approx 0$, we have $\Sigma_d^{AA/BB} \approx 0$.

## APPENDIX B: EMITTERS IN A DISSIPATIVE SSH BATH

In this section, we consider two-level QEs coupled to a dissipative SSH bath. In Appendix B 1, we will derive the single-QE dynamics and the shape of the BS in the topological band gap. In Appendix B 2, we will analyze two emitters and derive the interaction strength mediated by the BS, i.e., Eq. (9).

The total density matrix $\rho$ for the combined system of QEs and the SSH bath is now governed by the master equation

$$\dot{\rho} = -i[H_0, \rho] + \sum_j \frac{\gamma_b}{2}\mathcal{D}_{\rm b}[l_j]\rho + \sum_m \frac{\gamma_a}{2}\mathcal{D}_a[a_m]\rho, \tag{B1}$$

where the first dissipator is

$$\mathcal{D}_b[l_j]\rho = 2l_j\rho l_j^\dagger - \{l_j^\dagger l_j, \rho\} \tag{B2}$$

with $l_j = -ib_{A,j} + b_{B,j}$. The second dissipator is

$$\mathcal{D}_a[a_m]\rho = 2a_m\rho a_m^\dagger - \{a_m^\dagger a_m, \rho\}. \tag{B3}$$

The effective Hamiltonian $H_{\rm eff}$ of the master Eq. (B1) is written as

$$H_{\rm eff} = H_a' + H_b' + H_{ab}, \tag{B4}$$

with Hamiltonian $H_{ab}$ of the same form as in Eq. (A5). In Eq. (B4), the effective bath Hamiltonian $H_b'$ describes a nonreciprocal SSH model, i.e.,

$$H_b' = H_b + \frac{\gamma_b}{2}\sum_{j=0}^{N_b}[-i(b_{A,j}^\dagger b_{A,j} + b_{B,j}^\dagger b_{B,j}) \\ + (b_{A,j}^\dagger b_{B,j} - {\rm H.c})], \tag{B5}$$

with $H_b$ given by Eq. (A2). $H_b'$ exhibits complex-energy bands

$$\omega_b'(k) = -i\frac{\gamma_b}{2} \pm \sqrt{(J_1 + J_2\cos k)^2 + \left(J_2\sin k + i\frac{\gamma_b}{2}\right)^2}. \tag{B6}$$

In Eq. (B4), $H_a'$ is the effective emitter Hamiltonian, reading

$$H_a' = \Delta' \sum_{m=1}^{N_a} a_m^\dagger a_m, \tag{B7}$$

with $\Delta' = \Delta - i\gamma_a/2$.

Following Ref. [61], because $H_{\rm eff}$ conserves the number of excitations, the emitter dynamics can be exactly calculated through the Green function approach in a similar way as described in Appendix A. However, as we show in detail in Appendixes B 1 and B 2, the dissipative bath renders different analytic structures of the Green function.

### 1. Single Emitter

#### a. Dynamics

For a single emitter, its dynamics is described by $G_0(t) = \int (d\omega/2\pi) e^{-i\omega t} G_0(\omega)$, with

$$G_0(\omega) = \frac{1}{\omega - \Delta' - \Sigma_0(\omega)}. \tag{B8}$$

Here, the self-energy $\Sigma_0(\omega)$ is associated with the dissipative SSH bath with the spectrum in Eq. (B6). We obtain

$$\begin{aligned}\Sigma_0(\omega) &= \Omega^2 \int_{-\pi}^{\pi} \frac{dk}{2\pi} \frac{\omega + \frac{i\gamma_b}{2}}{[\omega - \omega_{b+}'(k)][\omega - \omega_{b-}'(k)]},\\ &= \Omega^2 \oint_{|z|=1} \frac{dz}{2\pi i z} \frac{\omega + \frac{i\gamma_b}{2}}{\left(\omega + \frac{i\gamma_b}{2}\right)^2 - F_0^+(z^{-1}) F_0^-(z)},\end{aligned} \tag{B9}$$

where we have introduced

$$F_j^{\pm}(z) = \left(J_1 \pm \frac{\gamma_b}{2} + J_2 z\right) z^j. \tag{B10}$$

The self-energy (B9) can be calculated using the residue theorem. The two poles, $z_\pm$, are found from solutions of the equation,

$$\begin{aligned}&-J_2\left(J_1 + \frac{\gamma_b}{2}\right) z^2 + \left[\left(\omega + \frac{i\gamma_b}{2}\right)^2 - \sigma_1\right] z - J_2\left(J_1 - \frac{\gamma_b}{2}\right)\\ &\quad = 0,\end{aligned} \tag{B11}$$

with $\sigma_1 = J_1^2 + J_2^2 - \gamma_b^2/4$. We find

$$z_\pm = \frac{\sigma_1 - \left(\omega + \frac{i\gamma_b}{2}\right)^2 \pm \Lambda'(\omega)}{-2J_1J_2 - J_2\gamma_b}, \tag{B12}$$

where we used the notation

$$\Lambda'(\omega) = \sqrt{\left[\left(\omega + \frac{i\gamma_b}{2}\right)^2 - \sigma_1\right]^2 - J_2^2\left(4J_1^2 - \gamma_b^2\right)}. \tag{B13}$$

Substituting Eq. (B12) into Eq. (B9) and applying the residue theorem, we obtain

$$\begin{aligned}\Sigma_0(\omega) &= -\frac{\Omega^2}{J_2\left(J_1 + \frac{\gamma_b}{2}\right)} \oint_{|z|=1} \frac{dz}{2\pi i} \frac{\omega + \frac{i\gamma_b}{2}}{(z - z_+)(z - z_-)},\\ &= \frac{\Omega^2\left(\omega + \frac{i\gamma_b}{2}\right)}{\Lambda'(\omega)} [\Theta_+(z_+) - \Theta_+(z_-)],\end{aligned} \tag{B14}$$

where $\Theta_+(z)$ is given by Eq. (A17).

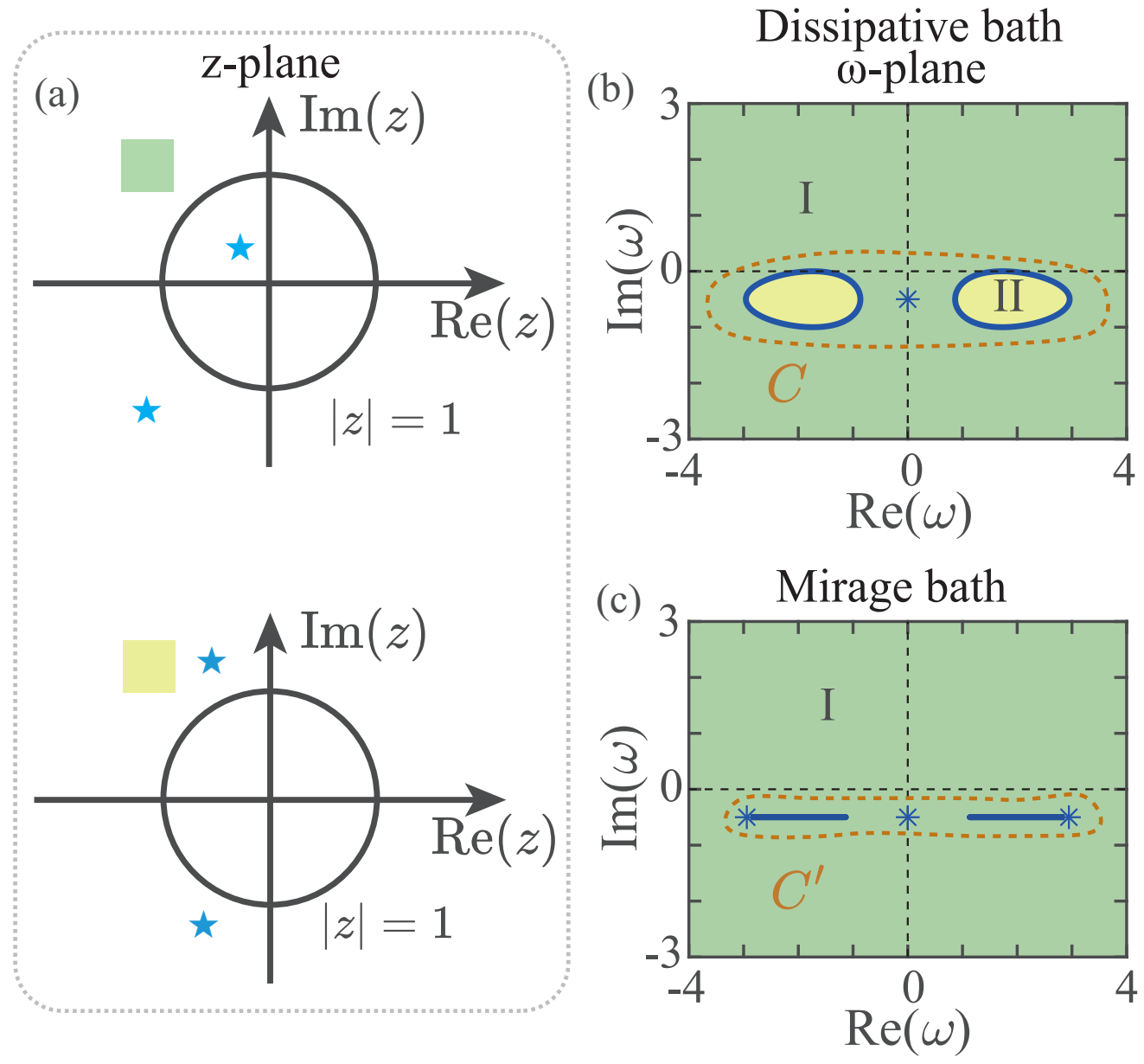

FIG. 12. Analytic structure of the self-energy and single-particle Green function for the dissipative SSH bath. (a) Distribution of the poles of the self-energy [see Eq. (B9)] in $z$ plane. For $\omega \in$ I [green region in panel (b)], one of the two poles $z_\pm$ (pentagram) of the self-energy is inside the unit circle, while the other is outside. For $\omega \in$ II [yellow region in panel (b)], both $z_\pm$ are inside (outside) the unit circle. (b) Analytic structure of $G_0(\omega)$ [see Eq. (B8)] in the complex frequency plane. The blue lines depict the branch loop $\omega = \omega_b'(k)$ [see Eq. (B6)]. The asterisk depicts the pole. The yellow (green) region denotes domains inside (outside) the loop. (c) Analytic structure of the fictitious Green function $G_f(\omega)$ associated with the mirage bath [see Eq. (C2)].

The analytic structure of the self-energy of the dissipative bath is crucially different from that of the closed bath [see Figs. 12(a) and 12(b)]. For the self-energy associated with the closed bath, there always exists one pole encircled by the unit circle. In contrast, for the dissipative bath, the number of poles encircled by the unit circle depends on the frequency $\omega$ [Figs. 12(a) and 12(b)]. (1) For $\omega \in$ I, only one pole is encircled. (2) For $\omega \in$ II, both poles fall inside (outside) the unit circle. Accordingly, the complex frequency domain can be separated into two disjoint domains, where Eq. (B14) acquires different expressions:

$$\Sigma_0(\omega) = \frac{\Omega^2\left(\omega + \frac{i\gamma_b}{2}\right)}{\Lambda'(\omega)} \operatorname{sign}(|z_-| - |z_+|), \quad \omega \in \text{I}, \tag{B15}$$

$$\Sigma_0(\omega) = 0, \quad \omega \in \text{II}. \tag{B16}$$

Thus, we arrive at Eq. (7).

#### b. BSs

Here, we derive the energy and the wavefunction of the BS formed in the complex-energy gap of the dissipative bath, following similar procedures as before.

From Eq. (B8), the energy of the BS is given by the pole equation

$$\omega_{\rm BS} - \Delta' - \Sigma_0(\omega_{\rm BS}) = 0. \tag{B17}$$

In the case $\Delta = 0$ and $\gamma_a = \gamma_b$, i.e., $\Delta' = -i\gamma_b/2$, it follows from Eq. (B16) that $\Sigma_0 = 0$. Therefore, in this case, there always exists a BS with $\omega_{\rm BS} = -i\gamma_b/2$.

The wavefunction for the $A$-BS and $B$-BS, respectively, are derived as follows:

(1) $A$-BS: we obtain

$$\begin{aligned} f_{A,j} &= \Omega\varphi_a \int_{-\pi}^{\pi} \frac{dk}{2\pi} \frac{\left(\omega_{\rm BS} + \frac{i\gamma_b}{2}\right)e^{ikj}}{[\omega_{\rm BS} - \omega'_{b+}(k)][\omega_{\rm BS} - \omega'_{b-}(k)]}, \\ &= -\frac{\Omega\varphi_a}{J_2\left(J_1 + \frac{\gamma_b}{2}\right)} \oint_{|z|=1} \frac{dz}{2\pi i} \frac{\left(\omega_{\rm BS} + \frac{i\gamma_b}{2}\right)z^j}{(z - z_+)(z - z_-)}, \end{aligned} \tag{B18}$$

where $\omega'_{b\pm}(k)$ is given by Eq. (B6) and $z_\pm$ is given by Eq. (B12). After applying the residue theorem, we have

$$f_{A,j} = \begin{cases} \frac{\Omega\varphi_a(\omega_{\rm BS} + \frac{i\gamma_b}{2})[z_+^j\Theta_+(z_+) - z_-^j\Theta_+(z_-)]}{\Lambda'(\omega_{\rm BS})}, & j \geqslant 0, \\ \frac{\Omega\varphi_a(\omega_{\rm BS} + \frac{i\gamma_b}{2})[z_-^j\Theta_-(z_-) - z_+^j\Theta_-(z_+)]}{\Lambda'(\omega_{\rm BS})}, & j < 0, \end{cases} \tag{B19}$$

where $\Theta_\pm(z)$ and $\Lambda'(\omega)$ have been defined in Eqs. (A17) and (B13), respectively.

Similarly, $f_{B,j}$ is calculated as

$$\begin{aligned} f_{B,j} &= \Omega\varphi_a \int_{-\pi}^{\pi} \frac{dk}{2\pi} \frac{\left(J_1 - \frac{\gamma_b}{2} + J_2 e^{ik}\right)e^{ikj}}{[\omega_{\rm BS} - \omega'_{b+}(k)][\omega_{\rm BS} - \omega'_{b-}(k)]} \\ &= \begin{cases} \frac{\Omega\varphi_a[F_j^-(z_+)\Theta_+(z_+) - F_j^-(z_-)\Theta_+(z_-)]}{\Lambda'(\omega_{\rm BS})}, & j \geqslant 0, \\ \frac{\Omega\varphi_a[F_j^-(z_-)\Theta_-(z_-) - F_j^-(z_+)\Theta_-(z_+)]}{\Lambda'(\omega_{\rm BS})}, & j < 0, \end{cases} \end{aligned} \tag{B20}$$

where $F_j^\pm(z)$ is given by Eq. (B10).

(2) $B$-BS: The amplitude of the photonic component $f_{A,j}$ is found to be

$$\begin{aligned} f_{A,j} &= \Omega\varphi_a \int_{-\pi}^{\pi} \frac{dk}{2\pi} \frac{\left(J_1 + \frac{\gamma_b}{2} + J_2 e^{-ik}\right)e^{ikj}}{[\omega_{\rm BS} - \omega'_{b+}(k)][\omega_{\rm BS} - \omega'_{b-}(k)]} \\ &= \begin{cases} \frac{\Omega\varphi_a[F_{-j}^+(z_+^{-1})\Theta_+(z_+) - F_{-j}^+(z_-^{-1})\Theta_+(z_-)]}{\Lambda'(\omega_{\rm BS})}, & j > 0, \\ \frac{\Omega\varphi_a[F_{-j}^+(z_-^{-1})\Theta_-(z_-) - F_{-j}^+(z_+^{-1})\Theta_-(z_+)]}{\Lambda'(\omega_{\rm BS})}, & j \leqslant 0. \end{cases} \end{aligned} \tag{B21}$$

Similarly, we can obtain $f_{B,j}$ as

$$\begin{aligned} f_{B,j} &= \Omega\varphi_a \int_{-\pi}^{\pi} \frac{dk}{2\pi} \frac{\left(\omega_{\rm BS} + \frac{i\gamma_b}{2}\right)e^{ikj}}{[\omega_{\rm BS} - \omega'_{b+}(k)][\omega_{\rm BS} - \omega'_{b-}(k)]} \\ &= \begin{cases} \frac{\Omega\varphi_a(\omega_{\rm BS} + \frac{i\gamma_b}{2})[z_+^j\Theta_+(z_+) - z_-^j\Theta_+(z_-)]}{\Lambda'(\omega_{\rm BS})}, & j > 0, \\ \frac{\Omega\varphi_a(\omega_{\rm BS} + \frac{i\gamma_b}{2})[z_-^j\Theta_-(z_-) - z_+^j\Theta_-(z_+)]}{\Lambda'(\omega_{\rm BS})}, & j \leqslant 0. \end{cases} \end{aligned} \tag{B22}$$

For $\Delta' = -i\gamma_b/2$, Eqs. (B19) and (B22) can be greatly simplified, as summarized in Table II. There, we see that, in contrast to the closed-bath case, the $A$-BS and $B$-BS now switch their chirality at different critical points. In the Figs. 13(a)–13(c), we plot the BSs for $\Delta' = -i\gamma_b/2$ across different regimes of the bath. Moreover, we also compare the analytical results of $f_{A/B,j}$ in A/B configurations with those obtained from the numerical diagonalization of $H_{\rm eff}$ with the bath size $N_b = 500$, for both $\Delta = 0$ and $\Delta \neq 0$, respectively. As shown in the Figs. 13(a)–13(c), a perfect agreement between the analytical and numerical results are found.

TABLE II. Analytical photonic BS in the dissipative SSH bath for $\Delta' = -i\gamma_b/2$ within the complex-energy band gap. Shown are the photonic distributions $f_{A/B,j}$ of the BS in real space. The top (bottom) table presents the expressions for the $A$-BS ($B$-BS), in different parameter regimes of the bath.

| | $A$-BS |
|---|---|
| $\lvert J_1 + \frac{\gamma_b}{2}\rvert > \lvert J_2\rvert$ | $f_{A,j} = 0$ |
| | $f_{B,j} = \begin{cases} -\frac{\Omega\varphi_a}{J_1 + \frac{\gamma_b}{2}}\left(-\frac{J_2}{J_1 + \frac{\gamma_b}{2}}\right)^j, & j \geqslant 0, \\ 0, & j < 0, \end{cases}$ |
| $\lvert J_1 + \frac{\gamma_b}{2}\rvert < \lvert J_2\rvert$ | $f_{A,j} = 0$ |
| | $f_{B,j} = \begin{cases} 0, & j \geqslant 0, \\ \frac{\Omega\varphi_a}{J_1 + \frac{\gamma_b}{2}}\left(-\frac{J_1 + \frac{\gamma_b}{2}}{J_2}\right)^{\lvert j\rvert}, & j < 0. \end{cases}$ |
| | $B$-BS |
| $\lvert J_1 - \frac{\gamma_b}{2}\rvert > \lvert J_2\rvert$ | $f_{A,j} = \begin{cases} 0, & j > 0, \\ -\frac{\Omega\varphi_a}{J_1 - \frac{\gamma_b}{2}}\left(-\frac{J_2}{J_1 - \frac{\gamma_b}{2}}\right)^{\lvert j\rvert}, & j \leqslant 0. \end{cases}$ |
| | $f_{B,j} = 0$ |
| $\lvert J_1 - \frac{\gamma_b}{2}\rvert < \lvert J_2\rvert$ | $f_{A,j} = \begin{cases} \frac{\Omega\varphi_a}{J_1 - \frac{\gamma_b}{2}}\left(-\frac{J_1 - \frac{\gamma_b}{2}}{J_2}\right)^j, & j > 0, \\ 0, & j \leqslant 0. \end{cases}$ |
| | $f_{B,j} = 0$ |

### 2. Two emitters

In this subsection, we consider two QEs in the band gap: the first QE is coupled to the sublattice $A$ at $j_1 = 0$, while the second QE is coupled to sublattice $B$ downstream at $j_2 = d > 0$. We are interested in the BS-mediated effective interaction between them, i.e.,

$$H_{\rm int} = \Sigma_d^{AB} a_1^\dagger a_2 + \Sigma_d^{BA} a_2^\dagger a_1. \tag{B23}$$

The goal of this section is to derive $\Sigma_d^{AB}$ and $\Sigma_d^{BA}$, which leads to Eq. (9).

The dynamics of the considered two QEs are described by the two-emitter Green function

$$G_d(\omega) = \frac{1}{\omega - \Delta' - \Sigma_d(\omega)}, \tag{B24}$$

where the self-energy matrix $\Sigma_d(\omega)$ is of the form (A24)

$$\Sigma_d(\omega) = \begin{pmatrix} \Sigma_0(\omega) & \Sigma_d^{AB}(\omega) \\ \Sigma_d^{BA}(\omega) & \Sigma_0(\omega) \end{pmatrix}.$$

However, different from the closed bath, one generally has $\Sigma_d^{AB} \neq \Sigma_d^{BA*}$ for a dissipative bath. The nondiagonal elements are explicitly calculated as

$$\begin{aligned} \Sigma_d^{AB}(\omega) &= \Omega^2 \int \frac{dk}{2\pi} \frac{\left(J_1 + \frac{\gamma_b}{2} + J_2 e^{-ik}\right)e^{-idk}}{[\omega - \omega'_{b+}(k)][\omega - \omega'_{b-}(k)]} \\ &= \Omega^2 \oint_{|z|=1} \frac{dz}{2\pi i z} \frac{F_d^+(z^{-1})}{\left(\omega + \frac{i\gamma_b}{2}\right)^2 - F_0^+(z^{-1})F_0^-(z)} \\ &= \begin{cases} \frac{\Omega^2[F_d^+(z_-^{-1})\Theta_-(z_-) - F_d^+(z_+^{-1})\Theta_-(z_+)]}{\Lambda'(\omega)}, & d \geqslant 0, \\ \frac{\Omega^2[F_d^+(z_+^{-1})\Theta_+(z_+) - F_d^+(z_-^{-1})\Theta_+(z_-)]}{\Lambda'(\omega)}, & d < 0, \end{cases} \end{aligned} \tag{B25}$$

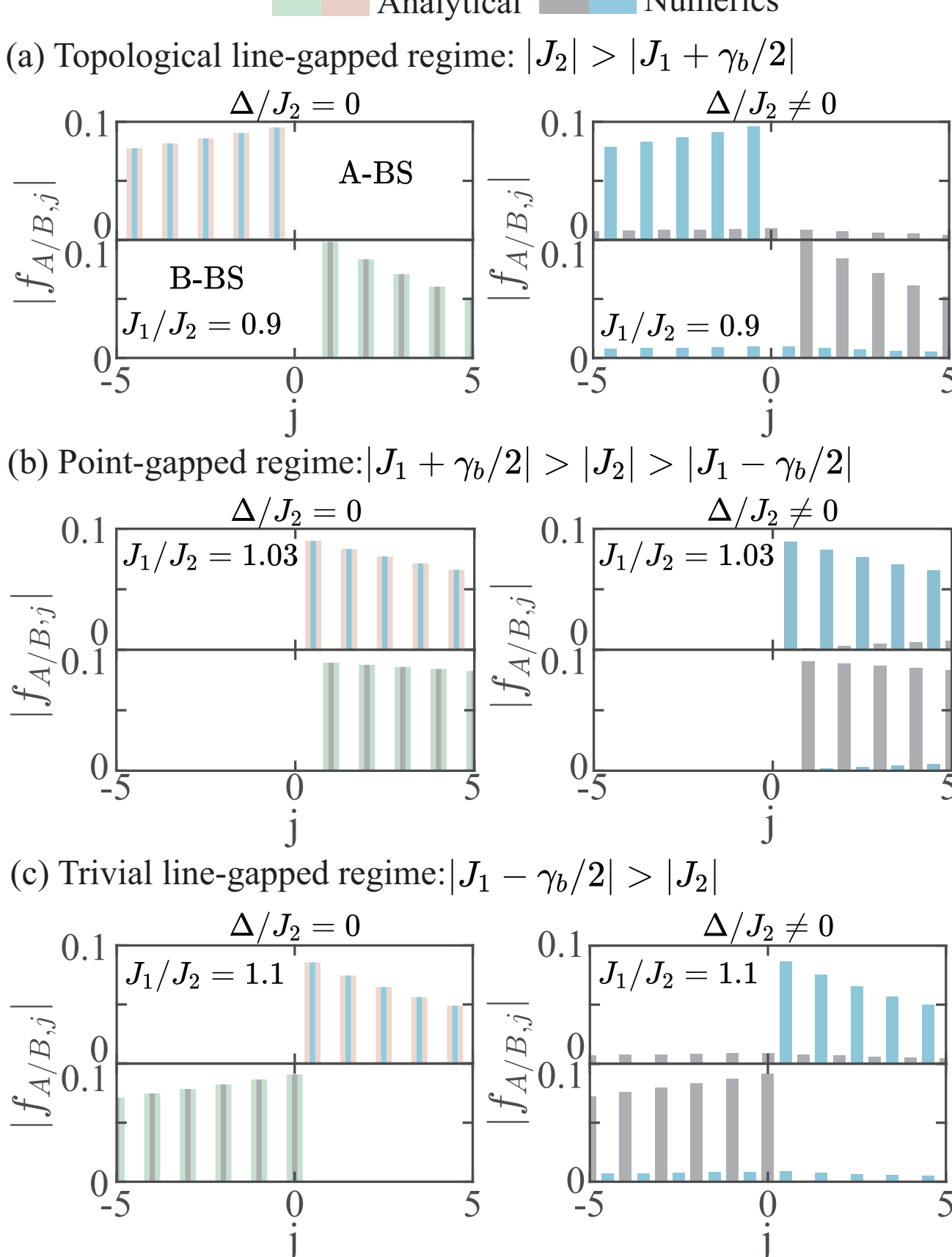

FIG. 13. Comparison between the analytic and numerical results for the photonic distribution $|f_{A/B,j}|$ of the BS in the dissipative SSH bath. We consider the emitter with $\Delta = 0$ in the left panel and $\Delta/J_2 = 0.02$ in the right panel, when $\gamma_a/J_2 = \gamma_b/J_2 = 0.1$ and $\Omega/J_2 = 0.1$. Results are shown for (a) $J_1/J_2 = 0.9$ in the topological line-gapped regime, (b) $J_1/J_2 = 1.03$ in the point-gapped regime, and (c) $J_1/J_2 = 1.1$ in the trivial line-gapped regime. Both $A$-BS and $B$-BS are illustrated. Analytical results are obtained from Table II. Numerical results are obtained through the numerical diagonalization of the effective Hamiltonian (B4) with a single QE and bath size $N_b = 500$.

where $z_\pm$ is given by Eq. (B12), $\Lambda'(\omega)$ is given by Eq. (B13), $\Theta_+(x)$ is given by Eq. (A17), and $F_j^\pm(x)$ is given by Eq. (B10). Similarly, we obtain

$$
\begin{aligned}
\Sigma_d^{BA}(\omega) &= \Omega^2 \int \frac{dk}{2\pi} \frac{\left(J_1 - \frac{\gamma_b}{2} + J_2 e^{ik}\right)e^{idk}}{[\omega - \omega'_{b+}(k)][\omega - \omega'_{b-}(k)]} \\
&= \begin{cases} \frac{\Omega^2[F_d^-(z_+)\Theta_+(z_+) - F_d^-(z_-)\Theta_+(z_-)]}{\Lambda'(\omega)}, & d \geqslant 0, \\ \frac{\Omega^2[F_d^-(z_-)\Theta_-(z_-) - F_d^-(z_+)\Theta_-(z_+)]}{\Lambda'(\omega)}, & d < 0, \end{cases}
\end{aligned} \tag{B26}
$$

The BS-mediated interaction is best illustrated for the choice $\Delta' = -i\gamma_b/2$ in the middle of the gap. In this case, the BS has energy $\omega_{\rm BS} = -i\gamma_b/2$. For the weak coupling, applying the single-pole approximation to the self-energy [(B25) and (B26)] with $\omega \approx -i\gamma_b/2$, we find

$$
\Sigma_d^{AB} = \begin{cases} -\frac{\Omega^2}{J_1 - \frac{\gamma_b}{2}}\left(-\frac{J_2}{J_1 - \frac{\gamma_b}{2}}\right)^d, & d \geqslant 0, \\ 0, & d < 0, \end{cases} \tag{B27}
$$

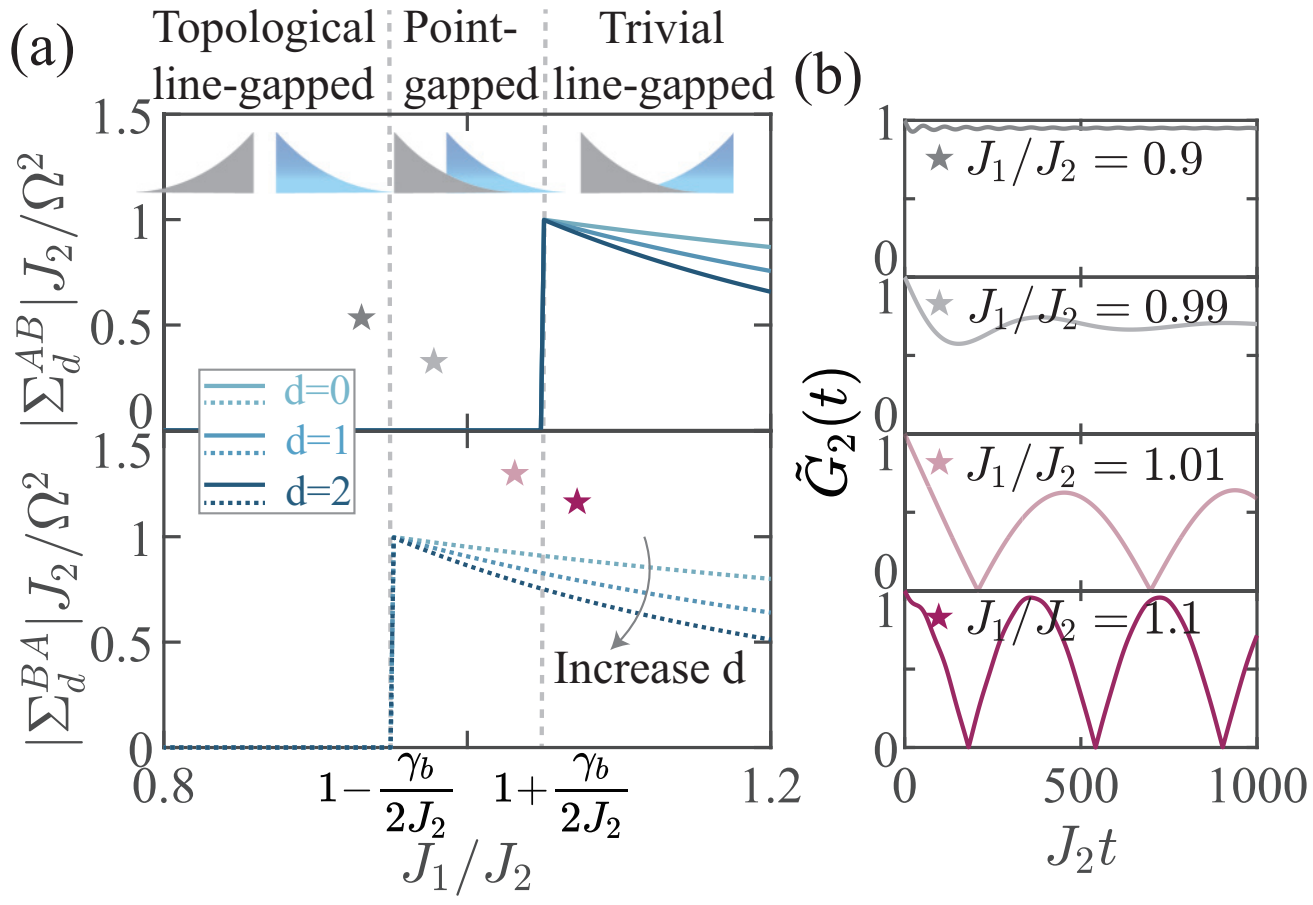

FIG. 14. BS-mediated interaction in the dissipative SSH bath for $\Delta = 0$, $\gamma_a/J_2 = \gamma_b/J_2 = 0.1$, and $\Omega/J_2 = 0.1$. The first QE ($a_1$) is coupled to sublattice $A$ at $j_1 = 0$, and the second QE ($a_2$) is coupled to sublattice $B$ at $j_2 = d$. (a) Absolute value of the dipolar coupling $\Sigma_d^{AB}$ (top panel) and $\Sigma_d^{BA}$ (bottom panel) as a function of $J_1/J_2$. Results are obtained from Eqs. (B27)–(B30). The insets show the shape of the photon $A$-BS (gray) and $B$-BS (blue) in the three distinct gap regimes of the bath, respectively. (b) Probability amplitude $\tilde{G}_2(t) = e^{\gamma_b t/2}|\langle 0|a_2(t)a_2^\dagger(0)|0\rangle|$ to find an initially excited second QE in the excited state, for $J_1/J_2 = 0.9$, $J_1/J_2 = 0.99$, $J_1/J_2 = 1.01$, and $J_1/J_2 = 1.1$. Results are shown for $d = 0$ and are obtained via the Green function approach. Other parameters are the same as panel (a).

for $|J_1 - \gamma_b/2| > |J_2|$, and

$$
\Sigma_d^{AB} = \begin{cases} 0, & d \geqslant 0, \\ \frac{\Omega^2}{J_1 - \frac{\gamma_b}{2}}\left(-\frac{J_1 - \frac{\gamma_b}{2}}{J_2}\right)^{|d|}, & d < 0, \end{cases} \tag{B28}
$$

for $|J_1 - \gamma_b/2| < |J_2|$. Similarly, for interaction $\Sigma_d^{BA}$, we have

$$
\Sigma_d^{BA} = \begin{cases} -\frac{\Omega^2}{J_1 + \frac{\gamma_b}{2}}\left(-\frac{J_2}{J_1 + \frac{\gamma_b}{2}}\right)^d, & d \geqslant 0, \\ 0, & d < 0, \end{cases} \tag{B29}
$$

for $|J_1 + \gamma_b/2| > |J_2|$, and

$$
\Sigma_d^{BA} = \begin{cases} 0, & d \geqslant 0, \\ \frac{\Omega^2}{J_1 + \frac{\gamma_b}{2}}\left(-\frac{J_1 + \frac{\gamma_b}{2}}{J_2}\right)^{|d|}, & d < 0, \end{cases} \tag{B30}
$$

for $|J_1 + \gamma_b/2| < |J_2|$.

In Fig. 14(a), we illustrate $\Sigma_d^{AB}$ and $\Sigma_d^{BA}$ as a function of $J_1/J_2$ for $\Delta' = -i\gamma_b/2$ under various $d$. We see that $\Sigma_d^{AB}$ and $\Sigma_d^{BA}$ witness different critical points. In Fig. 14(b), we assume $a_2$ is initially excited while $a_1$ is initially in the vacuum state, and plot the normalized probability to find $a_2$ in the excited state, $\tilde{G}_2(t) = e^{\gamma_b t/2}|\langle 0|a_2(t)a_2^\dagger(0)|0\rangle|$, across different regimes. Intriguingly, for $J_1/J_2 = 1.01$ in the point gap, the dynamics exhibits oscillations, clearly indicating the presence of interaction between the two QEs, as opposed to what is expected from the results $\Sigma_d^{AB} = 0$ in this regime. In other words, the dynamics in Fig. 14(b) does not reflect the topology of the dissipative bath.

Finally, we note that, when two QEs are coupled to the same sublattice, one can similarly calculate the interaction

strength, yielding

$$\Sigma_d^{AA/BB}(\omega) = \Omega^2 \int \frac{dk}{2\pi} \frac{\left(\omega + \frac{i\gamma_b}{2}\right)e^{idk}}{[\omega - \omega'_{b+}(k)][\omega - \omega'_{b-}(k)]}$$
$$= \begin{cases} \frac{\Omega^2(\omega+\frac{i\gamma_b}{2})[z_+^d\Theta_+(z_+)-z_-^d\Theta_+(z_-)]}{\Lambda'(\omega)}, & d \geqslant 0, \\ \frac{\Omega^2(\omega+\frac{i\gamma_b}{2})[z_-^d\Theta_-(z_-)-z_+^d\Theta_-(z_+)]}{\Lambda'(\omega)}, & d < 0. \end{cases} \tag{B31}$$

For $\Delta' = -i\gamma_b/2$, we have $\omega_{\rm BS} = -i\gamma_b/2$. For weak coupling, under the single-pole approximation $\omega \approx \omega_{\rm BS}$, we obtain $\Sigma_d^{AA/BB} = 0$, i.e., no interaction between the two QEs on the sublattices.

## APPENDIX C: THE MIRAGE BATH ON THE SECOND RIEMANN SHEET

Following Ref. [61], we show exactly the same emitter dynamics can be obtained from a mirage bath on the second Riemann sheet through analytic continuation [see Fig. 12(c)]. In Appendix C 1, we derive the single-QE dynamics and the BS associated with the mirage bath. In Appendix C 2, we derive the BS-mediated interaction [see Eq. (17)]. In Appendix C 3, we extend to the case with a string of emitters.

### 1. Single QE

According to Eq. (A6), mathematically, the single-emitter dynamics involves a contour integration along the contour $C \in \mathrm{I}$ [Fig. 12(b)], dashed curve], which cannot be contracted across the branch loop [Fig. 12(b), blue curve] in the first Riemann sheet (green region). Inside the branch loop ($\omega \in$ II), one has $\Sigma_0(\omega) = 0$. However, through analytic continuation, we can extend the self-energy $\Sigma_0(\omega)$ defined for $\omega \in$ I in Eq. (B15) to region II of the second Riemann sheet [Fig. 12(c)], i.e.,

$$G_0(t) = \oint_C \frac{d\omega}{2\pi} e^{-i\omega t} G(\omega) = \oint_{C'} \frac{d\omega}{2\pi} e^{-i\omega t} G_f(\omega). \tag{C1}$$

Here, $G_f(\omega)$ is the fictitious Green function associated with the mirage bath,

$$G_f(\omega) = \frac{1}{\omega - \Delta' - \Sigma_0^f(\omega)}, \tag{C2}$$

where the self-energy,

$$\Sigma_0^f(\omega) = \frac{\Omega^2\left(\omega + \frac{i\gamma_b}{2}\right)}{\Lambda'(\omega)} \mathrm{sign}(|z_-| - |z_+|), \quad \omega \in \mathrm{I, II}, \tag{C3}$$

with $\Lambda'(\omega)$ given by Eq. (B13), is the analytic continuation of $\Sigma_0(\omega)$ ($\omega \in$ I) in Eq. (B15) into the domain II of the second Riemann sheet.

We now explicitly prove Eq. (C3). The analytic continuation can be performed by deforming the integral contour of Eq. (B9) from $|z| = 1$ to $|z| = r$ with

$$r = \sqrt{\frac{J_1 - \frac{\gamma_b}{2}}{J_1 + \frac{\gamma_b}{2}}}. \tag{C4}$$

Specifically, we have

$$\Sigma_0^f(\omega) = \Omega^2 \int_{-\pi}^{\pi} \frac{dk}{2\pi} \frac{\omega + \frac{i\gamma_b}{2}}{[\omega - \omega'_{b+}(k)][\omega - \omega'_{b-}(k)]}$$
$$= \Omega^2 \oint_{|z|=1} \frac{dz}{2\pi i z} \frac{\omega + \frac{i\gamma_b}{2}}{\left(\omega + \frac{i\gamma_b}{2}\right)^2 - F_0^+(z^{-1})F_0^-(z)}$$
$$= \Omega^2 \oint_{|z|=r} \frac{dz}{2\pi i z} \frac{\omega + \frac{i\gamma_b}{2}}{\left(\omega + \frac{i\gamma_b}{2}\right)^2 - F_0^+(z^{-1})F_0^-(z)}$$
$$= \Omega^2 \oint_{|z'|=1} \frac{dz'}{2\pi i z'} \frac{\omega + \frac{i\gamma_b}{2}}{\left(\omega + \frac{i\gamma_b}{2}\right)^2 - F_0^+(r^{-1}z'^{-1})F_0^-(rz')}$$
$$= \Omega^2 \int_{-\pi}^{\pi} \frac{dk}{2\pi} \frac{\omega + \frac{i\gamma_b}{2}}{\left[\omega - \omega^f_{b+}(k)\right]\left[\omega - \omega^f_{b-}(k)\right]}, \tag{C5}$$

where $\omega_b^f(k)$ is interpreted as the spectrum of the mirage bath:

$$\omega_b^f(k) = -\frac{i\gamma_b}{2} \pm \sqrt{(\tilde{J}_1 + J_2 \cos k)^2 + J_2^2 \sin^2 k}, \tag{C6}$$

with the effective intrasite coupling

$$\tilde{J}_1^2 = J_1^2 - (\gamma_b/2)^2. \tag{C7}$$

Equation (C5) is further calculated as

$$\Sigma_0^f(\omega) = -\frac{\Omega^2}{\tilde{J}_1 J_2} \oint_{|z|=1} \frac{dz}{2\pi i} \frac{\omega + \frac{i\gamma_b}{2}}{(z - z_+)(z - z_-)},$$
$$= \frac{\Omega^2\left(\omega + \frac{i\gamma_b}{2}\right)}{\Lambda'(\omega)} \mathrm{sign}(|z_-| - |z_+|). \tag{C8}$$

Here, $z_\pm$ is the two roots of the equation

$$-\tilde{J}_1 J_2 z^2 + \left[\left(\omega + \frac{i\gamma_b}{2}\right)^2 - \tilde{J}_1^2 - J_2^2\right] z - \tilde{J}_1 J_2 = 0, \tag{C9}$$

giving

$$z_\pm = -\frac{\tilde{J}_1^2 + J_2^2 - \left(\omega + \frac{i\gamma_b}{2}\right)^2 \pm \Lambda'(\omega)}{2\tilde{J}_1 J_2}. \tag{C10}$$

Thus, the self-energy in region I of the first Riemann sheet and II of the second Riemann sheet has the unified expression, which proves Eq. (C3).

We now derive the BSs associated with the mirage bath. From Eq. (C2), the energy of the BS is given by the pole equation

$$\omega_{\rm BS} - \Delta' - \Sigma_0^f(\omega_{\rm BS}) = 0. \tag{C11}$$

For the case $\Delta = 0$ and $\gamma_a = \gamma_b$, i.e., $\Delta' = -i\gamma_b/2$, it follows from Eq. (C3) that $\Sigma_0^f = 0$. Thus, in this case, there exists one exact solution: $\omega_{\rm BS} = -i\gamma_b/2$.

The wavefunctions of the $A$-BS and $B$-BS, respectively, are calculated as follows.

TABLE III. Mirage bath: Analysis of BS wavefunction with energy $\omega_{\rm BS} = -i\gamma_b/2$ at $\Delta = 0$ and $\gamma_a = \gamma_b$, while a QE is placed at $j = 0$ in the mirage bath. $\varphi_a$ is the QE wavefunction, which can be obtained by normalization.

| | $A$-BS | $B$-BS |
|---|---|---|
| $\left\|\sqrt{J_1^2 - \frac{\gamma_b^2}{4}}\right\| > \|J_2\|$ | $f_{A,j} = 0$ | $f_{B,j} = 0$ |
| | $f_{B,j} = \begin{cases} -\frac{\Omega\varphi_a}{\sqrt{J_1^2-\gamma_b^2/4}}\Big(-\frac{J_2}{\sqrt{J_1^2-\gamma_b^2/4}}\Big)^j, & j \geqslant 0, \\ 0, & j < 0 \end{cases}$ | $f_{A,j} = \begin{cases} 0, & j > 0, \\ -\frac{\Omega\varphi_a}{\sqrt{J_1^2-\gamma_b^2/4}}\Big(-\frac{J_2}{\sqrt{J_1^2-\gamma_b^2/4}}\Big)^{\|j\|}, & j \leqslant 0 \end{cases}$ |
| $\left\|\sqrt{J_1^2 - \frac{\gamma_b^2}{4}}\right\| < \|J_2\|$ | $f_{A,j} = 0$ | $f_{B,j} = 0$ |
| | $f_{B,j} = \begin{cases} 0, & j \geqslant 0, \\ \frac{\Omega\varphi_a}{\sqrt{J_1^2-\gamma_b^2/4}}\Big(-\frac{\sqrt{J_1^2-\gamma_b^2/4}}{J_2}\Big)^{\|j\|}, & j < 0 \end{cases}$ | $f_{A,j} = \begin{cases} \frac{\Omega\varphi_a}{\sqrt{J_1^2-\gamma_b^2/4}}\Big(-\frac{\sqrt{J_1^2-\gamma_b^2/4}}{J_2}\Big)^j, & j > 0, \\ 0, & j \leqslant 0 \end{cases}$ |

(1) $A$-BS: Using Eqs. (C6) and (C10), the amplitude of photonic component $f_{A,j}$ is written as

$$
\begin{aligned}
f_{A,j} &= \Omega\varphi_a \int_{-\pi}^{\pi} \frac{dk}{2\pi} \frac{\big(\omega_{\rm BS} + \frac{i\gamma_b}{2}\big)e^{ikj}}{\big[\omega_{\rm BS} - \omega_{b+}^f(k)\big]\big[\omega_{\rm BS} - \omega_{b-}^f(k)\big]} \\
&= -\frac{\Omega\varphi_a}{\tilde{J}_1 J_2} \oint_{|z|=1} \frac{dz}{2\pi i} \frac{\big(\omega_{\rm BS} + \frac{i\gamma_b}{2}\big)z^j}{(z - z_+)(z - z_-)},
\end{aligned}
\tag{C12}
$$

where $\varphi_a$ is obtained from normalization. Using the residue theorem and using Eq. (A17), we write

$$
f_{A,j} = \begin{cases} \frac{\Omega\varphi_a(\omega_{\rm BS}+\frac{i\gamma_b}{2})[z_+^j\Theta_+(z_+)-z_-^j\Theta_+(z_-)]}{\Lambda'(\omega_{\rm BS})}, & j \geqslant 0, \\ \frac{\Omega\varphi_a(\omega_{\rm BS}+\frac{i\gamma_b}{2})[z_-^j\Theta_-(z_-)-z_+^j\Theta_-(z_+)]}{\Lambda'(\omega_{\rm BS})}, & j < 0. \end{cases}
\tag{C13}
$$

In a similar fashion, we obtain

$$
\begin{aligned}
f_{B,j} &= \Omega\varphi_a \int_{-\pi}^{\pi} \frac{dk}{2\pi} \frac{(\tilde{J}_1 + J_2 e^{ik})e^{ikj}}{\big[\omega_{\rm BS} - \omega_{b+}^f(k)\big]\big[\omega_{\rm BS} - \omega_{b-}^f(k)\big]} \\
&= \begin{cases} \frac{\Omega\varphi_a[\tilde{F}_j(z_+)\Theta_+(z_+)-\tilde{F}_j(z_-)\Theta_+(z_-)]}{\Lambda'(\omega_{\rm BS})}, & j \geqslant 0, \\ \frac{\Omega\varphi_a[\tilde{F}_j(z_-)\Theta_-(z_-)-\tilde{F}_j(z_+)\Theta_-(z_+)]}{\Lambda'(\omega_{\rm BS})}, & j < 0, \end{cases}
\end{aligned}
\tag{C14}
$$

where we used

$$
\tilde{F}_j(z) = (\tilde{J}_1 + J_2 z)z^j.
\tag{C15}
$$

(2) $B$-BS: Following similar procedures, we obtain

$$
\begin{aligned}
f_{A,j} &= \Omega\varphi_a \int_{-\pi}^{\pi} \frac{dk}{2\pi} \frac{(\tilde{J}_1 + J_2 e^{-ik})e^{ikj}}{\big[\omega_{\rm BS} - \omega_{b+}^f(k)\big]\big[\omega_{\rm BS} - \omega_{b-}^f(k)\big]} \\
&= \begin{cases} \frac{\Omega\varphi_a[\tilde{F}_{-j}(z_+^{-1})\Theta_+(z_+)-\tilde{F}_{-j}(z_-^{-1})\Theta_+(z_-)]}{\Lambda'(\omega_{\rm BS})}, & j > 0, \\ \frac{\Omega\varphi_a[\tilde{F}_{-j}(z_-^{-1})\Theta_-(z_-)-\tilde{F}_{-j}(z_+^{-1})\Theta_-(z_+)]}{\Lambda'(\omega_{\rm BS})}, & j \leqslant 0. \end{cases}
\end{aligned}
\tag{C16}
$$

In addition, $f_{B,j}$ is calculated as

$$
\begin{aligned}
f_{B,j} &= \Omega\varphi_a \int_{-\pi}^{\pi} \frac{dk}{2\pi} \frac{\big(\omega_{\rm BS} + \frac{i\gamma_b}{2}\big)e^{ikj}}{\big[\omega_{\rm BS} - \omega_{b+}^f(k)\big]\big[\omega_{\rm BS} - \omega_{b-}^f(k)\big]} \\
&= \begin{cases} \frac{\Omega\varphi_a(\omega_{\rm BS}+\frac{i\gamma_b}{2})[z_+^j\Theta_+(z_+)-z_-^j\Theta_+(z_-)]}{\Lambda'(\omega_{\rm BS})}, & j > 0, \\ \frac{\Omega\varphi_a(\omega_{\rm BS}+\frac{i\gamma_b}{2})[z_-^j\Theta_-(z_-)-z_+^j\Theta_-(z_+)]}{\Lambda'(\omega_{\rm BS})}, & j \leqslant 0. \end{cases}
\end{aligned}
\tag{C17}
$$

The expressions of the $A$-BS and $B$-BS are most transparent for $\Delta = 0$ and $\gamma_a = \gamma_b$, as summarized in Table. III and explicitly visualized in the left panels of Figs. 15(a) and 15(b). There, we see that the $A$-BS and $B$-BS in the mirage bath switch their chirality at the same phase-transition points, in contrast to those in the physical bath. In the Figs. 15(a) and 15(b), we also compare the analytical results of $f_{A/B,j}$ in A/B configurations with the results obtained from the numerical diagonalization of $H_{\rm eff}$ with the bath size $N_b = 500$, for both $\Delta = 0$ and $\Delta \neq 0$, respectively. Again, perfect agreement is found.

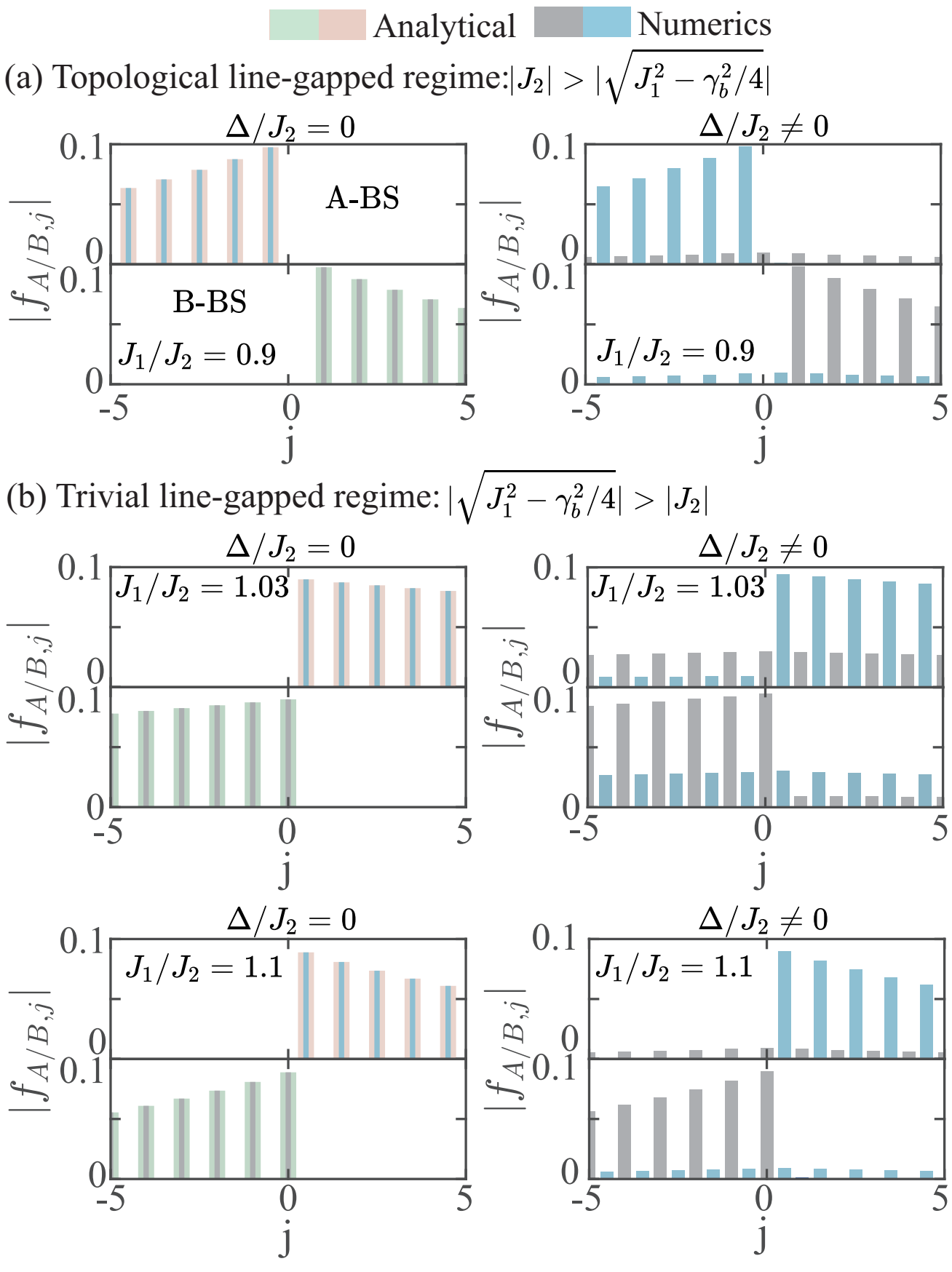

FIG. 15. Comparison between the analytic and numerical results for $|f_{A/B,j}|$ of the BS associated with the mirage bath on the second Riemann sheet. The emitter has $\Delta = 0$ in the left panel and $\Delta/J_2 = 0.02$ in the right panel. Same parameters are used as in Fig. 13. Results are shown for (a) topological and (b) trivial regimes of the mirage bath. Analytical results are obtained from Table III. Numerical results are obtained through numerical diagonalization of the fictitious Hamiltonian (C28) with a single QE and bath size $N_b = 500$.

#### 2. Two emitters

Here, we analytically continue the two-emitter's Green function from region $\omega \in$ I (green) into the region $\omega \in$ II (yellow region of Fig. 12) and show the emergence of the mirage bath. To this end, we apply the analytic continuation by deforming the integral contour of Eq. (B25) from $|z| = 1$ to $|z| = r$, with the notation (B10), we write

$$
\begin{aligned}
\Sigma^{AB}_{d,f}(\omega) &= \Omega^2 \int \frac{dk}{2\pi} \frac{\left(J_1 + \frac{\gamma_b}{2} + J_2 e^{-ik}\right) e^{-idk}}{[\omega - \omega'_{b+}(k)][\omega - \omega'_{b-}(k)]} \\
&= \Omega^2 \oint_{|z|=1} \frac{dz}{2\pi i z} \frac{F^+_d(z^{-1})}{\left(\omega + \frac{i\gamma_b}{2}\right)^2 - F^+_0(z^{-1}) F^-_0(z)} \\
&= \Omega^2 \oint_{|z|=r} \frac{dz}{2\pi i z} \frac{F^+_d(z^{-1})}{\left(\omega + \frac{i\gamma_b}{2}\right)^2 - F^+_0(z^{-1}) F^-_0(z)} \\
&= \Omega^2 \oint_{|z'|=1} \frac{dz'}{2\pi i z'} \frac{F^+_d(r^{-1} z'^{-1})}{\left(\omega + \frac{i\gamma_b}{2}\right)^2 - F^+_0(r^{-1} z'^{-1}) F^-_0(r z')} \\
&= \Omega^2 r^{-(d+1)} \int_{-\pi}^{\pi} \frac{dk}{2\pi} \frac{(\tilde{J}_1 + J_2 e^{-ik}) e^{-ikd}}{\left[\omega - \omega^f_{b+}(k)\right]\left[\omega - \omega^f_{b-}(k)\right]}.
\end{aligned} \tag{C18}
$$

In the last line, we have transformed back to the variable $k$ through $z' = e^{ik}$ and used Eq. (C7).

Similarly, we obtain

$$
\begin{aligned}
\Sigma^{BA}_{d,f}(\omega) &= \Omega^2 \int \frac{dk}{2\pi} \frac{\left(J_1 - \frac{\gamma_b}{2} + J_2 e^{ik}\right) e^{idk}}{[\omega - \omega'_{b+}(k)][\omega - \omega'_{b-}(k)]} \\
&= \Omega^2 \oint_{|z|=1} \frac{dz}{2\pi i z} \frac{F^-_d(z)}{\left(\omega + \frac{i\gamma_b}{2}\right)^2 - F^+_0(z^{-1}) F^-_0(z)} \\
&= \Omega^2 \oint_{|z|=r} \frac{dz}{2\pi i z} \frac{F^-_d(z)}{\left(\omega + \frac{i\gamma_b}{2}\right)^2 - F^+_0(z^{-1}) F^-_0(z)} \\
&= \Omega^2 \oint_{|z'|=1} \frac{dz'}{2\pi i z'} \frac{F^-_d(r z')}{\left(\omega + \frac{i\gamma_b}{2}\right)^2 - F^+_0(r^{-1} z'^{-1}) F^-_0(r z')} \\
&= \Omega^2 r^{(d+1)} \int \frac{dk}{2\pi} \frac{(\tilde{J}_1 + J_2 e^{ik}) e^{ikd}}{\left[\omega - \omega^f_{b+}(k)\right]\left[\omega - \omega^f_{b-}(k)\right]}.
\end{aligned} \tag{C19}
$$

Note that, through comparisons between $\Sigma_d$ and $\Sigma_{d,f}$, we can write the effective total Hamiltonian with the first QE on sublattice $A$ at $j_1 = 0$ and the second QE on sublattice $B$ at $j_2 = d$ as

$$
\begin{aligned}
H^f_{\rm eff} &= \Delta' \sum_m a^\dagger_m a_m + \Omega[b^\dagger_{A,0} a_1 + r^{-(d+1)} b^\dagger_{B,d} a_2] \\
&\quad + \Omega[a^\dagger_1 b_{A,0} + r^{(d+1)} a^\dagger_2 b_{B,d}] + H^f_b,
\end{aligned} \tag{C20}
$$

where the effective mirage-bath Hamiltonian is

$$
\begin{aligned}
H^f_b &= \sum_j [\tilde{J}_1 b^\dagger_{A,j} b_{B,j} + J_2 b^\dagger_{A,j+1} b_{B,j} + \text{H.c.}] \\
&\quad - i\frac{\gamma_b}{2} \sum_j (b^\dagger_{A,j} b_{A,j} + b^\dagger_{B,j} b_{B,j}).
\end{aligned} \tag{C21}
$$

Now we explicitly calculate $\Sigma^{AB/BA}_{d,f}$. Applying the residue theorem to Eq. (C18), we obtain

$$
\begin{aligned}
\Sigma^{AB}_{d,f}(\omega) &= \Omega^2 r^{-(d+1)} \int_{-\pi}^{\pi} \frac{dk}{2\pi} \frac{(\tilde{J}_1 + J_2 e^{-ik}) e^{-ikd}}{\left[\omega - \omega^f_{b+}(k)\right]\left[\omega - \omega^f_{b-}(k)\right]} \\
&= \begin{cases} \frac{\Omega^2[\tilde{F}_d(z_-^{-1})\Theta_-(z_-) - \tilde{F}_d(z_+^{-1})\Theta_-(z_+)]}{r^{d+1}\Lambda'(\omega)}, & d \geqslant 0, \\ \frac{\Omega^2[\tilde{F}_d(z_+^{-1})\Theta_+(z_+) - \tilde{F}_d(z_-^{-1})\Theta_+(z_-)]}{r^{d+1}\Lambda'(\omega)}, & d < 0. \end{cases}
\end{aligned} \tag{C22}
$$

Similarly, Eq. (C19) is calculated as

$$
\begin{aligned}
\Sigma^{BA}_{d,f}(\omega) &= \Omega^2 r^{(d+1)} \int \frac{dk}{2\pi} \frac{(\tilde{J}_1 + J_2 e^{ik}) e^{ikd}}{\left[\omega - \omega^f_{b+}(k)\right]\left[\omega - \omega^f_{b-}(k)\right]} \\
&= \begin{cases} \frac{\Omega^2 r^{d+1}[\tilde{F}_d(z_+)\Theta_+(z_+) - \tilde{F}_d(z_-)\Theta_+(z_-)]}{\Lambda'(\omega)}, & d \geqslant 0, \\ \frac{\Omega^2 r^{d+1}[\tilde{F}_d(z_-)\Theta_-(z_-) - \tilde{F}_d(z_+)\Theta_-(z_+)]}{\Lambda'(\omega)}, & d < 0, \end{cases}
\end{aligned} \tag{C23}
$$

When $\Delta' = -i\gamma_b/2$, the BS has the complex energy $\omega_{\rm BS} = -i\gamma_b/2$. For the weak coupling, we calculate Eqs. (C22) and (C23) by making the single-pole approximation $\omega \approx \omega_{\rm BS}$. For $|\sqrt{J_1^2 - \gamma_b^2/4}| > |J_2|$, we find

$$
\Sigma^{AB}_{d,f} = \begin{cases} -\frac{\Omega^2}{J_1 - \frac{\gamma_b}{2}}\left(-\frac{J_2}{J_1 - \frac{\gamma_b}{2}}\right)^d, & d \geqslant 0, \\ 0, & d < 0, \end{cases} \tag{C24}
$$

$$
\Sigma^{BA}_{d,f} = \begin{cases} -\frac{\Omega^2}{J_1 + \frac{\gamma_b}{2}}\left(-\frac{J_2}{J_1 + \frac{\gamma_b}{2}}\right)^d, & d \geqslant 0, \\ 0, & d < 0. \end{cases} \tag{C25}
$$

For $|\sqrt{J_1^2 - \gamma_b^2/4}| < |J_2|$, we find

$$
\Sigma^{AB}_{d,f} = \begin{cases} 0, & d \geqslant 0, \\ \frac{\Omega^2}{J_1 - \frac{\gamma_b}{2}}\left(-\frac{J_1 - \frac{\gamma_b}{2}}{J_2}\right)^{|d|}, & d < 0, \end{cases} \tag{C26}
$$

$$
\Sigma^{BA}_{d,f} = \begin{cases} 0, & d \geqslant 0, \\ \frac{\Omega^2}{J_1 + \frac{\gamma_b}{2}}\left(-\frac{J_1 + \frac{\gamma_b}{2}}{J_2}\right)^{|d|}, & d < 0. \end{cases} \tag{C27}
$$

In Fig. 16, we illustrate $\Sigma^{AB}_{d,f}$ and $\Sigma^{BA}_{d,f}$ as a function of $J_1/J_2$ for various $d > 0$ (i.e., $a_2$ is to the right of $a_1$) when $\Delta' = -i\gamma_b/2$. The results indicate the absence of interaction in the topological regime $J_1/J_2 < \sqrt{1 + \gamma_b^2/(4J_2^2)}$ of the mirage bath, and the presence of interaction otherwise. This explains the dynamical behavior observed in Fig. 14(b), which reflects the topology of the mirage bath, instead of the physical bath.

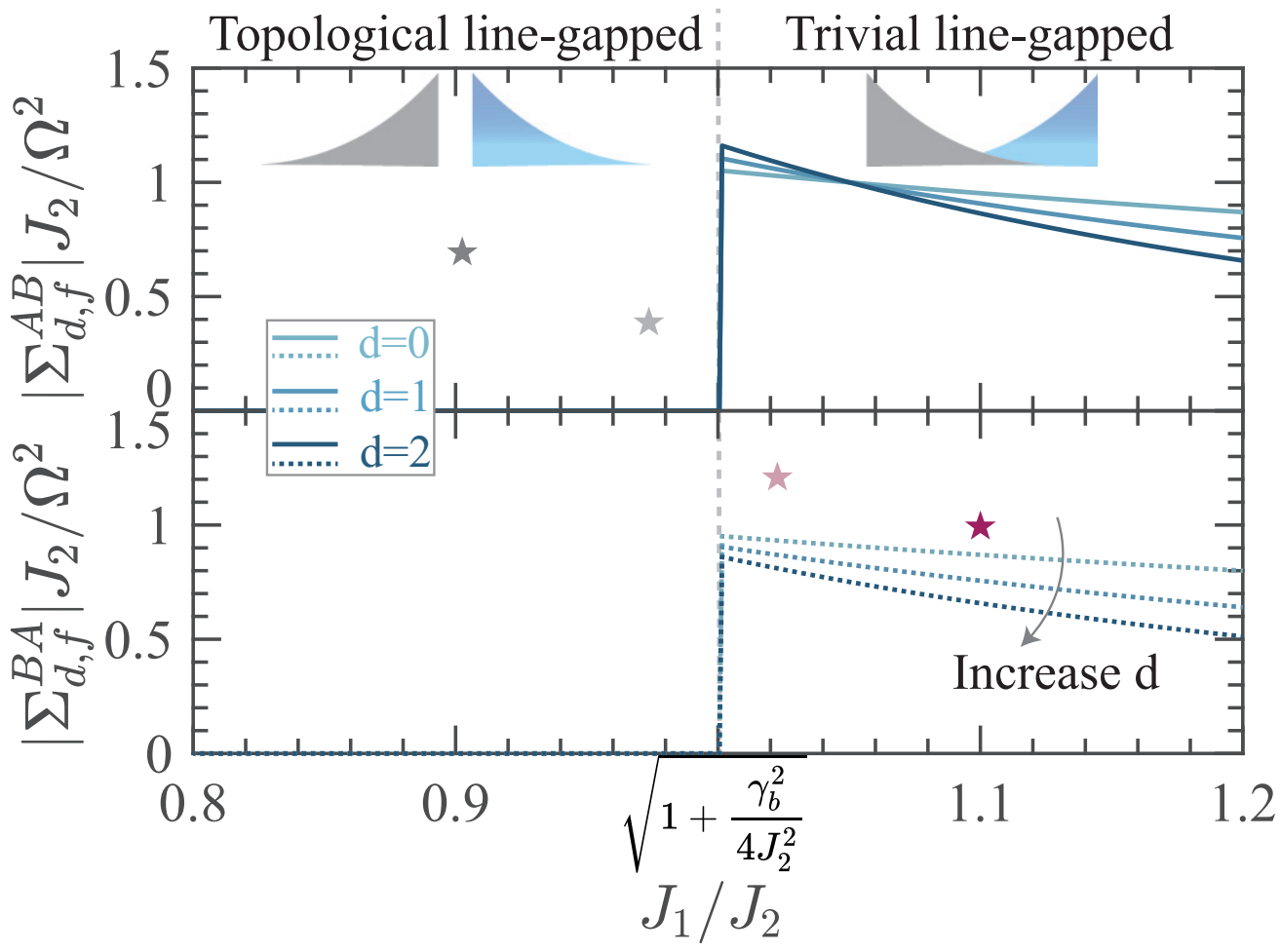

FIG. 16. Dipole-dipole interaction mediated by photon BS of the mirage bath on the second Riemann sheet for $\Delta = 0$, $\gamma_a/J_2 = \gamma_b/J_2 = 0.1$, and $\Omega/J_2 = 0.1$. $a_1$ is coupled to sublattice $A$ at $j_1 = 0$, and $a_2$ is coupled to sublattice $B$ at $j_2 = d$. Absolute value of the dipolar coupling $\Sigma_{d,f}^{AB}$ (top panel) and $\Sigma_{d,f}^{BA}$ (bottom panel) as a function of $J_1/J_2$. The insets show the shapes of $A$-BS (gray) and $B$-BS (blue) in the topological and trivial line-gapped regimes of the mirage bath. Results are obtained from Eqs. (C24)–(C27).

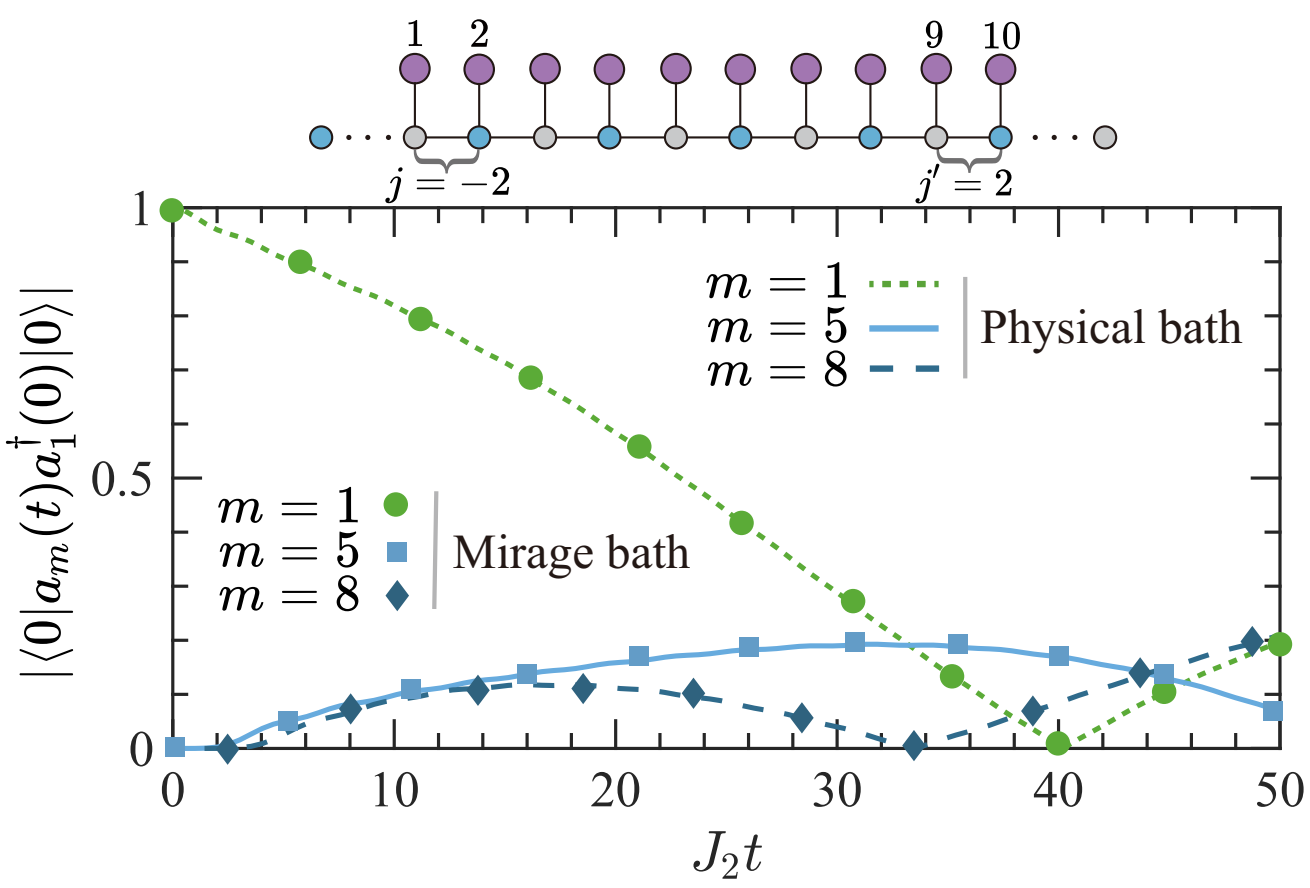

FIG. 17. Emission dynamics of a chain of ten QEs coupled to a dissipative photonic bath. Initially, the leftmost QE is excited, while all other QEs are in the vacuum state. Results of $|\langle 0|a_m(t)a_1^\dagger(0)|0\rangle|$ are shown for $m = 1, 5$, and 8. The numerical results obtained from the original total Hamiltonian (B4) associated with the physical bath are compared with that obtained using the total Hamiltonian (C28) associated with the mirage bath. In both cases, the bath size is $N_b = 2000$. For other parameters, $J_1/J_2 = 1.1$, $\gamma_a/J_2 = \gamma_b/J_2 = 0.05$, $\Omega/J_2 = 0.2$, and $\Delta/J_2 = 0$.

### 3. String of QEs

For an ensemble of QEs, their full dynamics is equivalently given by the effective Hamiltonian.

$$H_{\rm eff}^f = \sum_{m=1}^{N_a} \Delta' a_m^\dagger a_m + \sum_{m=1}^{N_a} \Omega\big[\xi_\alpha^m \alpha_{j_m}^\dagger a_m + \left(\xi_\alpha^m\right)^{-1} a_m^\dagger \alpha_{j_m}\big] + H_b^f, \tag{C28}$$

where $\xi_A^m = [\sqrt{(J_1 - \gamma_b/2)/(J_1 + \gamma_b/2)}]^{-j_m}$ and $\xi_B^m = [\sqrt{(J_1 - \gamma_b/2)/(J_1 + \gamma_b/2)}]^{-j_m - 1}$, where $H_b^f$ is given by Eq. (C21). In Fig. 17, we compare the dynamics of the QEs for the mirage bath and the physical bath.

## APPENDIX D: TWO EXCITATIONS

In this section, we extend our studies from the two-level QEs with single excitation to the case of a highly nonlinear QE with two excitations. Specifically, we consider the total density matrix $\rho$ for the combined system of a nonlinear QE and the SSH bath is governed by the master equation

$$\dot{\rho} = -i[H_a'' + H_b + H_{ab}, \rho] + \sum_j \frac{\gamma_b}{2}\mathcal{D}_{\rm b}[l_j]\rho + \sum_m \frac{\gamma_a}{2}\mathcal{D}_a[a_m]\rho, \tag{D1}$$

where the emitter Hamiltonian takes the form

$$H_a'' = \Delta a^\dagger a + \frac{U}{2} a^{\dagger 2} a^2 + \varepsilon(a^\dagger e^{-i\omega_d t} + \text{H.c.}). \tag{D2}$$

In Eq. (D2), $U$ characterizes the strength of on-site Kerr interaction, and $\varepsilon$ and $\omega_d$ are, respectively, the strength and frequency of the driving field. We focus on the case with a weak driving field; i.e., $\varepsilon$ is much smaller than the spectral gap of the undriven system.

The goal of this section is to derive second-order correlation function $g^{(2)}(\tau)$ in the steady state of a weakly driven emitter. Following the formalism in Ref. [61], in Appendix D 1 we present key steps for deriving the dynamics of an undriven emitter in the two-excitation subspace, and in Appendix D 2 we calculate the second-order correlation function $g^{(2)}(\tau)$ in the steady state of a weakly driven emitter, leading to Fig. (8).

### 1. Spontaneous emission of two excitations

In this section, we assume the absence of driving field (i.e., $\varepsilon = 0$) and derive the emitter's dynamics in the two-excitation subspace, which is determined by the two-particle retarded Green function

$$D(t) = -i\tfrac{1}{2}\langle 0|a^2(t)a^{\dagger 2}(0)|0\rangle. \tag{D3}$$

In the frequency domain, the two-particle Green function can be written as

$$D(\omega) = \frac{1}{\Pi^{-1}(\omega) - U}, \tag{D4}$$

with the function

$$\Pi(\omega) = i\int \frac{d\omega'}{2\pi} G_0(\omega')G_0(\omega - \omega'). \tag{D5}$$

Here, $G_0(\omega)$ is the single-emitter Green function in the single-excitation subspace given by Eq. (B8).

Since $G_0(\omega)$ exhibits a branch loop [see Fig. 12(b)], the two-particle Green function (D4) exhibits the branch area in the frequency plane, making the computation of the dynamics complicated. However, by using the mirage bath, the computation can be much simplified. We have

$$D(t) = \int \frac{d\omega}{2\pi} D(\omega) e^{-i\omega t} = \int \frac{d\omega}{2\pi} D_f(\omega) e^{-i\omega t}. \tag{D6}$$

Here, $D_f(\omega)$ is the two-particle Green function associated with the mirage bath, given by

$$D_f(\omega) = \frac{1}{\Pi_f^{-1}(\omega) - U}, \tag{D7}$$

where we have

$$\Pi_f(\omega) = i \int \frac{d\omega'}{2\pi} G_f(\omega') G_f(\omega - \omega'). \tag{D8}$$

Using Eq. (D6), we obtain the plot in Fig. 8(c).

### 2. Second-order correlation function $g^{(2)}(\tau)$

Here, we calculate the steady-state correlation function $g^{(2)}(\tau)$ of a weakly driven emitter:

$$g^{(2)}(\tau) = \frac{1}{n^2} \mathrm{Tr}[a^\dagger a^\dagger(\tau) a(\tau) a \rho_{\rm ss}], \tag{D9}$$

with $n = \mathrm{Tr}(a^\dagger a \rho_{\rm ss})$ being the first-order correlation function, in the steady state $\rho_{\rm ss}$ of the master equation (D1). According to Ref. [61], one has

$$g^{(2)}(\tau) = |1 + \overline{\Pi}_f(\tau) T(2\omega_d)|^2, \tag{D10}$$

where the scattering matrix $T(2\omega_d) = [U^{-1} - \Pi_f(2\omega_d)]^{-1}$ and $\overline{\Pi}_f(\tau)$ follows from Eq. (D8), giving

$$\overline{\Pi}_f(\tau) = i \int \frac{d\omega'}{2\pi} G_f(\omega_d + \omega') G_f(\omega_d - \omega') e^{-i\omega'\tau}. \tag{D11}$$

Here, $G_f(\omega)$ is the single-particle Green function associated with the mirage bath, as shown in Eq. (C2). Using Eq. (D10), we obtain the plot shown in Fig. 8(d). The statistics of photons is quantified by

$$g^{(2)}(0) = \left| \frac{1}{1 - U\Pi_f(2\omega_d)} \right|^2. \tag{D12}$$

## APPENDIX E: PHYSICAL BATH WITH OBC VERSUS MIRAGE BATH

As remarked in Ref. [61] and the main text, the mirage bath (PBC) has the identical bulk spectrum with—and is topologically equivalent to—that of the physical bath subjected to OBCs. However, the mirage bath and the physical bath under OBC are different baths, as they exhibit completely different eigenstates. In this section, we show that the BSs associated with the two baths exhibit different properties.

In Fig. 18, we compare the photonic distributions of the BS in the real space for the mirage bath and the OBC bath. Specifically, $f_{A/B,j}$ associated with the BS in the mirage bath is obtained from Table III, while that associated with the OBC bath is numerically obtained from diagonalization. We see that the chirality of the photon BS in the mirage bath faithfully reflects the topology of the mirage bath (red curves). In marked contrast, the photon BS in OBC bath always resides on one side of the emitter (green curves), irrespective of $J_1/J_2$. For the OBC bath, the bath topology is encoded in the fact that in the topological phase, the BS exponentially decays from the right boundary of the bath, while in the trivial phase, the BS exponentially decays from the QE.

Note that for the OBC physical bath, in general, it is difficult to analytically calculate the BS. However, for the special case $\Delta' = -i\gamma_b/2$, the effective emitter-bath Hamiltonian contains a dark state with the eigenenergy $-i\gamma_b/2$, whose wavefunction can be analytically derived following the same spirit as Ref. [23]. For the $A$-BS, the wavefunction of this dark state is obtained as

$$f_{A,j} = 0,$$
$$f_{B,j} = \begin{cases} -\frac{\Omega\varphi_a}{J_1 + \frac{\gamma_b}{2}} \left( -\frac{J_2}{J_1 + \frac{\gamma_b}{2}} \right)^j, & j \geqslant 0, \\ 0, & j < 0. \end{cases} \tag{E1}$$

In a similar manner, for the $B$-BS, the dark state with the energy $-i\gamma_b/2$ is obtained as

$$f_{A,j} = \begin{cases} 0, & j > 0, \\ (-1)^{|j|+1} \frac{\Omega\varphi_a}{J_1 - \frac{\gamma_b}{2}} \left( \frac{J_2}{J_1 - \frac{\gamma_b}{2}} \right)^{|j|}, & j \leqslant 0, \end{cases}$$
$$f_{B,j} = 0, \tag{E2}$$

in agreement with the numerical results.

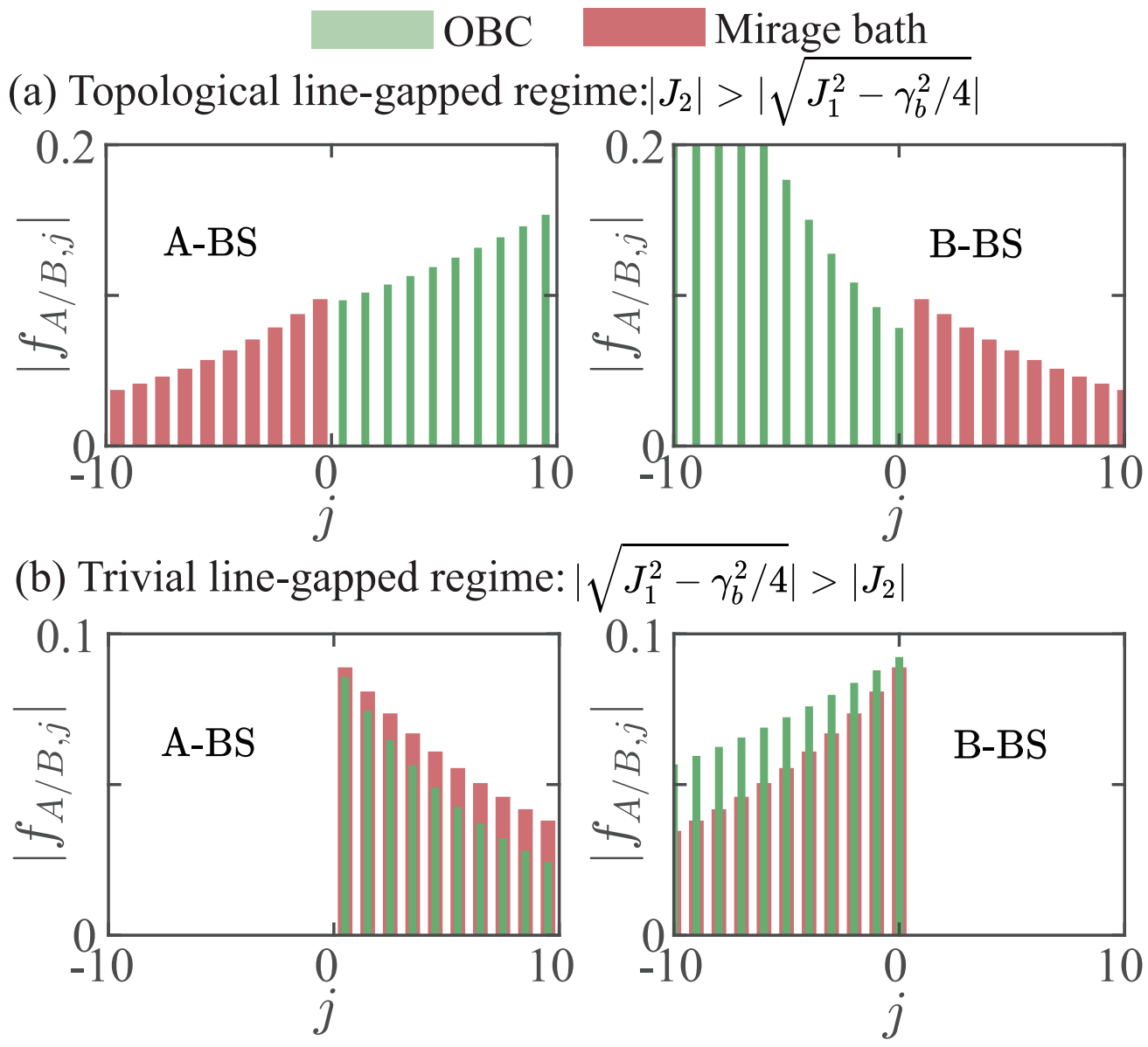

FIG. 18. Comparison between the photonic distribution of the BS in the mirage bath (PBC) and the physical bath subjected to OBC. $|f_{A/B,j}|$ are shown for (a) $J_1/J_2 = 0.9$ and (b) $J_1/J_2 = 1.1$, when $\Delta/J_2 = 0$, $\Omega/J_2 = 0.1$, and $\gamma_a/J_2 = \gamma_b/J_2 = 0.1$. For the mirage bath, the results of $|f_{A/B,j}|$ are obtained from Table III. For the OBC physical bath, $|f_{A/B,j}|$ is obtained from numerical diagonalization with $N_b = 20$.

## APPENDIX F: EMITTERS IN DIFFERENT POINT GAPS

In the main text, the emphasis is in the point-gap regime where the two spectral loops merge into one. In this section, we extend our analysis to a different point-gap scenario as illustrated in Fig. 19(a); i.e., we consider the line-gapped regimes illustrated in Fig. 19(a), where two spectral loops are separated by a line gap, and assume the emitter's transition frequency $\Delta'$ lies in, say, the right loop (point gap). We show that, still, the emitter dynamics can be understood via the mirage bath.

A concrete example is illustrated in Fig. 19(a), where we assume $\Delta/J_2 = 1$, $\gamma_a/J_2 = 0.1$, and $\gamma_b/J_2 = 0.3$. First, we choose $J_1/J_2 = 0.8$ in the topological line-gapped regime of the dissipative bath and calculate the BS associated with the physical bath using Eqs. (B19)–(B22), as shown in the left panel of Fig. 19(b). On the right panel, we show the photon BS of the mirage bath from Eqs. (C13) to (C17). We see that while the photon BS of the physical bath is obviously chiral, that of the mirage bath is not.

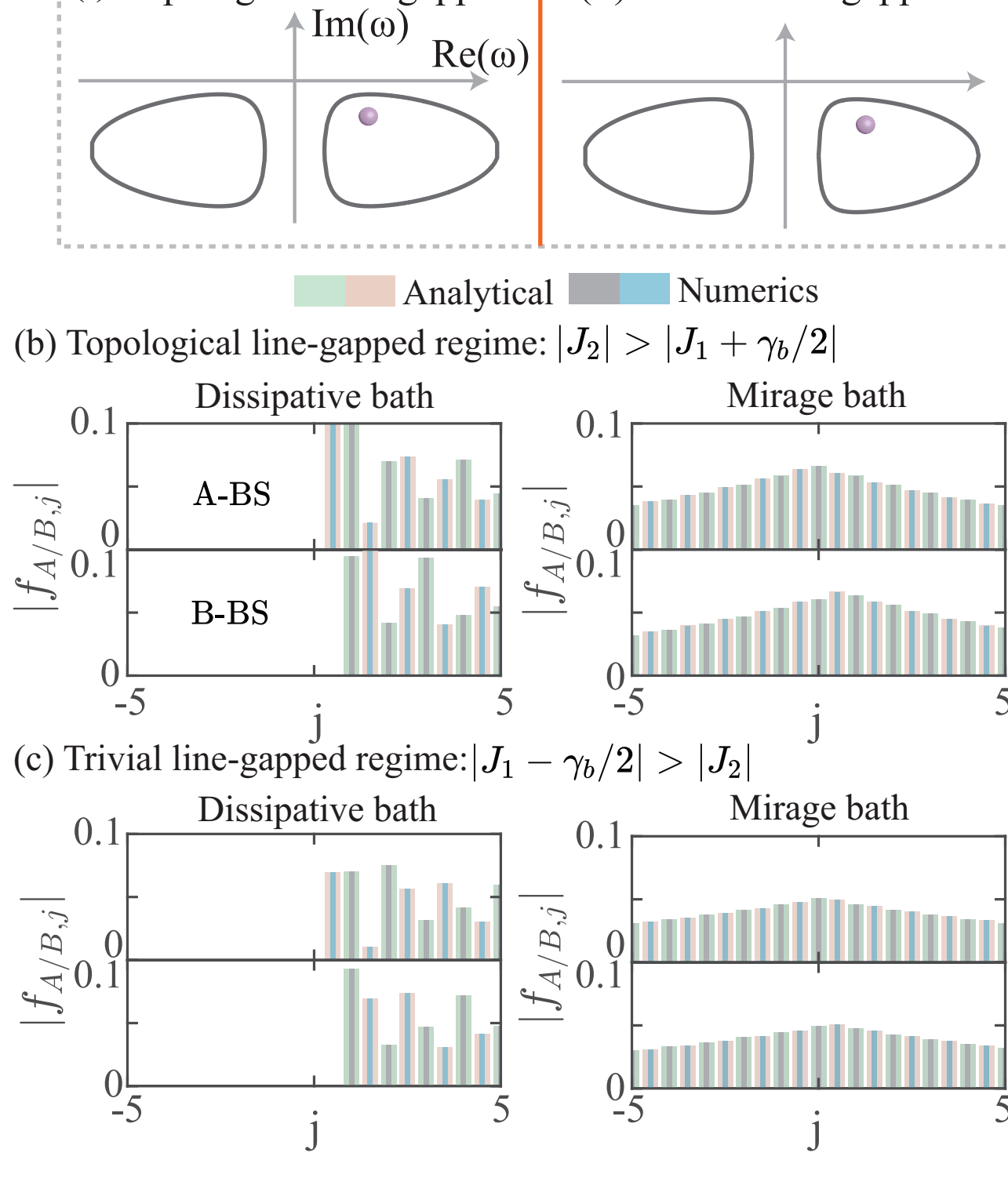

FIG. 19. Emitter in the right point gap of the dissipative bath: (a) Emitter in the right point gap of the dissipative bath in the (i) topological line-gapped regime and (ii) trivial line-gapped regime. (b), (c) Photonic distribution $|f_{A/B,j}|$ of the BS in the dissipative bath (left panel) and mirage bath (right panel), for (b) $J_1/J_2 = 0.8$ in topological regime and (c) $J_1/J_2 = 1.2$ in the trivial regime. We consider $\Omega/J_2 = 0.1$, $\Delta/J_2 = 1$, $\gamma_a/J_2 = 0.1$, and $\gamma_b/J_2 = 0.3$. Both analytical and numerical results are shown. Analytical results are obtained using Eqs. (B19)–(B22) for dissipative bath and Eqs. (C13)–(C17) for mirage bath. Numerical results are obtained through the numerical diagonalization of the effective Hamiltonian (B4) and (C28) with a single QE and bath size $N_b = 500$.

## APPENDIX G: BATH CORRELATION

Here, we prove that a mirage bath emerges even in the absence of emitters.

We describe the free propagation of a single bath excitation using the bath's correlation function for times $t > 0$:

$$C_{sj,s'j'}(t) = -i\langle 0|b_{sj}(t)b^{\dagger}_{s'j'}(0)|0\rangle, \tag{G1}$$

where $s, s'$ represents the sublattice index $A/B$ within unit cell $j, j'$. For the bath with PBC, the Hamiltonian given by Eq. (B5) in the momentum space is expressed as $H'_b(k) = \sum_{k,s,s'} b^{\dagger}_{s,k} H'_{b,ss'}(k) b_{s',k}$ with $b_{s,k} = \sum_j b_{sj} e^{-ikj}/\sqrt{N_b}$. Here, for each $k$, the dimension of $H'_b(k)$ is the number of sublattices.

The bath's correlation function (G1) can be derived as

$$C_{sj,s'j'}(t) = \int \frac{d\omega}{2\pi} e^{-i\omega t} \frac{1}{N_b} \sum_k \left[ \frac{e^{ik(j-j')}}{\omega - H'_b(k)} \right]_{ss'}. \tag{G2}$$

In the continuum limit, the bath's correlation function

$$C_{sj,s'j'}(t) = \int \frac{d\omega}{2\pi} e^{-i\omega t} \Sigma^{ss'}_{jj'}(\omega) \tag{G3}$$

is exactly the Fourier transform of the "self-energy"

$$\Sigma_{jj'}(\omega) = \int \frac{dk}{2\pi} \frac{e^{ik(j-j')}}{\omega - H'_b(k)} \tag{G4}$$

derived in Appendix B. As a result, the same analytic continuation procedure can be applied directly to the propagator $C_{sj,s'j'}(t)$, leading to

$$C_{sj,s'j'}(t) = \int \frac{d\omega}{2\pi} e^{-i\omega t} \Sigma^{ss'}_{jj',f}(\omega) \tag{G5}$$

with

$$\Sigma_{jj',f}(\omega) = \int \frac{dk}{2\pi} \frac{e^{ik(j-j')}}{\omega - H^f_b(k)}, \tag{G6}$$

where $H^f_b$ is given by Eq. (C21). This representation, completely governed by the mirage bath, yields identical dynamics to the original bath, even though the mirage-bath's spectrum differs. Thus, the mirage bath acts as a dual of the original bath, independent of the presence of emitters.

## APPENDIX H: DERIVATION OF MIRAGE-BATH'S SPECTRUM

Here, we derived the mirage-bath's spectrum $\omega^f_b(\theta)$ using the Sylvester elimination method.

Based on the conditions given in Eq. (26), we derive the characteristic equation

$$\begin{aligned} \lambda(\omega, \bar{z}_w) &= 0, \\ \lambda(\omega, \bar{z}_w e^{i\theta}) &= 0 \end{aligned} \tag{H1}$$

governing the spectrum of the mirage bath, where $\lambda(\omega, z) = \sum_{l=0}^{m} c_l(\omega) z^l$. Then, by applying the Sylvester elimination method to these two equations for eliminate $\bar{z}_w$, we obtain the Sylvester matrix

$$R = \begin{pmatrix} c_m & c_{m-1} & c_{m-2} & \cdots & c_0 & 0 & \cdots & 0 \\ 0 & c_m & c_{m-1} & c_{m-2} & \cdots & c_0 & \cdots & 0 \\ \vdots & \vdots & \vdots & \vdots & \vdots & \vdots & \vdots & \vdots \\ 0 & 0 & \cdots & 0 & c_m & c_{m-1} & \cdots & c_0 \\ c'_m & c'_{m-1} & c'_{m-2} & \cdots & c'_0 & 0 & \cdots & 0 \\ 0 & c'_m & c'_{m-1} & c'_{m-2} & \cdots & c'_0 & \cdots & 0 \\ \vdots & \vdots & \vdots & \vdots & \vdots & \vdots & \vdots & \vdots \\ 0 & 0 & \cdots & 0 & c'_m & c'_{m-1} & \cdots & c'_0 \end{pmatrix} \tag{H2}$$

with $c_l' = c_l e^{ij\theta}$ $(l = 0, ..., m)$. The spectrum of the mirage bath is obtained by substituting Sylvester matrix $R$ into Eq. (27).

## APPENDIX I: TOPOLOGICAL INVARIANT OF THE MIRAGE BATH

In this section, we elucidate the computation of the mirage-bath's topological invariant, constructed through analytic continuation of the corresponding physical bath invariant. To demonstrate the efficacy of this methodological framework, we analyze three representative examples illustrating its broad applicability to complex topological systems.

### 1. Nonreciprocal SSH bath

As an initial case study, we consider an SSH bath with nonreciprocal coupling, whose Hamiltonian is written as

$$H_{\rm b}' = \sum_j \left[\left(J_1 + \frac{\gamma_b}{2}\right) b_{A,j}^\dagger b_{B,j} + \left(J_1 - \frac{\gamma_b}{2}\right) b_{B,j}^\dagger b_{A,j} - i\frac{\gamma_b}{2} b_{A,j}^\dagger b_{A,j} - i\frac{\gamma_b}{2} b_{B,j}^\dagger b_{B,j}\right] + \sum_j (J_2 b_{A,j+1}^\dagger b_{B,j} + \text{H.c.}). \tag{I1}$$

The Hamiltonian in momentum space is written as

$$H_b'(k) = -i\frac{\gamma_b}{2} I + (J_1 + J_2\cos k)\sigma_x + \left(J_2 \sin k + i\frac{\gamma_b}{2}\right)\sigma_y, \tag{I2}$$

where $I$ is the identity matrix, and $\sigma_x$, $\sigma_y$ are Pauli matrices. The associated Green function is defined as

$$G_b(k,\omega) = \frac{1}{\omega - H_b'(k)}. \tag{I3}$$

Following the Green's function formalism for topological characterization [76], we calculate the winding number $\nu$ through the contour integral:

$$\begin{aligned}\nu &= \lim_{\omega\to\omega_{\rm EP}} \int \frac{dk}{4\pi i} \text{tr}[\sigma_z G_b(k,\omega)^{-1} \partial_k G_b(k,\omega)] \\ &= \lim_{\omega\to\omega_{\rm EP}} \oint_{|z|=1} \frac{dz}{4\pi i} \text{tr}[\sigma_z G_b(z,\omega)^{-1} \partial_z G_b(z,\omega)] \\ &= \oint_{|z|=1} \frac{dz}{4\pi i}\left[\frac{1}{z + \frac{J_1 - \frac{\gamma_b}{2}}{J_2}} - \frac{1}{z + \frac{J_2}{J_1 + \frac{\gamma_b}{2}}} + \frac{1}{z}\right],\end{aligned} \tag{I4}$$

where $z = e^{ik}$ and $\omega_{\rm EP} = -i\gamma_b/2$ denote the exceptional point frequency. when $J_1/J_2 = 1/6$ and $\gamma_b/J_2 = 4/3$ in the topological line-gap regime, three poles of Eq. (I4): $|(J_1 - \gamma_b/2)/J_2| = 1/2 < 1$ and 0 within the unit circle, while $|J_2/(J_1 + \gamma_b/2)| = 6/5 > 1$ lies outside the unit circle. Then, by applying the residue theorem, we obtain the topological invariant $\nu = 1$. Similarly, when $J_1/J_2 = 1$, $\gamma_b/J_2 = 4/3$ in the point-gap regime, the topological invariant $\nu = 0.5$ and when $J_1/J_2 = 2$, $\gamma_b/J_2 = 4/3$ in the trivial line-gap regime, the topological invariant $\nu = 0$.

We now calculate the topological invariant $\nu_f$ of the mirage bath. By selecting a contour $C_z$ in the $z$ plane that encloses two poles—analogous to the line-gapped regime of the physical bath—we obtain the expression for the mirage-bath's topological invariant

$$\nu_f = \oint_{|z|=C_z} \frac{dz}{4\pi i}\left[\frac{1}{z + \frac{J_1 - \frac{\gamma_b}{2}}{J_2}} - \frac{1}{z + \frac{J_2}{J_1 + \frac{\gamma_b}{2}}} + \frac{1}{z}\right], \tag{I5}$$

where the pole at $z = 0$ remains consistently within the contour. Therefore, when $|(J_1 - \gamma_b/2)/J_2| > |J_2/(J_1 + \gamma_b/2)|$, the mirage-bath's topological invariant $\nu_f = 0$, and when $|(J_1 - \gamma_b/2)/J_2| < |J_2/(J_1 + \gamma_b/2)|$, the topological invariant $\nu_f = 1$. The phase boundary between these regimes is precisely demarcated by the equality condition

$$\left|\frac{J_1 - \frac{\gamma_b}{2}}{J_2}\right| = \left|\frac{J_2}{J_1 + \frac{\gamma_b}{2}}\right|. \tag{I6}$$

### 2. Nonreciprocal SSH bath with NNN coupling

As a second case study, we extend our analysis to a nonreciprocal SSH bath with NNN coupling. The system is described by the non-Hermitian Hamiltonian:

$$H_{\rm b}' = \sum_j (J_2 b_{A,j+1}^\dagger b_{B,j} + J_3 b_{A,j}^\dagger b_{B,j+1} + \text{H.c.}) + \sum_j \left[\left(J_1 + \frac{\gamma_b}{2}\right) b_{A,j}^\dagger b_{B,j} + \left(J_1 - \frac{\gamma_b}{2}\right) b_{B,j}^\dagger b_{A,j} - i\frac{\gamma_b}{2} b_{A,j}^\dagger b_{A,j} - i\frac{\gamma_b}{2} b_{B,j}^\dagger b_{B,j}\right]. \tag{I7}$$

The Hamiltonian in momentum space is written as

$$H_b'(k) = -i\frac{\gamma_b}{2} I + (J_1 + J_2\cos k + J_3\cos k)\sigma_x + \left(J_2\sin k - J_3 \sin k + i\frac{\gamma_b}{2}\right)\sigma_y. \tag{I8}$$

Applying the Green's function formalism, we compute the topological invariant

$$\begin{aligned}\nu &= \lim_{\omega\to\omega_{\rm EP}} \int \frac{dk}{4\pi i} \text{tr}[\sigma_z G_b(k,\omega)^{-1} \partial_k G_b(k,\omega)] \\ &= \oint_{|z|=1} \frac{dz}{4\pi i}\left[\frac{1}{z - z_1} + \frac{1}{z - z_2} - \frac{1}{z - z_3} - \frac{1}{z - z_4}\right],\end{aligned} \tag{I9}$$

where $z = e^{ik}$, $z_1, z_2$ are two solutions of equation $J_2 z^2 + (J_1 - \gamma_b/2)z + J_3 = 0$ with $|z_1| \leqslant |z_2|$ and $z_3, z_4$ are two solutions of equation $J_3 z^2 + (J_1 + \gamma_b/2)z + J_2 = 0$ with $|z_3| \leqslant |z_4|$. When $J_1/J_2 = 0.1$, $J_3/J_2 = 0.2$, and $\gamma_b/J_2 = 4/3$ in the topological line-gap regime, four poles of Eq. (I9) can be classified into two types: $|z_1| = |z_2| \approx 0.4472 < 1$ within the unit circle, while $|z_3| = |z_4| \approx 2.2361 > 1$ lies outside the unit circle. In this case, there are two poles within the unit circle. Similarly, the topological invariant $\nu = 1$ is obtained by using the residue theorem.

We now calculate the topological invariant $\nu_f$ of the mirage bath. By appropriately selecting the contour $C_z$ to enclose two poles of the Green function $G_b(k,\omega)$—analogous to the line-gapped regime of the physical bath—we obtain the mirage-bath's topological invariant

$$\nu_f = \oint_{|z|=C_z} \frac{dz}{4\pi i}\left[\frac{1}{z-z_1} + \frac{1}{z-z_2} - \frac{1}{z-z_3} - \frac{1}{z-z_4}\right]. \tag{I10}$$

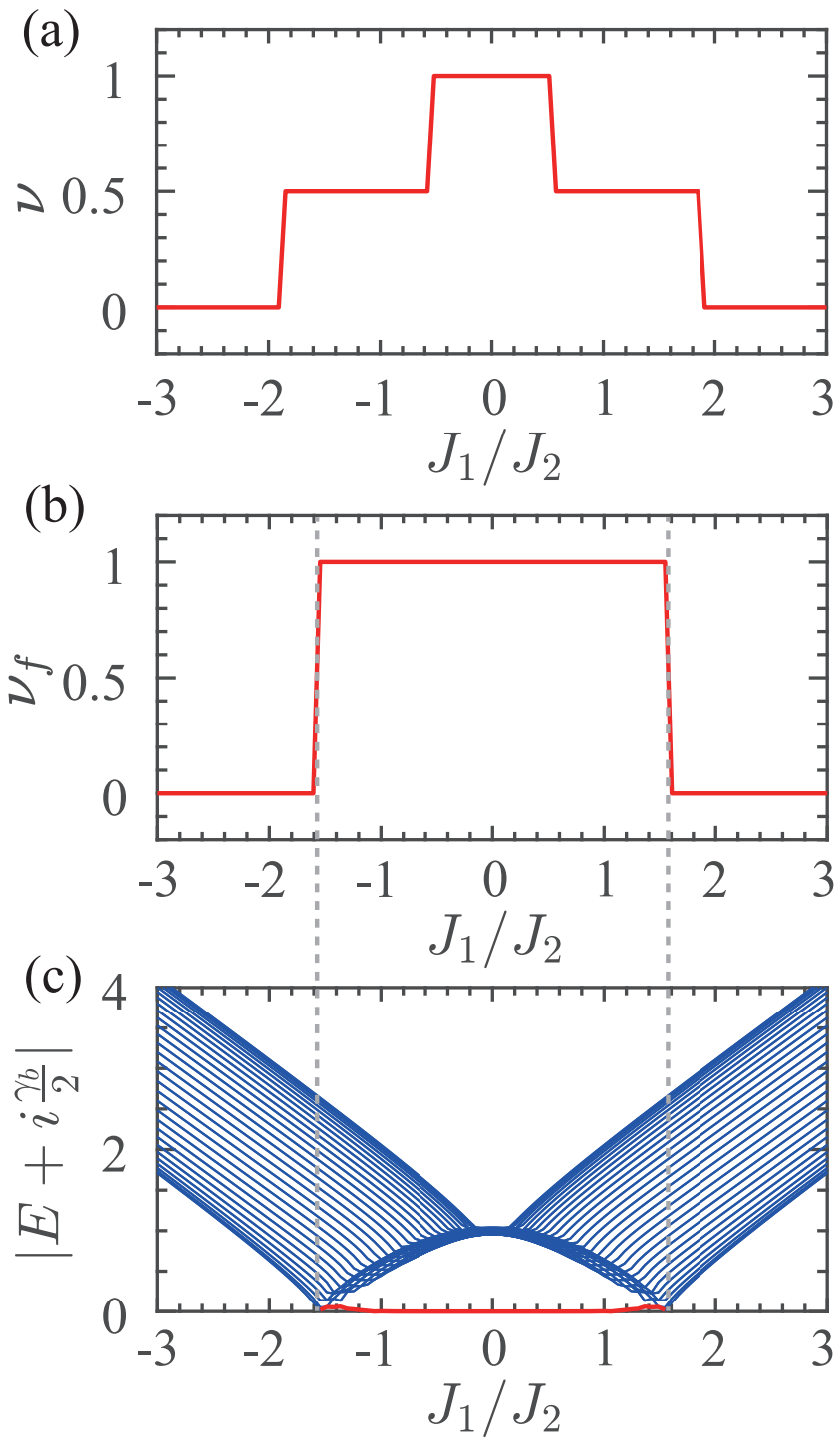

FIG. 20. Topological phase digram of the mirage bath obtained from analytic continuation. The physical bath is described by nonreciprocal SSH model with NNN coupling (I7). (a) topological index $\nu$ of the physical bath [see Eq. (I9)] as a function of $J_1/J_2$. (b) topological index $\nu_f$ of the mirage bath [see Eq. (I10)] as a function of $J_1/J_2$. (c) Spectrum of the physical bath when subjected to OBC as a function of $J_1/J_2$. Results are obtained from numerical diagonalization of the tight-binding Hamiltonian with a size $N_b = 30$ under OBC. The zero-mode line is shown in red. For other parameters, $J_3/J_2 = 0.2$ and $\gamma_b/J_2 = 4/3$.

By applying the residue theorem, we obtain that when $|z_1| \leqslant |z_2| < |z_3| \leqslant |z_4|$, the mirage-bath's topological invariant $\nu_f = 1$, when $|z_3| \leqslant |z_4| < |z_1| \leqslant |z_2|$, the topological invariant $\nu_f = -1$, and for other cases, the topological invariant $\nu_f = 0$. Therefore, the resultant phase boundaries for topological transitions are determined by the conditions:

$$|z_2| = |z_3| \quad \text{or} \quad |z_1| = |z_4|. \tag{I11}$$

The topological invariant of the mirage bath is shown in Fig. 20.

### 3. A more complex extended SSH bath

As a third case study, we investigate a more complex extended SSH bath. The system is described by the non-Hermitian Hamiltonian:

$$\begin{aligned} H_b' = &\sum_j \left[\left(J_1 + \frac{\gamma_b}{2}\right) b_{A,j}^\dagger b_{B,j} + \left(J_1 - \frac{\gamma_b}{2}\right) b_{B,j}^\dagger b_{A,j}\right] \\ &+ \sum_j (J_2 b_{A,j+1}^\dagger b_{B,j} + J_3 b_{A,j}^\dagger b_{B,j+1} + J_4 b_{A,j}^\dagger b_{B,j+2} \\ &+ \text{H.c.}) - i\frac{\gamma_b}{2}\sum_j (b_{A,j}^\dagger b_{A,j} + b_{B,j}^\dagger b_{B,j}). \end{aligned} \tag{I12}$$

The Hamiltonian in momentum space is written as

$$\begin{aligned} H_b'(k) = &-i\frac{\gamma_b}{2} I + (J_1 + J_2\cos k + J_3 \cos k + J_4 \cos 2k)\sigma_x \\ &+ \left(J_2 \sin k - J_3 \sin k - J_4 \sin 2k + i\frac{\gamma_b}{2}\right)\sigma_y. \end{aligned} \tag{I13}$$

By the same method, we calculate the topological invariant

$$\begin{aligned} \nu &= \lim_{\omega\to\omega_{\text{EP}}} \int \frac{dk}{4\pi i} \text{tr}[\sigma_z G_b(k,\omega)^{-1} \partial_k G_b(k,\omega)] \\ &= \oint_{|z|=1} \frac{dz}{4\pi i}\left[\frac{1}{z-z_1} + \frac{1}{z-z_2} + \frac{1}{z-z_3} \right. \\ &\quad \left. - \frac{1}{z-z_4} - \frac{1}{z-z_5} - \frac{1}{z-z_6} - \frac{1}{z}\right], \end{aligned} \tag{I14}$$

where $z = e^{ik}$, $z_1, z_2, z_3$ are three solutions of equation $J_2 z^3 + (J_1 - \gamma_b/2)z^2 + J_3 z + J_4 = 0$ with $|z_1| \leqslant |z_2| \leqslant |z_3|$, and $z_4, z_5, z_6$ are three solutions of equation $J_4 z^3 + J_3 z^2 + (J_1 + \gamma_b/2)z + J_2 = 0$ with $|z_4| \leqslant |z_5| \leqslant |z_6|$. When $J_1/J_2 = 0.3$, $J_3/J_2 = 0.2$, $J_4/J_2 = 0.1$, and $\gamma_b/J_2 = 0.4$ in the topological line-gap regime, the magnitudes of the seven poles corresponding to Eq. (I11) are as follows: $|z_1| \approx 0.3487 < 1$, $|z_2| = |z_3| \approx 0.5355 < 1$, $|z_4| = 2 > 1$, $|z_5| = |z_6| \approx 2.2361 > 1$, and origin $z = 0 < 1$. In this case, there are four poles within the unit circle. Similarly, the topological invariant $\nu = 1$ is obtained by using the residue theorem.

We now calculate the topological invariant $\nu_f$ of the mirage bath. Similarly, by selecting a contour $C_z$ that encloses four poles—analogous to the line-gapped regime of the physical bath—we obtain the expression for the mirage-bath's topological invariant

$$\begin{aligned} \nu_f = &\oint_{|z|=C_z} \frac{dz}{4\pi i}\left[\frac{1}{z-z_1} + \frac{1}{z-z_2} + \frac{1}{z-z_3} \right. \\ &\left. - \frac{1}{z-z_4} - \frac{1}{z-z_5} - \frac{1}{z-z_6} - \frac{1}{z}\right], \end{aligned} \tag{I15}$$

where the pole at $z = 0$ remains consistently within the contour. The other three poles surrounded by the contour have the following possibilities: (1) $z_1, z_2, z_3$, (2) $z_1, z_2, z_4$, (3) $z_1, z_4, z_5$, and (4) $z_4, z_5, z_6$. Using the residue theorem, the mirage-bath's topological invariant corresponding to these cases are, respectively, obtained as (1) $\nu_f = 1$, (2) $\nu_f = 0$, (3) $\nu_f = -1$, and (4) $\nu_f = -2$. The resultant phase boundaries for topological transitions are determined by the conditions:

$$|z_3| = |z_4| \quad \text{or} \quad |z_2| = |z_5| \quad \text{or} \quad |z_1| = |z_6|. \tag{I16}$$

## APPENDIX J: GREEN FUNCTIONS OF OTHER BATHS

Here, we present the Green functions for three distinct non-Hermitian baths: the Hatano-Nelson bath, the Trefoil bath, and the nonreciprocal SSH bath with NNN coupling.

### 1. Hatano-Nelson bath

We present a comprehensive analysis of the effective Hamiltonian framework for non-Hermitian emitter-bath systems, focusing on the HN bath architecture. The effective total Hamiltonian is written as

$$H_{\text{eff}} = H'_a + H'_b + H_{ab}, \tag{J1}$$

where Hamiltonians $H'_a$ and $H_{ab}$ are of the same form as in Eqs. (B7) and (A5). The effective bath Hamiltonian $H'_b$ is written as

$$H'_b = \sum_j \left[\left(J_1 + \frac{\gamma_b}{2}\right) b_j b^\dagger_{j+1} + \left(J_1 - \frac{\gamma_b}{2}\right) b^\dagger_{j+1} b_j - i\gamma_b b^\dagger_j b_j\right], \tag{J2}$$

which generates complex-energy bands

$$\omega'_b(k) = 2J_1 \cos k + i\gamma_b \sin k - i\gamma_b. \tag{J3}$$

For a single emitter, the dynamics is described by the Green function

$$\begin{aligned} G_0(t) &= \int \frac{d\omega}{2\pi} e^{-i\omega t} G_0(\omega) \\ &= \int \frac{d\omega}{2\pi} e^{-i\omega t} \frac{1}{\omega - \Delta' - \Sigma_0(\omega)}, \end{aligned} \tag{J4}$$

where the self-energy $\Sigma_0(\omega)$ is derived from the HN bath spectrum [Eq. (J3)] via integration:

$$\begin{aligned} &\Sigma_0(\omega) \\ &= \Omega^2 \int_{-\pi}^{\pi} \frac{dk}{2\pi} \frac{1}{\omega - \omega'_b(k)} \\ &= \Omega^2 \oint_{|z|=1} \frac{dz}{2\pi i} \frac{1}{-\left(J_1 + \frac{\gamma_b}{2}\right) z^2 + (\omega + i\gamma_b) z - \left(J_1 - \frac{\gamma_b}{2}\right)} \\ &= \Omega^2 \oint_{|z|=1} \frac{dz}{2\pi i} \frac{1}{-\left(J_1 + \frac{\gamma_b}{2}\right)(z - z_+)(z - z_-)}, \end{aligned} \tag{J5}$$

where $z = e^{ik}$ and

$$z_\pm = \frac{-(\omega + i\gamma_b) \pm \sqrt{(\omega + i\gamma_b)^2 - 4J_1^2 + \gamma_b^2}}{-2\left(J_1 + \frac{\gamma_b}{2}\right)}. \tag{J6}$$

By implementing analytic continuation through contour deformation, we construct a mirage bath that replicates the emitter dynamics of the physical HN bath. The mirage Green function is written as $G_f(\omega) = 1/[\omega - \Delta' - \Sigma^f_0(\omega)]$, where the mirage self-energy $\Sigma^f_0(\omega)$ is obtained by deforming the integral contour of Eq. (J5) from $|z| = 1$ to $|z| = \sqrt{(J_1 - \gamma_b/2)/(J_1 + \gamma_b/2)}$. This yields

$$\begin{aligned} &\Sigma^f_0(\omega) \\ &= \Omega^2 \oint_{|z|=\sqrt{\frac{J_1 - \frac{\gamma_b}{2}}{J_1 + \frac{\gamma_b}{2}}}} \frac{dz}{2\pi i} \frac{1}{-\left(J_1 + \frac{\gamma_b}{2}\right) z^2 + (\omega + i\gamma_b) z - \left(J_1 - \frac{\gamma_b}{2}\right)} \\ &= \Omega^2 \oint_{|z|=1} \frac{dz}{2\pi i} \frac{1}{-\tilde{J}_1 z^2 + (\omega + i\gamma_b) z - \tilde{J}_1}, \end{aligned} \tag{J7}$$

where $\tilde{J}_1^2 = J_1^2 - \gamma_b^2/4$. The mirage-bath spectrum is then derived as

$$\omega^f_b(k) = 2\tilde{J}_1 \cos k - i\gamma_b. \tag{J8}$$

The corresponding virtual heat bath Hamiltonian is written as

$$H^f_b = \sum_j [\tilde{J}_1 b_j b^\dagger_{j+1} + \tilde{J}_1 b^\dagger_{j+1} b_j - i\gamma_b b^\dagger_j b_j]. \tag{J9}$$

### 2. Trefoil bath

The effective bath Hamiltonian of the Trefoil bath is written as

$$\begin{aligned} H'_b = &\sum_j (J_1 b^\dagger_{A,j} b_{B,j} + J_2 b^\dagger_{A,j} b_{A,j+1} + \text{H.c.}) \\ &+ \sum_j \left[\left(J_3 + \frac{\gamma_b}{2}\right) b^\dagger_{A,j} b_{A,j+3} + \left(J_3 - \frac{\gamma_b}{2}\right) b^\dagger_{A,j+3} b_{A,j} \right. \\ &\left. - i\gamma_b b^\dagger_{A,j} b_{A,j}\right], \end{aligned} \tag{J10}$$

which generates complex-energy bands given by

$$\begin{aligned} \omega'_b(k) = &J_2 \cos k + J_3 \cos 3k + i\frac{\gamma_b}{2} \sin 3k - i\frac{\gamma_b}{2} \\ &\pm \sqrt{\left(J_2 \cos k + J_3 \cos 3k + i\frac{\gamma_b}{2} \sin 3k - i\frac{\gamma_b}{2}\right)^2 + J_1^2}. \end{aligned} \tag{J11}$$

For a single emitter coupled to this bath, the self-energy $\Sigma_0(\omega)$ is associated with the Trefoil bath spectrum in Eq. (J11). We obtain

$$\begin{aligned} \Sigma_0(\omega) &= \Omega^2 \int_{-\pi}^{\pi} \frac{dk}{2\pi} \frac{\omega}{[\omega - \omega'_{b+}(k)][\omega - \omega'_{b-}(k)]} \\ &= \Omega^2 \oint_{|z|=1} \frac{dz}{2\pi i} \frac{\omega z^2}{\lambda(\omega, z)}, \end{aligned} \tag{J12}$$

where

$$\begin{aligned} \lambda(\omega, z) = &-\left(J_3 + \frac{\gamma_b}{2}\right)\omega z^6 - J_2 \omega z^4 + \left(\omega^2 - J_1^2 + i\gamma_b \omega\right) z^3 \\ &- J_2 \omega z^2 - \left(J_3 - \frac{\gamma_b}{2}\right)\omega. \end{aligned} \tag{J13}$$

By implementing analytic continuation through contour deformation, we obtain the mirage self-energy

$$\Sigma^f_0(\omega) = \Omega^2 \oint_{|z|=C_z} \frac{dz}{2\pi i} \frac{\omega z^2}{\lambda(\omega, z)}, \tag{J14}$$

where the integration contour $C_z$ is chosen to preserve the number of poles within the path. The spectrum of the mirage bath is derived by applying the elimination method detailed in Appendix H to the polynomial [Eq. (J13)].

#### 3. Nonreciprocal SSH bath with NNN coupling

The effective bath Hamiltonian of nonreciprocal SSH bath with NNN coupling is written as

$$H_b' = \sum_j (J_2 b_{B,j}^\dagger b_{A,j+1} + J_3 b_{A,j}^\dagger b_{B,j+1} + \text{H.c.}) + \sum_j \left[\left(J_1 + \frac{\gamma_b}{2}\right) b_{A,j}^\dagger b_{B,j} + \left(J_1 - \frac{\gamma_b}{2}\right) b_{B,j}^\dagger b_{A,j} - i\frac{\gamma_b}{2}(b_{A,j}^\dagger b_{A,j} + b_{B,j}^\dagger b_{B,j})\right], \tag{J15}$$

yielding complex-energy bands given by

$$\omega_b'(k) = -i\frac{\gamma_b}{2} \pm \sqrt{J_1 + \frac{\gamma_b}{2} + J_2 e^{-ik} + J_3 e^{ik}} \times \sqrt{J_1 - \frac{\gamma_b}{2} + J_2 e^{ik} + J_3 e^{-ik}}. \tag{J16}$$

We consider two QEs: the first QE is coupled to the sublattice $A$ at $j_1 = 0$, while the second QE is coupled to sublattice $B$ downstream at $j_2 = d \geqslant 0$. The total effective Hamiltonian is written as

$$H_{\text{eff}} = \Delta' \sum_{m=1}^{2} a_m^\dagger a_m + \Omega(b_{A,0}^\dagger a_1 + b_{B,d}^\dagger a_2 + \text{H.c.}) + H_b'. \tag{J17}$$

The dynamics of two QEs is determined by the two-emitter Green function $G_d(\omega) = 1/[\omega - \Delta - \Sigma_d(\omega)]$, which has the same form as the Eq. (A23). The self-energy matrix $\Sigma_d$ is characterized by the following elements:

$$\Sigma_0(\omega) = \Omega^2 \oint_{|z|=1} \frac{dz}{2\pi i} \frac{(\omega + i\frac{\gamma_b}{2})z}{\lambda(\omega, z)},$$
$$\Sigma_d^{AB}(\omega) = \Omega^2 \oint_{|z|=1} \frac{dz}{2\pi i} \frac{[(J_1 + \frac{\gamma_b}{2})z + J_2 + J_3 z^2] z^{-d}}{\lambda(\omega, z)},$$
$$\Sigma_d^{BA}(\omega) = \Omega^2 \oint_{|z|=1} \frac{dz}{2\pi i} \frac{[(J_1 - \frac{\gamma_b}{2})z + J_2 z^2 + J_3] z^{d}}{\lambda(\omega, z)}, \tag{J18}$$

where

$$\lambda(\omega, z) = -J_2 J_3 z^4 - \left[J_2\left(J_1 + \frac{\gamma_b}{2}\right) + J_3\left(J_1 - \frac{\gamma_b}{2}\right)\right] z^3 + \left[\left(\omega + i\frac{\gamma_b}{2}\right)^2 - J_1^2 + \frac{\gamma_b^2}{4} - J_2^2 - J_3^2\right] z^2 - \left[J_2\left(J_1 - \frac{\gamma_b}{2}\right) + J_3\left(J_1 + \frac{\gamma_b}{2}\right)\right] z - J_2 J_3. \tag{J19}$$

Similarly, by applying the elimination method detailed in Appendix H to $\lambda(\omega, z)$, we derive the spectrum of the mirage bath corresponding to the nonreciprocal SSH model with NNN coupling.